\documentclass[twocolumn]{aastex63}
\usepackage{amsmath, amssymb, graphics, bm, graphicx,color,ulem}
\usepackage{appendix}
\usepackage{verbatim}
\usepackage{xcolor}
\usepackage{hyperref}

\newcommand{\orcid}[1]{\href{https://orcid.org/#1}{\textcolor[HTML]{A6CE39}{\aiOrcid}}}

\definecolor{mygray}{gray}{0.5}
\definecolor{deepgreen}{RGB}{20, 130, 22}
\definecolor{peach}{RGB}{180,70,100}
\definecolor{red2}{RGB}{250,0,200}

\newcommand{\y}{\color{red2}}
\newcommand{\x}{\sout}
\newcommand{\w}{\color{blue}}

\newcommand{\mathsym}[1]{{}}
\newcommand{\unicode}[1]{{}}

\newcommand{\sg}{\sigma}
\newcommand{\ag}{\alpha}
\newcommand{\bg}{\beta}
\newcommand{\cg}{\gamma}

\newcommand{\dg}{\delta}
\newcommand{\Dg}{\Delta}
\newcommand{\eps}{\epsilon}
\newcommand{\kg}{\kappa}

\newcommand{\Om}{\Omega}
\newcommand{\om}{\omega}
\newcommand{\lam}{\lambda}
\newcommand{\Lam}{\Lambda}
\newcommand{\pd}{\partial}
\newcommand{\im}{{\rm i}}
\newcommand{\der}{\text{d}}

\newcommand{\cs}{c_{\rm s}}
\newcommand{\vphi}{\varphi}
\newcommand{\brunt}{Brunt-V\"ais\"al\"a }

\newcommand{\zg}{\zeta}

\newcommand{\br}{{\bm r}}

\newcommand{\kp}{k_\perp}

\renewcommand{\la}{\langle}
\newcommand{\ra}{\rangle}

\newcommand{\bxi}{{\bm \xi}}
\newcommand{\bdel}{{\bm \nabla}}

\newcommand{\bzg}{{\bm \zeta}}
\newcommand{\btau}{{\bm \tau}}
\newcommand{\bchi}{{\bm \chi}}

\newcommand{\hr}{{\bm {\hat r}}}

\newcommand{\Ms}{M_\star}
\newcommand{\Rs}{R_\star}
\newcommand{\Is}{I_\star}
\newcommand{\Ls}{L_\star}

\newcommand{\mus}{\mu_\star}

\newcommand{\kgs}{\kappa_\star}
\newcommand{\Oms}{\Omega_\star}

\newcommand{\Msun}{{\rm M}_\odot}
\newcommand{\Rsun}{{\rm R}_\odot}

\newcommand{\btimes}{{\bm \times}}
\newcommand{\bcdot}{{\bm \cdot}}

\newcommand{\xir}{\xi_r}
\newcommand{\xip}{\xi_\perp}

\newcommand{\cC}{\mathcal{C}}
\newcommand{\cR}{\mathcal{R}}
\newcommand{\cM}{\mathcal{M}}
\newcommand{\bF}{{\bm f}}

\newcommand{\Eorb}{E_{\rm orb}}
\newcommand{\dEorb}{\dot E_{\rm orb}}
\newcommand{\Jorb}{J_{\rm orb}}

\newcommand{\Es}{E_\star}
\newcommand{\kr}{k_r}
\newcommand{\Porb}{P_{\rm orb}}
\newcommand{\days}{{\rm days}}

\newcommand{\zero}{{(0)}}
\newcommand{\one}{{(1)}}

\newcommand{\e}{{\rm e}}

\newcommand{\be}{\begin{equation}}
\newcommand{\ee}{\end{equation}}
\allowdisplaybreaks

\graphicspath{{./}{figures/}}

\begin{document}

\title{Tidal Circularization of Binaries by Resonance Locking I: The Importance of the Pre-Main-Sequence}

\correspondingauthor{J.~J. Zanazzi}
\email{jzanazzi@cita.utoronto.ca, wu@astro.utoronto.ca}

\author[0000-0002-9849-5886]{J.~J. Zanazzi}
\affiliation{Canadian Institute for Theoretical Astrophysics, University of Toronto, 60 St. George Street, Toronto, ON M5S 3H8, Canada}

\author[0000-0003-0511-0893]{Yanqin Wu}
\affiliation{Department of Astronomy and Astrophysics, University of Toronto, Toronto, ON M5S 3H4, Canada}

\keywords{binary stars--- tides -- stellar oscillations -- Young Stellar objects -- stellar evolution}

\begin{abstract}
Although tidal dissipation in binary stars has been studied for over a century,  theoretical predictions  have yet to match the observed properties of binary populations.  This work quantitatively examines the recent proposal of tidal circularization by resonance locking, where tidal dissipation arises from  resonances between the star's natural oscillation frequencies and harmonics of the orbital frequency, and where resonances are `locked' for an extended period of time due to concurrent stellar evolution.  We focus on tidal resonances with axi-symmetric gravity-modes, and examine binaries with primary masses from one to two solar masses.  We find that orbital evolution via resonance locking occurs primarily during the star's pre-main-sequence phase, with the main-sequence phase contributing negligibly. Resonance locking, { ignoring non-linearity,} can circularize binaries with peri-centre distances out to  $\sim 10$ stellar radii,
corresponding to circular periods of $\sim 4-6$ days. 
{ However, we find resonantly excited gravity-modes will become nonlinear in stellar cores, which prevents them from reaching their full, linear amplitudes.  We estimate that such a `saturated resonance lock' reduces the circularization period by about a third, but resonance locking remains much more effective than the cumulative actions of equilibrium tides.  In a companion paper, we examine recent binary data to compare against theory.}
%which can aonly circularize binaries out to about half that distance.  
%While a circularization period of $4$ days falls much below values reported in literature (up to $\sim 15$ days),  we show 
\end{abstract}

\section{Introduction}

%{\y here is my borg number: ORCID: 0000-0003-0511-0893. do you have one? i really need it because of my chinese name.}  {\jj included them, but image is not displayed for some reason}

%{\jj I took out the discussion for tidal evolution in things other than binary stars.  The introduction we have is already pretty long, and discussing other things not only adds to the length of the introduction, but sidelines the motivation for studying tidal evolution in binary stars.}

How tidal dissipation drives orbital and rotational evolution in binary star systems has been studied for centuries \citep[e.g.][]{Darwin(1879)}, but its final resolution still eludes us \citep[for reviews, see][]{Mazeh(2008),Ogilvie(2014)}.

Tidal circularization is observed in binaries. Short period binaries tend to be more circular \citep[e.g.][]{MayorMermilliod(1984),DuquennoyMayor(1991),Latham92,Latham(2002),Mathieu2004}. The most systematic study to date is that by \citet{MeibomMathieu(2005)}, who found that as stars age from Pre-Main-Sequence (PMS) to late Main-Sequence (MS), the ``characteristic circularization period'' ($P_{\rm circ}$, below which orbits are mostly circular) generally increases with age. While binaries in young open clusters (less than a few hundred million years) are circularized out to $P_{\rm circ} \sim 8$ days, older clusters appear to exhibit longer circularization periods, perhaps out to $P_{\rm circ} \sim 15$ days. Unfortunately, the data used to establish such a trend are sparse. Each cluster contains only about a dozen or so spectroscopic binaries, making the measurement of circularization periods difficult and susceptible to sample variance. In a companion paper to the current one \citep{ZanazziWu(2020)}, we will re-examine the issue of circularization periods using the much expanded binary samples from  APOGEE and \textit{Kepler}.

%\x{Although the eccentricitites $e$ and orbital periods $\Porb$ of binary stars have been observationally constrained through various methods throughout the $20^{\rm th}$ century, came up with the first quantitative, emperical measure of tidal circualrizaiton in binary star populations.  Fitting a damped exponential to the $e$ and $\Porb$ data for binaries in stellar clusters, \cite{MeibomMathieu(2005)} were the first to calculate the circularization period $P_{\rm circ}$ of a binary population, which is the orbital period below which all binaries have nearly-circular orbits.}
%\cite{MeibomMathieu(2005)} where the first to quantitatively test tidal theory predictions on binary star populations, and calculated the circularization periods of eccentric binaries of eccentric binaries in clusters.  
%\x{\cite{MeibomMathieu(2005)} found binaries with ages less than a few hundred million years lied on circular orbits out to $P_{\rm circ} \approx 8 \ {\rm days}$, with older clusters exhibiting longer circularization periods with age, out to $P_{\rm circ} \approx 15 \ {\rm days}$ when the cluster age exceeds a few billion years.}
%Because the tidal theories of Zahn have great difficulty circularizing old ($> 1 \ {\rm Gyr}$) populations out to $P_{\rm circ} \approx 15 \ \days$ \citep[e.g.][]{ClaretCunha(1997),MeibomMathieu(2005),Mazeh(2008)}, this has been recognized as an outstanding problem in the field. 

%{\y data seems compelling, but little theoretical success}
%{\y Eq. tide and dynamical tide, why each failed}

A large number of theoretical studies have been undertaken to explain the \citet{MeibomMathieu(2005)} results, and all have difficulties accounting for the long circularization periods in old binaries. In the following, we will review a few of these theoretical studies that are influential and are relevant for our study. 
%{\y Mazeh(2008) section 4.5 is a nice and short review. I rewrote your earlier version because: eq. tide exists in both fully convective and radiative region, it's the zero freq. response; dynamical tide is a totally different matter, it's non-zero freq.; } {\jj 100\% agree with rewrites.}

%\x{have difficulty circularizing old ($> {\rm few} \ {\rm Gyr}$) binaries out to $P_{\rm circ} \approx 15 \ {\rm day}$, explaining the results of \cite{MeibomMathieu(2005)} has been recognized as an outstanding problem in astrophysics \citep[e.g.][]{ClaretCunha(1997),Mazeh(2008)}. 
Zahn and collaboraters \citep{Zahn(1966),Zahn(1989),ZahnBouchet(1989), Zahn(2008)} developed the theory of equilibrium tides, where the star is assumed to react hydrostatically to the binary tidal forcing, and a lag time (or angle) is introduced to account for the internal dissipation of this response. The dissipation leads to energy and angular momentum transfer between the binary orbit and the star.  This has been shown to be most effective in stars with surface convection zones where turbulent eddies can provide kinematic viscosity. 

% {\y NO. THERE CAN BE 0-FREQUENCY RESPONSE IN RADIATIVE ZONE TOO. IT'S AN APPROXIMATION. THERE IS NOTHING SPECIAL ABOUT BRUNT. WE CAN TALK ABOUT IT TO CLEAR IT UP.} {\y I TOOK A LOOK AT \citet{ClaretCunha(1997)} and DECIDE TO DISCARD IT. NOT MUCH VALUE and MOST LIKELY WRONG USE Of ZAHN.} {\jj OK, agree with deletions of these sentences}
 
\citet{ZahnBouchet(1989)} pointed out that, during the pre-main-sequence, this process is more efficient because stars are more distended and have more extensive convection zones.  They found that binaries can be circularized out to $P_{\rm circ} \sim 8$ days before they reach the main-sequence. While this appears to explain the circularization data for young clusters, later studies have called into question some of their assumptions 
(see Sec.~\ref{sec:EqTides} for a discussion).  In addition, Zahn's equilbrium tide theory has difficulties explaining the observed spin rates of some binary stars \citep{Meibom(2006),Marilli(2007),Lurie(2017),Mazeh(2008)}, an area not considered in this work.
However, the equilibrium tide theory does enjoy an undisputed success in the case of evolved stars.
Observations show that post-MS binaries are circularized out to $\sim 100 \ {\rm days}$ \citep[e.g.][]{MayorMermilliod(1984),Bluhm(2016),PriceWhelan(2020)}, in accord with predictions from Zahn's theory \citep{Verbunt,Price2}.

%\x{Zahn's tidal theories have difficulty explaining the observed rotation rates of binary star populations.  Rotation period measurements of stars in binaries do show most binaries with orbital periods less than 10 days rotate synchronously \citep[e.g.][]{Meibom(2006),Marilli(2007),Lurie(2017)}.  However, numerous binaries with ages longer than the synchronization time have significantly asynchronous rotation \citep[e.g.][]{Meibom(2006),Marilli(2007),Mazeh(2008)}.  Moreover, a number of non-synchronously rotating binaries have been observed, with orbital periods significantly less than 10 days \citep{Lurie(2017)}.}

%\x{ In contrast, because both low and high mass stars have similar strucures during their Pre-Main-Sequence (PMS) evolution, they both circularize out to $P_{\rm circ} \approx 8 \ {\rm day}$ due to equilibrium tidal dissipation \citep{ZahnBouchet(1989)}, although this process depends sensitively on the turbulent viscosity, whose effectiveness is still debated }

 %{\y Resonance locking, evolution of the theory}
 %{\y for solar-type, Witte did something (grey line in Meibom), but also explained/confirmed in Fuller+Lai, need comment}

Another parallel line of investigation is called the theory of dynamical tides.
While the equilibrium tide assumes the star reacts instantaneously to the forcing, this theory accounts for the tidal excitation of dynamical modes. In stars with radiative interiors, gravity-modes are the most relevant \citep[e.g][]{Zahn(1975), SavonijePapaloizou(1983), Terquem(1998), GoodmanDickson(1998)}, while inertial-modes may also play a role in stars that are rotating sufficiently fast \citep[e.g.][]{SavonijePapaloizou(1997),OgilvieLin(2007),IvanovPapaloizou(2007)}.
% {\y I used references from Mazeh} 
In particular, when the frequency of tidal forcing lies close to a natural oscillation frequency of the host (a tidal resonance), the tidal response is strong and its associated dissipation brings about rapid orbital evolution. However, because the resonance width is normally very narrow, the binary orbit quickly shrinks through a resonance, and  the latter  has little lasting dynamical impact \citep{Terquem(1998),GoodmanDickson(1998)}. %Dynamical tides has not found to be 
%significant for tidal dissipation in solar-type binaries 

To address this issue, a variant of the dynamical tide theory, `resonance locking,' has been developed. { \citet{SavonijePapaloizou(1983),SavonijePapaloizou(1984)}}; \citet{WitteSavonije(1999b),Fuller(2016),Fuller(2017)} suggested that, when the mode in resonance increases in frequency (due to evolution in either stellar structure or stellar spin), in conjunction with the orbital frequency, the resonance may be maintained for an extended period, during which the binary continue to ``surf'' the resonance and experience significant orbital evolution. It is argued to be important for both early  type
\citep{WitteSavonije(1999b),WitteSavonije(2001),FullerLai(2012a),Burkart(2012)} 
and late type \citep{WitteSavonije(2002),SavonijeWitte(2002)} main-sequence binaries.

A tidal resonance can be expressed as
\begin{equation}
k \Omega \approx \sigma = \omega + m \Oms\, ,
\label{eq:pre-empt}
\end{equation}
where $k \Omega$ is the $k$-order harmonic of the binary orbital frequency, and $\sigma$ the mode frequency in the inertial frame. The latter is related to the mode frequency in the star's rotating frame ($\omega$), the mode's azimuthal quantum number ($m$), and the star's spin rate ($\Oms$). Resonance locking requires that, as the orbit evolves, $k \dot \Omega \approx \dot \sigma$. The latter can be caused by stellar evolution ($\dot \omega$), or by changes in the host spin ($\dot \Omega_\star$), two possibilities addressed respectively by
\citet{Fuller(2016),Fuller(2017)} and \citet{WitteSavonije(1999b),FullerLai(2012b)}.

%{\y unique contribution of this work} 
Despite the promise of resonance locking, it is surprising that no quantitative works have been performed to study its efficacy in circularizing solar-type binaries, and to compare the predictions against observations. In this work and an associated paper, we perform exactly these tasks. We study resonance locking by tracking how a star's oscillation frequencies evolve over its lifetime \citep[similar to][]{Fuller(2016),Fuller(2017)}. For the first time, we follow the evolution both during the pre-main-sequence and the main-sequence.  We find that the pre-main-sequence phase plays a disproportionately large role, while the main-sequence phase contributes negligibly to tidal evolution. This is in stark contrast to the trend reported by \citet{MeibomMathieu(2005)}.

%where locks are sustained by the internal structure evolution of the host, as opposed to spin evolution of the host \citep[e.g.][]{WitteSavonije(1999b),FullerLai(2012b)}.
This result naturally stimulates us to re-examine the observed data. We are aided by the large influx of spectrocopic and transiting data \citep[e.g.][]{Milliman(2014),VanEylen(2016),Triaud(2017),Windemuth(2019),PriceWhelan(2020)}, and we present our results in Paper II \citep{ZanazziWu(2020)}. With these two papers, we hope to establish that: 1) resonance locking works primarily during the PMS for solar-type binaries, and 2) resonance locking during the PMS { may help} explain the observed circularization periods. 

{While this paper is under review, the  referee pointed out an important caveat in our work. The resonantly excited gravity-modes in our calculations may well reach such large amplitudes in the stellar radiative cores that the waves may overturn, a process that has been discussed by 
\citet{GoodmanDickson(1998),BarkerOgilvie(2010),BarkerOgilvie(2011),Su}. This is indeed born out by our estimates. As many of the gravity-modes that can resonantly lock onto the tidal perturbation are non-linearly damped and saturated at lower amplitudes compared to our linear results, the efficacy of resonance locking is called into question. We dedicate a new section to discuss this aspect. The overall picture remains, at this moment, not as clear as we have hoped.
}

%{\y organization}
The structure of this paper is as follows.  Section~\ref{sec:GMode_rev} reviews the properties of gravity-modes (g-modes) in stars, the formalism describing their tidal excitation. We also describe how we calculate the g-mode properties over the host star's lifetime.  Section~\ref{sec:ModeLock} discusses the conditions a binary must satisfy to maintain resonance lock, as well as the binary's orbital evolution under resonance locking.  Section~\ref{sec:Results} presents our results on the tidal evolution of binary star populations under resonance locking. Section~\ref{sec:discuss} confirms the un-importance of equilibrium tides, discusses the theoretical uncertainties of our analysis, and observational predictions. Section~\ref{sec:Conc} summarizes our work.

\section{Gravity Modes: Properties, Evolution, and Tidal Excitation}
\label{sec:GMode_rev}

Before examining how the evolution of stellar binaries can drive rapid tidal circularization through resonance locking, we will review the properties of tidally excited gravity modes (g-modes).  
Under a tidal forcing $-\bdel U(\br, t)$
produced by a binary companion of mass $\Ms'$ orbiting at a semi-major axis $a$ and eccentricity $e$, the displacement in the host star (with mass $\Ms$, radius $\Rs$ and rotation frequency $\Oms$), $\bxi(\br,t)$ evolves as
\be
\frac{\pd^2 \bxi}{\pd t^2} + \mathcal{R} \bcdot \bxi  = {\bF}   -\bdel U,
\label{eq:bxi_ev}
\ee
where $\mathcal{R}$ is an operator acting on $\bxi$  due to restoring force inside the star, and $\bF$ represents any (weak) frictional force.  We restrict ourselves to gravity-modes, where the only contribution to $\mathcal{R}$ is the buoyancy force. It is reasonable to ignore pressure-modes (restored by pressure) as their eigenfrequencies lie well above the tidal forcing frequency.  We also assume that the tidal response is linear, that the host star's spin axis is aligned with the binary's orbital plane, and that tidal dissipation in the companion star is negligible.  We further assume the stellar spin rate is much lower than the g-mode  oscillation frequencies  ($\Oms \ll \om_\ag$) so we can ignore the Coriolis force. We also  ignore another branch of oscillations, the so-called inertial modes. The validity of these approximations will be discussed in Section~\ref{sec:discuss}.

The above equation simplifies for adiabatic, free  oscillations into
\be
{\w -}\om_\ag^2 \bxi_\ag + \mathcal{R} \bcdot \bxi_\ag = 0\, ,
\label{eq:bxi_ag}
\ee
where $\omega_\ag$ is the frequency for eigen-mode $\ag$.  We can express the effect of the weak friction force ${\bm f}$ by a damping rate, using the so-called quasi-adiabatic approximation \citep[see e.g.][]{FriedmanSchutz(1978a),FriedmanSchutz(1978b),Christensen-Dalsgaard(2014)}
\begin{equation}
    \gamma_\ag =  
- \frac{\im}{2\om_\ag} \frac{\la \bxi_\ag | \bF \, \e^{\im  \omega_\ag t}\ra}{\la \bxi_\ag | \bxi_\ag \ra }
\, .
\label{eq:cg_ag}
\end{equation}
 where $\bxi_\ag$ is the adiabatic eigenfunction from equation~\eqref{eq:bxi_ag}. Here, $\gamma_\ag > 0$ stands for the rate of energy (not amplitude) dissipation, and $\gamma_\ag \ll \omega_\ag$. The bracket stands for the volume integral, $\la {\bm u} | {\bm v} \ra \equiv \int {\bm u}^* \bcdot {\bm v} \rho \der V$, and the factor of $2$ arises when one converts the energy damping rate to amplitude damping rate. % {\y I am a bit fuzzy about this  -- do we need  the factor of $2$? is the text correct?} {\jj Yes, this is correct, and worded better than I would have worded it.}

Since the tidally forced perturbation is assumed to be linear, we can decompose the total displacement into a  sum over adiabatic eigenfunctions $\bxi_\ag$,
\begin{equation}
    {\bxi (\br,t)} = \sum_\ag c_\ag(t)\, \e^{-\im  \omega_\ag t}\,  {\bxi_\ag (\br)}\, ,\label{eq:decomp}
\end{equation}
where the mode amplitude $c_\ag$ varies over a timescale much longer than the mode frequency. As a result, we can approximately express
\begin{equation}
\dot \bxi(\br,t)
\simeq \sum_\ag - \im \omega_\ag \, c_\ag(t) \e^{-\im \omega_\ag t}  \bxi_\ag(\br)\, .
\label{eq:bxi_exp}
\ee

Substituting these above expressions into equation~\eqref{eq:bxi_ev}, we obtain the  amplitude equation that describes the evolution of $c_\ag(t)$ \citep[e.g.,][]{Schenk(2002),Burkart(2014),Fuller(2017)}
%{\y may need other references? I got different signs from you. in particular, no factor of $2$ in rhs, but factor of 2 for $\gamma$. The latter can be removed if you remove the $2$ in eq. 4. re-check?}
%{\jj The original expression is correct, derived in new appendix}
\begin{equation}
\dot c_\ag + (\im \om_\ag + \cg_\ag)  c_\ag = \frac{\im}{2 \om_\ag} \frac{\la \bxi_\ag | -\bdel U \, \e^{\im \omega_\ag t} \ra}{\la \bxi_\ag | \bxi_\ag \ra} \, .
\label{eq:c_ag_ev}
\end{equation}
We review the derivation of equation~\eqref{eq:c_ag_ev} in Appendix~\ref{app:Schenk_rev}.  In particular, when the Coriolis force  is also present, orthogonality between eigenmodes is called into question. However, equations \eqref{eq:c_ag_ev} and \eqref{eq:cg_ag} remain valid.

The solution to this in-homogeneous equation describes the evolution of the tidal response, and through it, the tidal evolution of the binary orbit. In the following, we consider the quantities $\omega_\ag$, $\gamma_\ag$ and the tidal overlap integral (right-hand-side of eq.~[\ref{eq:c_ag_ev}]), separately. In particular, since we are interested in tidal evolution that proceeds in the same timescale as stellar evolution,  we pay special attention to how these quantities change as the star ages. 

In the following, we suppress the mode index $\ag$ for notational simplicity.

\subsection{Gravity Mode Frequencies}

 This section calculates the g-mode eigen-frequencies for our stellar models (see eq.~[\ref{eq:bxi_ag}]). Since this exercise is  standard, we move most of the relevant algebra 
to Appendix~\ref{sec:gmodes}, and only list a few relevant results here. Again, we ignore the effects of rotation (Coriolis force and centrifugal force), except the effect of mode frequency shifting by the rotating frame.

 Of relevance for later discussions is the structure of a gravity mode: it propagates (is oscillatory) in the star's radiative region where its frequency lies below the local \brunt frequency $N$, and  is evanescent  in the convective region. For modes of high radial order ($n \gg 1$), the local dispersion relation (eq.~[\ref{eq:kr_rad}]) can be integrated across the star to yield a global dispersion relation
\be
\om \simeq \frac{\sqrt{\ell(\ell+1)}}{\pi n} \int_0^{\Rs} \frac{N(r)}{r} \der r,
\label{eq:om_WKB2}
\ee
where $\omega$ is mode frequency in a non-rotating frame, $n$ is the number of zero nodes in its radial displacement, and $\ell$ refers to the spherical degree when the angular dependence is described by the spherical harmonics function ($Y_{\ell m}$).

To normalize the mode eigenfunction, we adopt \citep[e.g.][]{Weinberg(2012)}
\be
2\la \bxi| \bxi\ra \equiv 2 \int_0^{\Rs} \left[ \xi_{r}^2 + \ell(\ell+1) \xi_{\perp}^2 \right] \rho r^2 \der r = 1\, .
\label{eq:bxi_norm2}
\ee
Here, $\xi_r$ and $\xi_\perp$ are the radial and horizontal displacements, respectively.  Under this normalization, the displacements do not have the dimension of length, and the mode energy  (in the star's rotating frame) is $E = \epsilon = \omega^2$. 
%{\y we use $\epsilon$ below differently, I think} {\jj No, $\eps$ is used consistently throughout this work.}

\begin{figure}
\begin{center}
\includegraphics[scale=0.73]{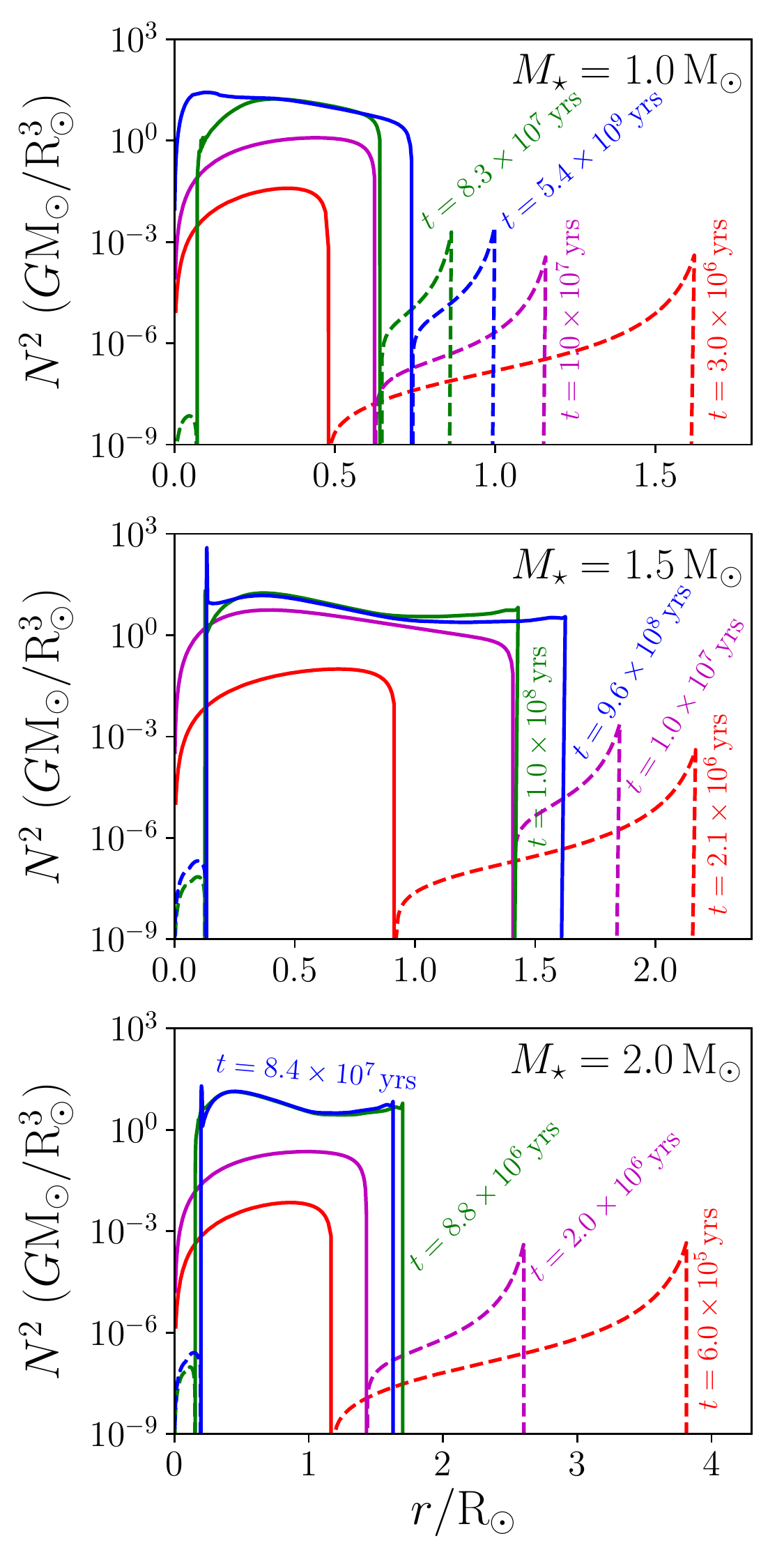}
\caption{
%{\y first the main purpose of the figure} 
Internal stratification inside stars, as they age through the pre-main-sequence and main-sequence phases.
Positive (solid) and negative (dashed) values of the Brunt-V\"ais\"al\"a frequency $N^2$ are plotted as functions of radius $r$, for the stellar masses $\Ms$ and ages $t$ indicated.  G-modes propagate where $\om^2 \leq N^2$. 
 \label{fig:N2_ev}
}
\end{center}
\end{figure}

\begin{figure}
\begin{center}
\includegraphics[width=\columnwidth]{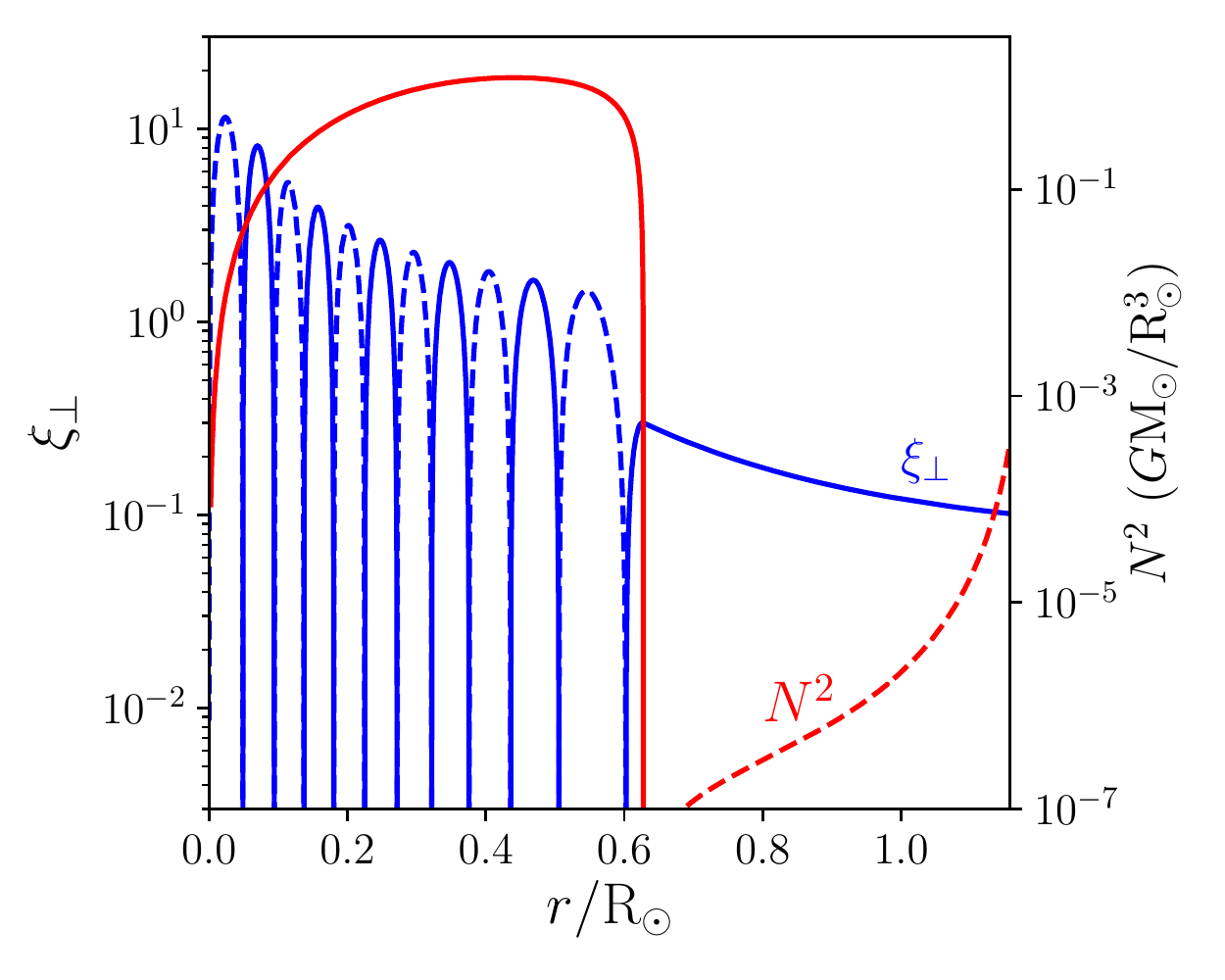}
\caption{The radial dependence of a $\ell=2, n=11$ g-mode's horizontal displacement ($\xi_{\perp}$)  in a $\Ms = 1.0 \, \Msun$ pre-main sequence star (age $1.04 \times 10^7\, {\rm yrs}$) is shown in blue. The \brunt frequency $N^2$  is shown in red, with positive (negative) values denoted by solid  (dashed) curves.  The magnitude scale for the horizontal displacement $\xip$ is  arbitrary.
The g-mode propagates where $\omega^2 \leq N^2$, and it becomes evanescent towards the bottom of the convection zone.  {Here, the g-mode frequency is $\om = 0.115 \, (G \Msun/\Rsun^3)^{1/2} = 0.99 \, (2\pi/{\rm day})$.}  The displacement amplitude remains roughly constant in the convection zone.  As the mode propagates into the stellar inner core, its amplitude rises and its wavelength shortens.
\label{fig:Gmode_prof}
}
\end{center}
\end{figure}

%\subsection{Gravity Mode Evolution %with Stellar Age}
%\label{sec:Gmode_evol}

As a star ages and its stratification changes, its g-mode frequencies also evolve.  Figure~\ref{fig:N2_ev} exhibits the radial profiles for the \brunt  frequency ($N^2$) for three different stellar masses, at different stages of the star's Pre-Main Sequence (PMS) and Main Sequence (MS) evolution.  During the PMS, which lasts from a few to a few tens of Myrs (see Fig.~\ref{fig:ModeTracks}), all stellar types have radiative cores and convective envelopes, with the magnitudes of $N^2$
roughly independent of the star's mass. 
The magnitude of  $N^2$ in the radiative regions increases with time as the star contracts and its core becomes more stably stratified. 
%{\jj deleted problematic opacity sentence}
This causes the g-mode frequencies to increase significantly during the PMS (eq.~[\ref{eq:om_WKB2}]).

After stars reach the MS, their structures diverge, with the surface convection zone disappearing in the more massive models ($\Ms \gtrsim 1.4 \ \Msun$). Over much of the MS, the magnitudes of $N^2$ in the radiative regions rise only mildly (if at all), in contrast to the strong increase in the PMS  phase. This leads to a slight increase in g-mode frequencies.

The radial structure of a representative g-mode is plotted in Figure \ref{fig:Gmode_prof},  where we see that the g-mode has an `evanescent tail' in the surface convection zone ($r \gtrsim 0.6 \, \Rsun$), and its horizontal displacement rapidly oscillates in the interior. The former is important for coupling to the smooth tidal potential, while the latter is relevant for the dissipation of these modes.

\subsection{Damping Rates}
\label{sec:DampRates}

Here, we evaluate how fast the mechanical energy from g-modes can be converted into heat (mode damping rates).
The dissipation causes the tidal displacement to have a phase lag relative to the tidal forcing, exerting a back-reaction torque on the perturber, which in turn drives the tidal evolution of the binary system.  

Dissipation of gravity-modes proceeds through three possible channels: turbulent damping inside the convection zone, radiative damping caused by photon diffusion, and last, the leakage of mode energy from the upper atmosphere. The detailed expressions for these rates are enumerated in Appendix \ref{sec:gamma}, where we also obtain order-of-magnitude estimates. Here, we present these estimates and compare them against numerical results (Fig. \ref{fig:ModeProps}).

Because a g-mode propagates almost adiabatically across the star, over (under) densities are hot (cold) spots relative to the stellar background.  Radiative diffusion works to diffuse these temperature inhomogeneities, damping the mode at a rate
\be
\cg_{\rm diff} \sim \frac{\Ls}{\Ms \Rs^2 \om^2} \Lam_{\rm core},
\label{eq:cg_diff_est}
\ee
where 
\be
\Lam_{\rm core} = \frac{\Ms \Rs^2}{\left( {\rho H r^4} \right)_{r=r_{\rm c}}}
\ee
parameterizes the characteristic mass concentration in the star's radiative core (evaluated at the characteristic radius $r_{\rm c}$, see Appendix \ref{subsec:radiative}), and $\Ls$ is the stellar luminosity.  The central concentration enters the estimate because the g-mode energy is mostly removed by radiative diffusion in the stellar centre, where its wavelength is short and its displacement is large.  The parameter $\Lam_{\rm core}$ increases with age as the star becomes more centrally-concentrated.  For a solar-mass star, $\Lam_{\rm core}
 \sim 10^2 - 10^3$ during the PMS, and $\sim 10^3 - 10^4$ during the MS. 
As a result, the damping rate increases as the star evolves from the PMS to the MS.
Examining the top panel of Figure~\ref{fig:ModeProps}, we see estimate~\eqref{eq:cg_diff_est} does well at reproducing the scaling of $\cg_{\rm diff}$ with frequency, as well as its dependence on stellar age.  Higher frequency modes have lower radial order and longer wavelengths. They suffer significantly weaker radiative dissipation.
%Because the WKB envolope of the g-mode peaks strongly near the stellar center ($r \lesssim 0.1 \, \Rsun$), we expect our estimate to require an enhancement of order $\bar \cg_{\rm diff} \sim 10^2- 10^3$ to match the numerically-computed $\cg_{\rm diff}$ values.  The top panel of Figure~\ref{fig:ModeProps} displays our results for $\cg_{\rm diff}$.  Although estimate~\eqref{eq:cg_diff_est} predicts the scaling of $\cg_{\rm diff}$ with frequency $f = \om/2\pi$ well, it is unable to accurately predict the magnitude of $\cg_{\rm diff}$ due to the variation of the core's stratification along the PMS and MS evolution.

Turbulent damping arises from convective eddies interacting viscously with the g-mode, causing a dissipation rate of order (see eq.~[\ref{eq:cg_turb_est_app}])
\begin{equation}
\gamma_{\rm turb} \sim 
\frac{\ell^2 }{3n\om} \left( \frac{z_{\rm cvz}^{2/3} \Ls^{2/3}}{ \Rs^2 M_{\rm cvz}^{2/3}} \right)
\, ,
    \label{eq:cg_turb_est}
\end{equation}
%{\y I changed eq. a bit. check to see if agree} {\jj yup agree with change}
where $z_{\rm cvz}$ and $M_{\rm cvz}$ are the depth and mass of the surface convection zone, respectively. The bottom panel of Figure~\ref{fig:ModeProps} compares this estimate with our numerical results and shows reasonable agreement. Compared to the radiative damping rates, the turbulent damping rates decrease mildly as the star ages, and do not depend sensitively on mode frequency.

The g-mode perturbation may propagate in a thin region above the photosphere if the mode 
 frequency lies below the so-called ``acoustic cut-off frequency.''  %\x{This process causes a small amount of the g-modes energy to leak out of the star, damping the g-mode at a rate of order
%\begin{equation}
%    \gamma_{\rm leak} \sim \frac{\om}{n \ell} \times \left( \frac{M_{\rm ph}}{M_{\rm cvz}} \right) \times \left( \frac{\om}{\om_{\rm ac,2}} \right),
%a    \label{eq:cg_leak_est}
%\end{equation}
%where $M_{\rm ph}$ is the mass in the star's photosphere, $\om_{\rm ac,2} = \Gamma_1 \cs/(2 \Rs)$ is the acoustic cutoff frequency, $\Gamma_1 = (\pd \ln p/\pd \ln \rho)_S$ is the adiabatic exponent, and $\cs^2 = (\der p/\der \rho)_S$ is the adiabatic sound speed.  For the g-modes of interest (periods of order a few days),$\cg_{\rm leak} \ll \cg_{\rm diff}, \cg_{\rm turb}$, and is negligible.}
{Our  calculation in Appendix~\ref{subsec:leakage} shows this to be negligible compared to other sources of  damping,  for g-modes with  periods of order a few days.}

Comparing the top and bottom panels of. Figure~\ref{fig:ModeProps}, we see radiative diffusion dominates over turbulent viscosity ($\cg_{\rm diff} \gtrsim \cg_{\rm turb}$) over the PMS and MS evolution of the star, with the exception of modes of the lowest radial orders ($n \lesssim 10$).  For all g-modes of concern, mode dissipation is weak, relative to the spacing between consecutive g-modes ($\gamma \ll \omega/n$).  Therefore, our assumption of individual, discrete modes is appropriate \citep{GoodmanDickson(1998)}.

\begin{figure}
\begin{center}
\includegraphics[width=\columnwidth]{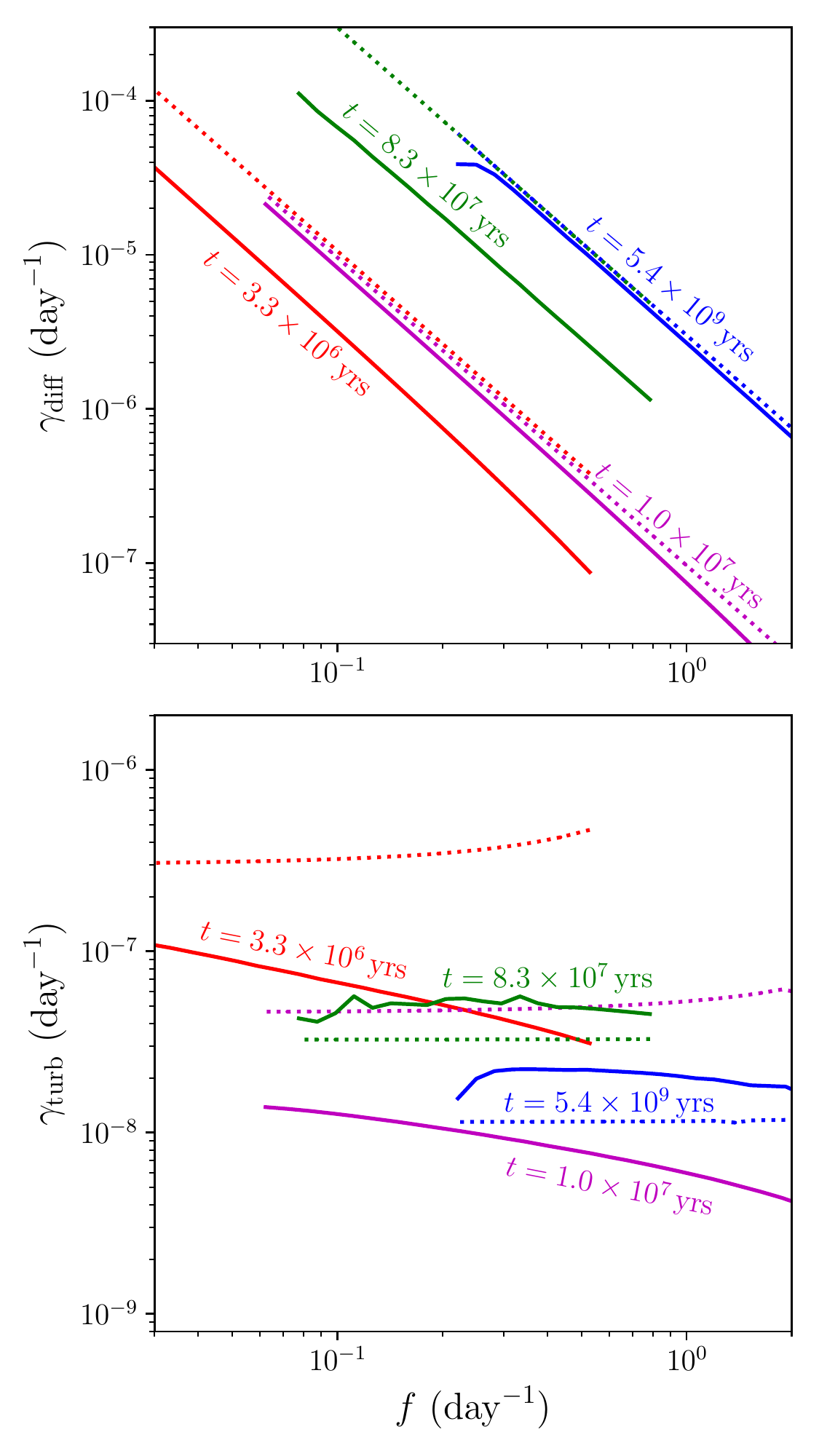}
\caption{G-mode damping rates as functions of mode frequencies ($f = 1/P$), in a sun-like star at the ages as indicated. 
Here $\ell=2$, and we consider g-mode frequencies with radial nodes $n=3-200$ during the PMS ($t \lesssim 4 \times 10^7 {\rm yrs}$), and $n=50-500$ during the main-sequence ($t \gtrsim 4\times 10^7 \, {\rm yrs}$).
The top panel shows damping by radiative diffusion, and the lower panel that by turbulent viscosity.
The numerical results are in solid lines, while our analytical estimates (eqs.~[\ref{eq:cg_diff_est}] \&~[\ref{eq:cg_turb_est}]) are in dotted lines. 
%from radiative diffusion $\cg_{\rm diff}$ (top panel) and turbulent viscosity $\cg_{\rm turb}$ (bottom panel), for our \texttt{MESA} solar-type star model at the ages indicated.  
%We use equations~\eqref{eq:cg_diff_num} and~\eqref{eq:cg_turb_num} for our numerical calculations of the damping rates, with 
We have adopted $\Lam_{\rm core} = 300$ ($3000$) during the PMS (MS), respectively, when evaluating equation~\eqref{eq:cg_diff_est}. 
%{\w interpretation: during PMS, modes are more weakly damped (?) but more strongly coupled to the tide; 
%jaggedness of the last two models arise from heavy cancellation?}{\y figure caption to be edited after resolving differences}
%{\jj Just looked at these values, and have $\om/n \gtrsim 10^{-3} \ {\rm day}^{-1}$ for all frequencies investigated here, so can safely ignore continuum effects from Goodman \& Dickson (1998).  It is a bit messy to include $\om/n$ in these plots, because need to greatly extend y-axis}
\label{fig:ModeProps} }
\end{center}
\end{figure}

\begin{figure}
\begin{center}
\includegraphics[width=\columnwidth]{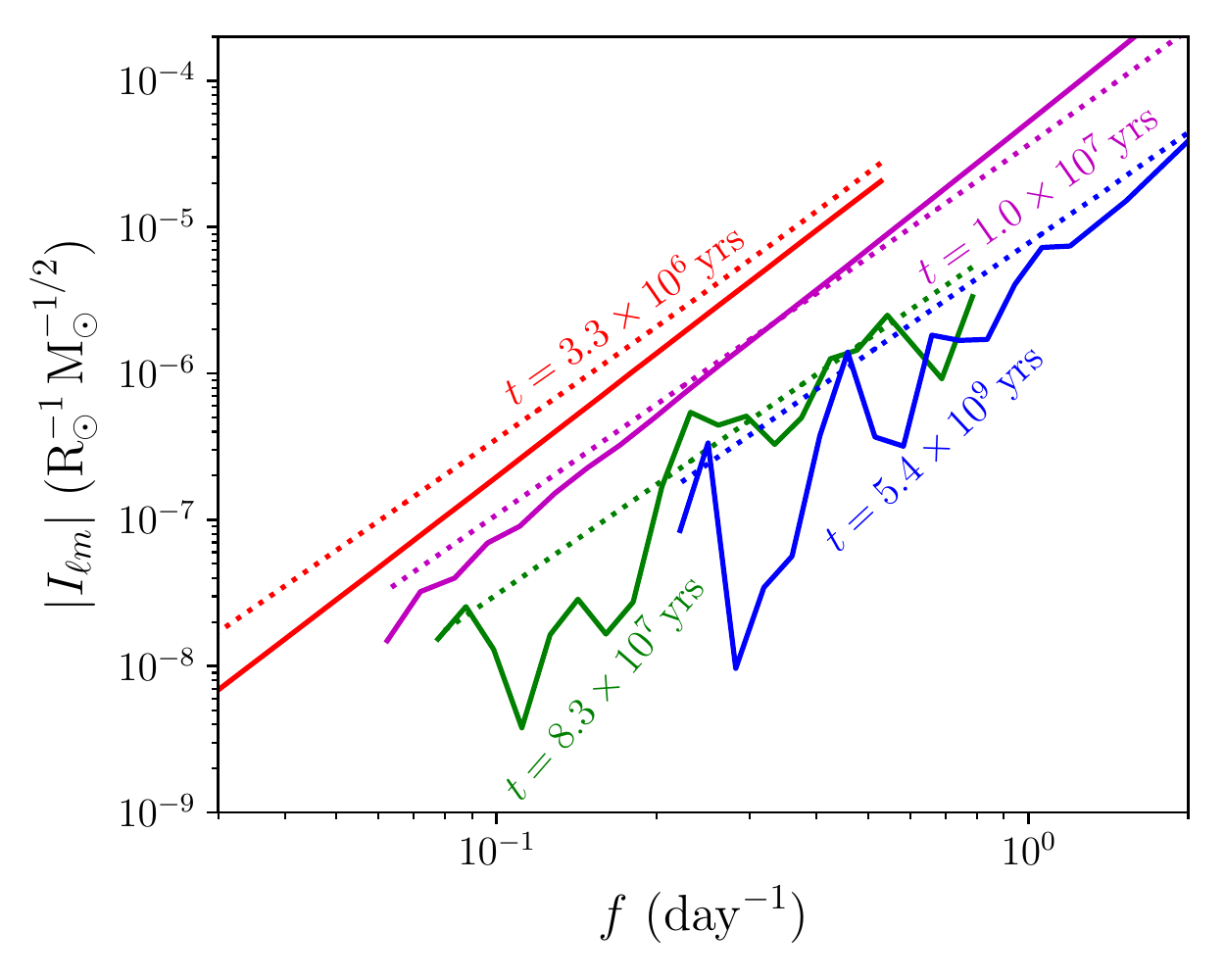}
\caption{
Numerical results (solid lines) and analytic estimates (dotted lines, equation~\ref{eq:I_est}) of the tidal overlap integral $I_{\ell m}$ (eq.~[\ref{eq:I_alm}]), as functions of the g-mode frequency $f = P^{-1} = \om/2\pi$, for a Sun-like star at different ages. 
%\x{ with $\bar \cg_{\rm rd} = 0.007$. } {\y you mean ${\bar I}=10^{-2}$? may neglect here}  
Here $\ell=2$, and we consider g-modes with radial order $n=3-200$ during the PMS,
($t \lesssim 4 \times 10^7 {\rm yrs}$), 
and $n=50-500$ during the MS. ($t \gtrsim 4\times 10^7 \, {\rm yrs}$).
%{\y simple interpretation}
We see that tidal coupling decreases steeply with radial order, and at the same period, tidal coupling is the largest during the PMS when the surface convection zone is deeper.
\label{fig:I_lm} 
}
\end{center}
\end{figure}

\subsection{Tidal Excitation}

The tidal potential from the binary companion acting on the host star may be decomposed into spherical harmonics and a Fourier series as \citep{PressTeukolsky(1977)}
\begin{align}
U(\br,t) = \ &-\frac{G \Ms'}{a} \sum_{\ell=2}^\infty \sum_{m=-\ell}^\ell \left( \frac{r}{a} \right)^{\ell+1}  W_{\ell m}
\nonumber \\
&\times \sum_{k=-\infty}^{\infty} X_{k \ell m}(e) \e^{-\im \om_{km} t} { Y_{\ell m}(\theta, \vphi)}\, ,
\label{eq:U_tide}
\end{align}
where
\begin{align}
    &W_{\ell m} = \frac{4\pi}{2\ell + 1} Y^*_{\ell m} \left( \frac{\pi}{2},0 \right)\, , \\
    &X_{k \ell m}(e) = \frac{1}{\pi} \int_0^\pi \frac{\der E}{(1 - e \cos E)^\ell}
    \nonumber \\
    &\times \cos \left[ k (E - e \sin E) - 2 m \tan^{-1} \left( \sqrt{ \frac{1+e}{1-e} } \tan \frac{E}{2} \right) \right]\, , \\
    &\om_{km} = k \Om - m \Oms\, ,
    \label{eq:om_km}
\end{align}
and $\Om = \sqrt{G(\Ms + \Ms')/a^3}$ is the orbital frequency/mean-motion.  From now on, sums over $k$ will implicitly be from $-\infty$ to $\infty$, $\ell$ from $2$ to $\infty$, and $m$ from $-\ell$ to $\ell$, unless otherwise noted. 
%\x{ See \cite{PressTeukolsky(1977)} for $W_{\ell m}$ expressed in terms of $\ell$,$m$.} {\y isn't it all defined here? something else missing?}  
We focus on the dominant $\ell = 2$ component in our numerical results, but the following discussion applies to all $\ell$ values.

%{\y a whole section here moved forward. commented out now}
%{To calculate how the star's g-modes evolve under external forcing, we review the derivations given by \cite{FriedmanSchutz(1978a),FriedmanSchutz(1978b),Schenk(2002)}.  Under the tidal forcing $-\bdel U$, the host star's displacement $\bxi(\br,t)$ evolves as
%\be
%\frac{\pd^2 \bxi}{\pd t^2} + %\mathcal{C} \bcdot \bxi = -\bdel U,
%\label{eq:bxi_ev}
%\ee
%where $\mathcal{C}$ is an operator acting on $\bxi$ which incorporates the restoring force of buoyancy.  Without the tidal forcing, (time independent) eigenvectors $\bxi_\ag(\br)$ with eigenfrequencies $\om_\ag$ satisfy
%\be
%\om_\ag^2 \bxi_\ag + \mathcal{C} %\bcdot \bxi_\ag = 0,
%\label{eq:bxi_ag}
%\ee
%which is explicitly decomposed in terms of $\xi_{r;\ag}$ and $\dg p_\ag/\rho = \om_\ag^2 r \xi_{\perp;\ag}$ in equations~\eqref{eq:dxirdr} and~\eqref{eq:dhdr}.  To calculate $\bxi(\br,t)$, we perform the phase-space expansion of \cite{Schenk(2002)}:
%\be
%\left[\begin{array}{c}
%\bxi(\br,t) \\
%\dot \bxi(\br,t)
%\end{array} \right]
%= \sum_\ag c_\ag(t)
%\left[ \begin{array}{c}
%\bxi_\ag(\br) \\
%-\im \om_\ag \bxi_\ag(\br)
%\end{array} \right].
%\label{eq:bxi_exp}
%\ee}
%{\y It's orthogonality + $c$ changes much slower than $\omega$. Not complicated, except when there is Coriolis.}
%\x{Using subtle arguments related to the orthogonality of eigenvectors of equation~\eqref{eq:bxi_ag} (including Coriolis forces), equation~\eqref{eq:bxi_exp} may be substituted into~\eqref{eq:bxi_ev} to obtain the amplitude evolution $c_\ag(t)$  \citep{Schenk(2002)}:}

Using this decomposition of the tidal potential, equation~\eqref{eq:c_ag_ev} is recast as
\begin{equation}
\dot c + (\im \om + \cg) c = \im \om \sum_{k \ell m} U_{k \ell m} \e^{-\im \om_{km} t} \, ,
\label{eq:c_ag_ev2}
\end{equation}
where  
\begin{align}
U_{k \ell m} &= \frac{\Ms'}{\Ms} \left( \frac{\Rs}{a} \right)^{\ell + 1} \left( \frac{\Es}{\eps} \right) W_{\ell m} I_{\ell m} X_{k \ell m} \, ,
\label{eq:U_aklm} \\
I_{\ell m} &= \frac{\la \bxi | \bdel(r^\ell Y_{\ell m}) \ra}{\Ms \Rs^\ell} 
= \frac{1}{\Ms \Rs^\ell} \int \dg \rho \, r^{\ell + 2} \der r \, .
\label{eq:I_alm}
\end{align}
Here, $\Es = G \Ms^2/\Rs$,  and $\gamma$ is the mode damping rate  (Sec.~\ref{sec:DampRates}).

The in-homogeneous solution to equation~\eqref{eq:c_ag_ev2} is
\be
c(t) = \sum_{k \ell m} \frac{\om}{\om - \om_{km} - \im \cg} U_{k \ell m} \e^{-\im \om_{km} t}\, .
\label{eq:c_ag_inhomo}
\ee
This solution assumes that changes in the frequencies $(\om,\om_{km},\cg)$ are sufficiently slow that the mode amplitude $c(t)$ can respond adiabatically. On the other hand, the dissipation needs to damp the mode amplitude quickly, compared with variations in $(\om,\om_{km},\cg)$, so that no prior information is retained during the evolution of $c(t)$. In Appendix~\ref{app:adiabatic}, we further justify the assumption of adiabaticity for our problem, and argue for adiabadicity conditions contrary to that suggested in
\citet{Burkart(2014)},  who implicitly assumed negligible damping rates in their derivation.
%{\y what is $t$? can this be compressed w/ the previous sentence? if it's hard to say clearly, maybe leave it out all together} {\jj let me know if this is clearer, otherwise fine with removing statement}
%does not apply to binaries circularizing via resonance locking (and most astrophysical applications of interest).}
Adiabaticity guarantees the success of resonance locking, a central topic in this work.

Near a resonance ($\om \approx \om_{km}$), one tidal component $(k,m)$ dominates, and solution~\eqref{eq:c_ag_ev} simplifies to
%{\y not sure why repeat}  {\jj Define $C$ for $C^{\rm max}$ below, because not clear if the maximum of c, which is the sum of all eigenmodes, has a maximum right at resonance}
\be
c(t) 
\simeq C(t)
\equiv 
\frac{\om}{\om - \om_{km} - \im \cg} U_{k \ell m} \e^{-\im \om_{km} t}.
\label{eq:c_aklm}
\ee
%{\y is $c$ capital or lower case? check consistency} {\jj capital, changed, thanks for the catch}
The maximum amplitude %$c_{\ag k \ell m}^{\rm max}$
of $C$ occurs at exact resonance ($\om = \om_{km}$):
\be
C^{\rm max} = \im \frac{\om}{\cg} U_{k \ell m} \e^{-\im \om_{km} t}.
\label{eq:c_agklm^max}
\ee

%{\y mode doesn't have complext conjugate. rephrase?}
Each eigen-mode has energy $E_{\rm mode}$ and angular momentum $J_{\rm mode}$ in the inertial frame given by \citep{FriedmanSchutz(1978a),FriedmanSchutz(1978b),Burkart(2014),Fuller(2017)}
\be
E_{\rm mode} = \frac{\sg \eps}{\om} |c|^2,
\hspace{5mm}
J_{\rm mode} = \frac{m \eps}{\om} |c|^2,
\label{eq:EJmodes}
\ee
where
\be
\sg = \om + m \Oms
\ee
is the mode frequency in the inertial frame. The mode energy and angular momentum also attain their maximum values at exact resonance:
\begin{equation}
E_{\rm mode}^{\rm max} = \frac{\eps \sg \om}{\cg^2} |U_{k \ell m}|^2\, ,
\hspace{5mm}
J_{\rm mode}^{\rm max} = \frac{m \eps \om}{\cg^2} |U_{k \ell m}|^2\, .
\label{eq:EJ}
\end{equation}

Numerically, the overlap integral $I_{\ell m}$ is delicate to calculate. To confirm numerical results, we estimate its magnitude here.  Because the Eulerian density perturbation $\dg\rho$ oscillates rapidly in a mode's propagating region, most of the contribution to the overlap integral arises from the evanescent zone, or the surface convection zone:\footnote{Here, we consider low-frequency modes which propagate to the bottom of the convection zone.}
\begin{equation}
    I_{\ell m} = \frac{1}{\Ms \Rs^\ell} \int \delta \rho \, r^{\ell+2} \der r \sim \frac{1}{\Ms} \int \left(\frac{\delta\rho}{\rho}\right) \rho r^2 dr  \, .
\end{equation}
 The Eulerian density perturbation is as expressed in equation~\eqref{eq:useful},
\begin{equation}
\frac{\delta\rho}{\rho} = \frac{r \omega^2}{\cs^2} \xi_\perp  + \frac{N^2}{g} \xi_r \simeq \frac{r \omega^2}{\cs^2} \xi_\perp \, ,
\label{eq:drho}
\end{equation}
where $\xi_\perp (\xi_r)$ is the tangential (radial) displacement  (see Appendix \S \ref{sec:gmodes}) and we have taken $N^2 \approx 0$ in the convection zone. Since $c_s^2 \sim g z \propto z$ near the surface, the integrand is the largest at the  radiative-convective boundary, or 
\begin{equation}
I_{\ell m} \sim \bar I \frac{M_{\rm cvz}}{M_\star} \left(\frac{r \omega^2 \xi_{\perp}}{c_{\rm s}^2}\right)_{\rm cvz}\, ,
    \label{eq:I_est}
\end{equation}
where quantities within the brackets are to be evaluated at the bottom of the convection zone. Furthermore, we provide an estimate for $\xi_\perp$ for our adopted normalization in equation~\eqref{eq:xihnorm}.
We have also inserted a numerical fudge-factor $\bar I$. As Figure \ref{fig:I_lm} shows, our numerical results agree with this analytical estimate, including its dependencies on stellar structure and mode frequency, as long as we adopt a constant fudge factor of  $\bar I \sim 10^{-2}$ ($\bar I = 0.007$ for this example).  
 %{\y interior cancellation doesn't matter, already discounted}

This expression indicates that the overlap is reduced at higher mode order $n$, and for stars with shallower convection zones. In particular, for stars without a surface convection zone ($\Ms \gtrsim 1.4 \, \Msun$ on the MS), the overlap is entirely contributed by the thin region above the first radial node (not accounted for here) and is very small.
This makes resonant excitation of g-modes difficult for early type main-sequence stars.

\section{Resonance Locking {-- linear regime}}
\label{sec:ModeLock}

Now we turn to the central concept in this work, resonance locking between internal modes and the tidal potential. We review conditions for resonance locking, and the resulting orbital evolution, building on previous works by
 \citet{Fuller(2016),Fuller(2017)}.
 %{\y contents moved to discussion section}
 
%{\y perhaps there are some formula/results that are new?} %{\jj Yes, in particular eq. (43) is new, a very general result.  Should highlight in some way.}

\subsection{Conditions for Stable resonance locking}
\label{sec:StableModeLock}

\begin{figure}
\begin{center}
\includegraphics[scale=0.7]{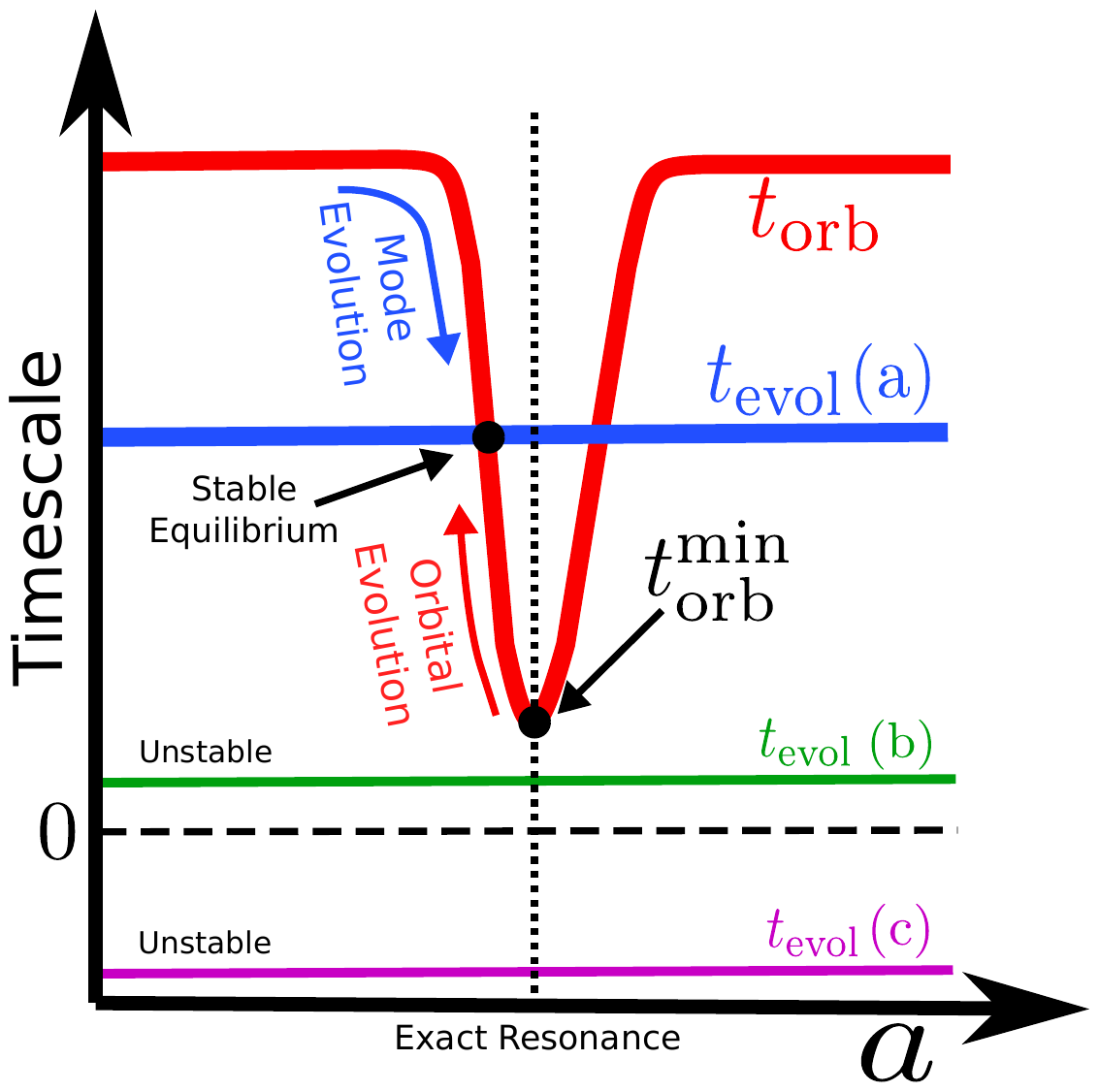}
\caption{
Cartoon figure illustrating the conditions for resonance locking. The timescales for orbital evolution ($t_{\rm orb} = \Om/\dot \Om$, red curve) and mode evolution ($t_{\rm evol} = \sg/\dot \sg$,  colored horizontal lines for different cases) 
are plotted as functions of the binary semi-major axis $a$.  
Under tidal dissipation, the orbit moves left-ward (orbital frequency increases). Stable resonance locking occurs when: the mode frequency also increases  with time, and the evolution timescale at exact resonance ($\sg = k \Om$, dotted line) satisfies $
t_{\rm orb}^{\rm min}\leq t_{\rm evol}$ .  Case (a) is stable, while cases (b) and (c) are unstable.  
}
\label{fig:ModeLock}
\end{center}
\end{figure}

When a mode is resonantly excited by the tidal potential, it converts orbital energy into heat and causes orbital decay. 
The relevant rates of energy and angular momentum change are
\citep{Weinberg(2012),Burkart(2014)}:
%{\y JJ, can you make sure eq. 28 is compatible w/ eq (31), namely, $\dot E_{\rm orb} = 2 \gamma E_{\rm mode} = 2 \gamma \epsilon c^2$?}  {\jj Checked, and works out to be $\dot E_{\rm orb} = 2 \gamma \epsilon (\sg/\om) |c|^2 = 2 \gamma E_{\rm mode}$, as expected (see eq. 22)}
\begin{align}
    \dot E_{\rm orb} &= -\sum_{k \ell m} 2 k \Om \eps U_{k \ell m} {\rm Im} \left( c \, \e^{\im \om_{km} t} \right),
    \label{eq:dotEorb_full} \\
    \dot J_{\rm orb} &= \sum_{k \ell m} 2 m \eps U_{k \ell m} {\rm Im} \left( c \, \e^{\im \om_{km} t} \right),
    \label{eq:dotJorb_full} \\
    \dot J_{\rm spin} &= \sum 2 \cg J_{\rm mode}.
    \label{eq:dotJspin_full}
\end{align}
Here, $\Eorb = -G \Ms \Ms'/(2a)$ is the orbital energy, $\Jorb = -2\sqrt{1-e^2} \Eorb/\Om$ is the orbital angular momentum, $J_{\rm spin} = \Is \Oms$ is the host star's spin angular momentum, with $\Is$ its moment of inertia. ${\rm Im}(X)$ denotes the imagninary part of complex number $X$.  The above expressions contain an implicit summation over all eign-modes, although typically only one mode is important at any given time.

The most important mode  (causing the largest changes in $\dot E_{\rm orb}$, $\dot J_{\rm orb}$) is one that lies near %{\y strictly speaking, this is not correct, there could be higher order resonances that are closer, but less effective.} {\jj agreed, changed} 
resonance,
$\omega \approx \om_{km} = k \Omega - m \Omega_*$,  or in the inertial frame  $\sigma = \omega + m \Omega_* \approx k \Omega$.
Under the action of this mode, binary orbit evolves over a timescale 
%{\y change definition from $t_a$ to $t_{\rm orb}$, no need for $t_a$ anywhere} {\jj OK}
$t_{\rm orb} = \Omega/{\dot \Omega} = - \frac{2}{3}\, a/{\dot a} =  - \frac{2}{3}\, [E_{\rm orb}/(2\gamma E_{\rm mode})]$.  %{\y kept the factor of $2\gamma$ as before, you sure?} {\jj yes} 
Here, $E_{\rm mode}$ is the energy of the excited mode (eq.~[\ref{eq:EJmodes}]).  At exact resonance ($\sg =  k \Omega$), the orbital decay is the fastest, and its associated timescale is at a minimum,
%{\y $\gamma > 0$ for damping, changed $t_{k\ell m}$ to $t_{\rm orb}$} {\jj OK}
\begin{align}
&t_{\rm orb}^{\rm min}
= -\frac{\Eorb}{3 \cg E_{\rm mode}^{\rm max}}
\nonumber \\
&=\frac{\cg}{6 \sg \om} \left( \frac{\Ms}{\Ms'} \right) \left( \frac{a}{\Rs} \right)^{2\ell+1} \left( \frac{\eps}{\Es} \right)
%\nonumber \\
W_{\ell m}^{-2} I_{\ell m}^{-2} X_{k \ell m}^{-2}(e)\, .
\label{eq:t_aklm^min}
\end{align}
%{\y scratched negative sign, ok? $\gamma > 0 $ for damping throughout paper, also factor of 4 turned into 6} {\jj sure}
%{\y  yet to check this expression.} {\jj have checked multiple times over past 1-2 years because very important expression for analysis, can guarrantee correct} 

Resonance locking  operates if $\dot \sg \simeq k {\dot \Omega} \le k {\dot \Omega^{\rm max}}$.
Let the timescale for mode frequency increase in the inertial frame be
\begin{equation}
    t_{\rm evol} \equiv  \sigma/{\dot \sigma}\, .    \label{eq:tevol}
\end{equation}
In general, $\sg$ is a function of both the stellar structure and the star's spin rate.\footnote{Even for zonal ($m=0$) modes, the Coriolis force may affect mode frequencies (see \S \ref{sec:discuss}).}
In this work, we ignore the spin contribution (see discussions in \S \ref{sec:discuss}), 
so $t_{\rm evol}$ is set entirely by stellar evolution.\footnote{Tidal heating is insignificant for altering the stellar structure, for the binary systems we consider.}
%{\y why mention giant planets? we are looking at stars. I added a footnote to be clear. We didn't just ignore it, it is insignificiant.} {\jj OK}
  A stable resonance lock then requires (Fig.~\ref{fig:ModeLock})
\begin{equation}t_{\rm evol} \geq t_{\rm orb}^{\rm min}\, ,
\label{eq:stable}
\end{equation}
where both the magnitudes and the signs of the timescales matter. 
Here, $t_{\rm orb}^{\rm min}$ depends on {both mode properties and binary properties.} It is shorter for modes that are more strongly coupled to the tidal potential, are more weakly damped,  {and for binaries that have shorter orbital periods, and higher eccentricities.}  

In addition, stable locking also requires $\cg$ to be large enough for the mode amplitude $c$ to evolve adiabatically (eq.~[\ref{eq:c_aklm}]):
\begin{equation}
\gamma^{-1} \ll  t_{\rm orb}^{\rm min}, t_{\rm evol}\, .
\label{eq:newadiabatic}
\end{equation}
Given our calculated mode damping rates (Fig. \ref{fig:ModeProps}), this requirement is always satisfied.

\subsection{Orbital Evolution with resonance locking}
\label{sec:OrbModeLock}

\begin{figure}
\begin{center}
\includegraphics[trim=8 0 5 0, clip,scale=0.495]{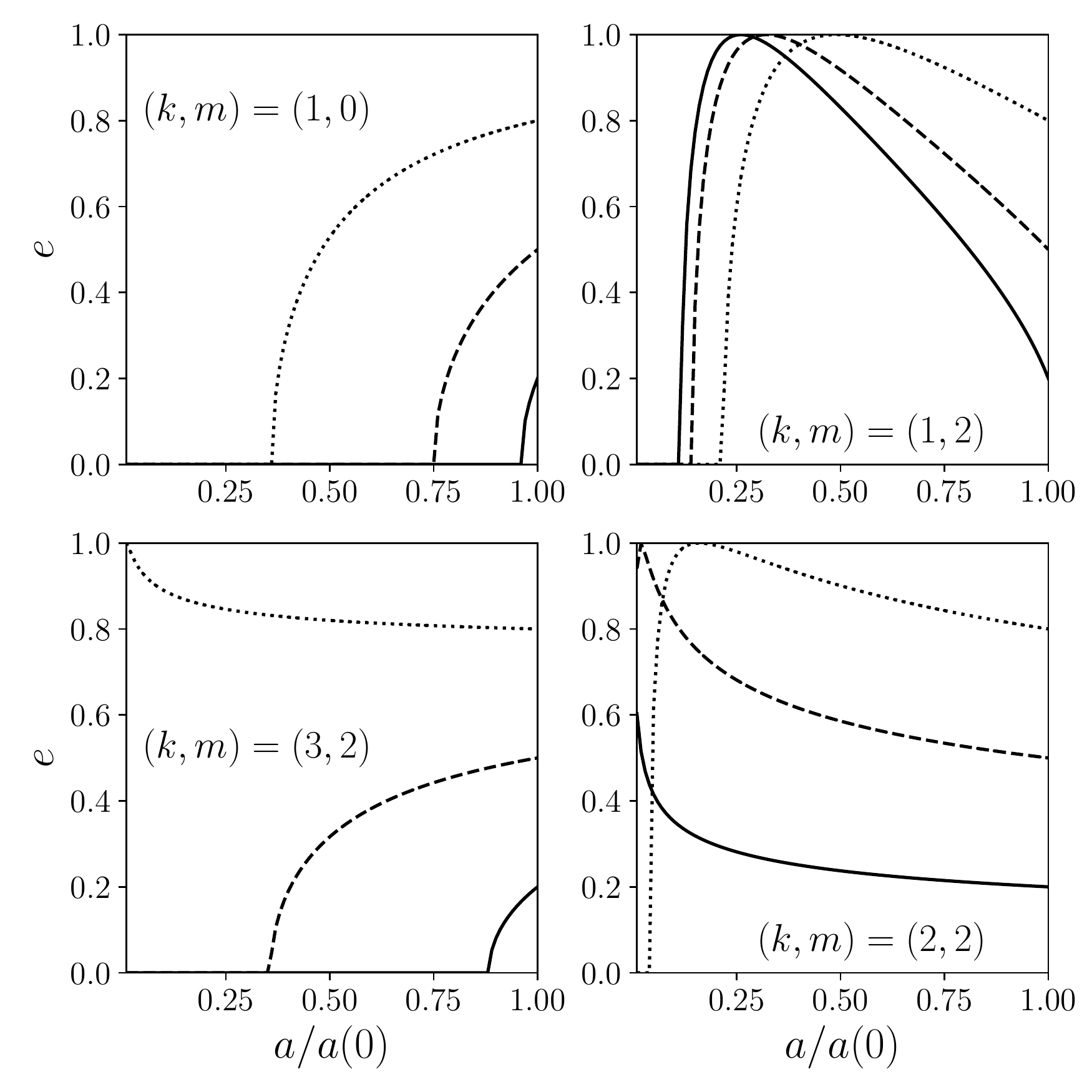}
\caption{
Evolution of binary eccentricity during inward migration for different resonant locks to tidal frequency components $\om_{km} = k \Om - m \Oms$, using equation~\eqref{eq:Lam_km}.  Different lines denote different initial eccentricities, with $e(0)=0.2$ (solid), $e(0) = 0.5$ (dashed), and $e(0) = 0.8$ (dotted).
\label{fig:EccTracks}
}
\end{center}
\end{figure}

As orbital energy is dissipated in the host star, the binary orbit shrinks. Meanwhile, the total angular momentum is conserved.  
This section derives the resultant evolution in eccentricity $e$ and semi-major axis $a$ during a resonance lock.

Conservation of angular momentum requires $\dot J_{\rm mode} = - \dot J_{\rm orb} - \dot J_{\rm spin}$. It may be shown using equations~\eqref{eq:c_agklm^max}, \eqref{eq:dotJorb_full}, and~\eqref{eq:dotJspin_full} that $\dot J_{\rm orb} \simeq - \dot J_{\rm spin}$ when $\om_{km} \simeq \om$, so all binary orbital angular momentum is transferred to the host star's spin angular momentum near resonance  ($\dot J_{\rm mode} \simeq 0$ when $\om \simeq \om_{km}$).  Because $E_{\rm mode}$ is related to $J_{\rm mode}$ via $J_{\rm mode} = - \frac{m}{\sg} E_{\rm mode}$, the evolution of orbital energy and angular momentum
 %{\y actually applies at all times} 
 is related as \citep{Fuller(2017)}
\be
\dot J_{\rm orb} = - \frac{m}{\sg} \dot E_{\rm orb}.
\ee

While resonance locking  operates,
\begin{align}
    k \dot \Om &\simeq \dot \sg = \frac{\pd \sg}{\pd \om} \dot \om + \frac{\pd \sg}{\pd \Oms} \dot \Omega_\star
    \nonumber \\
    &= \frac{\om}{t_{\rm mode}} + m B \frac{\Oms}{t_\star},
\end{align}
%{\y \x{I think this is incompatible w/ eq. (33). This should be $k \dot \Om = \frac{\om_\ag}{t_\ag} + m \frac{\Oms}{t_\star},$}}
where $t_\star = \Oms/\dot \Omega_\star$ is the host star's spin up/down timescale, $t_{\rm mode} = \om/\dot \om$ is the mode frequency evolution timescale (in the star's rotating frame), while the parameter
\be
B = \frac{1}{m} \frac{\pd \sg}{\pd \Oms} 
\ee
takes into account the dependence of the inertial frame frequency $\sg$ on $\Oms$ \citep[see, e.g.][]{Vick(2019)}. 

Since
${\dot \Om}/{\Om} = {3 \dot E_{\rm orb}}/({2 \Eorb})$,
the binary orbital energy evolves as 
\be
\frac{\dEorb}{\Eorb} = \frac{2}{3} \left[ \frac{1}{t_{\rm mode}} + \frac{m}{k} \left( \frac{B}{t_\star} - \frac{1}{t_{\rm mode}} \right) \frac{\Oms}{\Om} \right].
\label{eq:dotE_ModeLock}
\ee
To obtain an evolution equation for the binary eccentricity, we differentiate $\Jorb = -2 \sqrt{1-e^2} \Eorb/\Om$ to obtain
\be
\frac{\der e}{\der t} = - \frac{1-e^2}{2e} \left( 1 - \frac{m}{k\sqrt{1-e^2}} \right) \frac{\dEorb}{\Eorb}.
\label{eq:dedt}
\ee
Expressing 
%The binary semi-major axis evolution is given by
${\dot a}/a
= - {\dEorb}/{\Eorb},$
%\label{eq:dadt}
we integrate equation~\eqref{eq:dedt} 
to obtain the circularization track for binary evolution via resonance locking:
\be
a \left(\sqrt{1-e^2} - \frac{m}{k} \right)^2 = \text{constant}.
\label{eq:Lam_km}
\ee
The ratio $m/k$ is related to the ratio between the angular momentum  and energy a given mode extracts from the binary's orbit. Notice equation~\eqref{eq:Lam_km} is independent of $\Oms$.  For zonal modes ($m=0$), the mode angular momentum is zero  ($J_{\rm mode} = 0$), and equation~\eqref{eq:Lam_km} yields $a(1-e^2) \propto \Jorb^2 = \text{constant}$ for all $k$. 

Some example tracks for different $(k,m)$ are plotted in Figure~\ref{fig:EccTracks}.  We see that when $m=0$, $e$ always decreases with time, while $m\ne 0$ evolution frequently leads to more eccentric orbits \citep[as is noted by, e.g.][]{WitteSavonije(1999b),Fuller(2017)}.  

%{\y I prefer using zonal vs. sectoral, because it's better to use english words in text, and keep symbols to eq. as much as possible. it's a style preferrence that I learned.} {\jj Agree with style, I just thought with your previous comment you wanted this style changed only in appendix, not the main text.}
Fortunately, as we argue in Appendix~\ref{app:ModeLock_NotAxisSymm}, the relevant resonance lock is with the zonal modes, while locking with the sectoral modes ($m\neq 0$) drives negligible orbital evolution. This is related to the fact that the star's moment of inertia ($\Is$) is much smaller than the binary's moment of inertia ($\mu a^2$, where $\mu = \Ms \Ms'/(\Ms + \Ms')$ is the reduced mass of the binary).  The conservation of angular momentum then demands that the spin evolution proceeds at a much higher rate than the orbital evolution, $\dot \Omega_\star \sim (\mu a^2 / \Is) \dot \Omega \gg \dot \Omega$.  Meanwhile, resonance lock requires
\be
\dot \omega \simeq k \dot \Omega - m \dot \Omega_\star \simeq - m \dot \Omega_\star \, ,
\ee
or the sectoral mode frequency in the star's frame has to evolve rapidly. This is difficult to satisfy (App.~\ref{app:nonsec_evolspin}), unless
$k \gg m$.\footnote{Except for the highly eccentric systems like the `heart-beat' stars, one usually finds $k \not\gg m$. }
{We also consider when the binary spin has been nearly tidally synchronized with the binary orbit, for arbitrary eccentricity (App.~\ref{app:nonsec_syncspin}).  Because near tidal synchronization of the stellar spin happens when $k \Omega \approx m \Oms$, resonant locks occur only on high-order g-modes ($\om \approx k \Om - m \Oms \approx 0$ implies $n$ large, see eq.~[\ref{eq:om_WKB}]).  Hence, we find resonance locking is only possible for much higher-order g-modes when compared with zonal modes (locks occur when $\om \ll k \Omega/m$).}
So from now on, we focus exclusively on zonal modes, where tidal dissipation always lead to orbital circularization.

\section{{Application  to binaries -- linear study}}
\label{sec:Results}

This section applies the formalism developed above to test the effectiveness of resonance locking with zonal g-modes  ($m=0$) in  circularizing binary systems. 
We apply the Modules for Experiments in Stellar Astrophysics (\texttt{MESA})
\citep{Paxton(2011),Paxton(2013),Paxton(2015),Paxton(2018),Paxton(2019)} code to model the stellar evolution. We use the \texttt{pre\_ms\_to\_wd} test case within \texttt{MESA} to track stars with masses from $\Ms = 1.0, 1.5$, to $2.0 \, \Msun$, and from the beginning of the PMS to near the end of the MS.
%\footnote{Specifically, we vary the variable \texttt{initial\_mass} in  \texttt{inlist\_1.0} to be \texttt{initial\_mass=1.0,1.5,2.0}. }

%\subsection{Pre-amble}

%Here, we briefly take stock of what we have learned, in  anticipation of the full numerical results.

First, tidal circularization of stellar binaries by resonance locking usually requires mode frequencies to increase with time.
%{\y is this true generally for all ${k,m}$ component? or just some?} {\jj not generally true for all $k,m$, have changed statement}
This is possible for gravity-modes during both the PMS and the MS, with the PMS stars showing much more marked changes. This condition, however, precludes inertial modes as good candidates for resonance locking.

Second, resonance locking is more feasible for modes that are  strongly coupled to the tidal potential, and are weakly damped. Both of these are optimized during the PMS, where the stars are less heavily stratified and g-modes of the requisite periods have much lower radial orders.  

Third, as we have argued above, sectoral modes ($m\neq 0$) are likely un-important for the long-term evolution.  So we can focus only on the $m=0$ tidal potential. On the other hand, the relevant orbital harmonics ($k$) run from $1$ to some small integers, with the maximum value depending on the orbital eccentricity.

\subsection{Regions of Stable resonance locking}
\label{sec:Results_ModeTracks}

\begin{figure}
\begin{center}
\includegraphics[trim=10 0 5 0,width=\columnwidth]{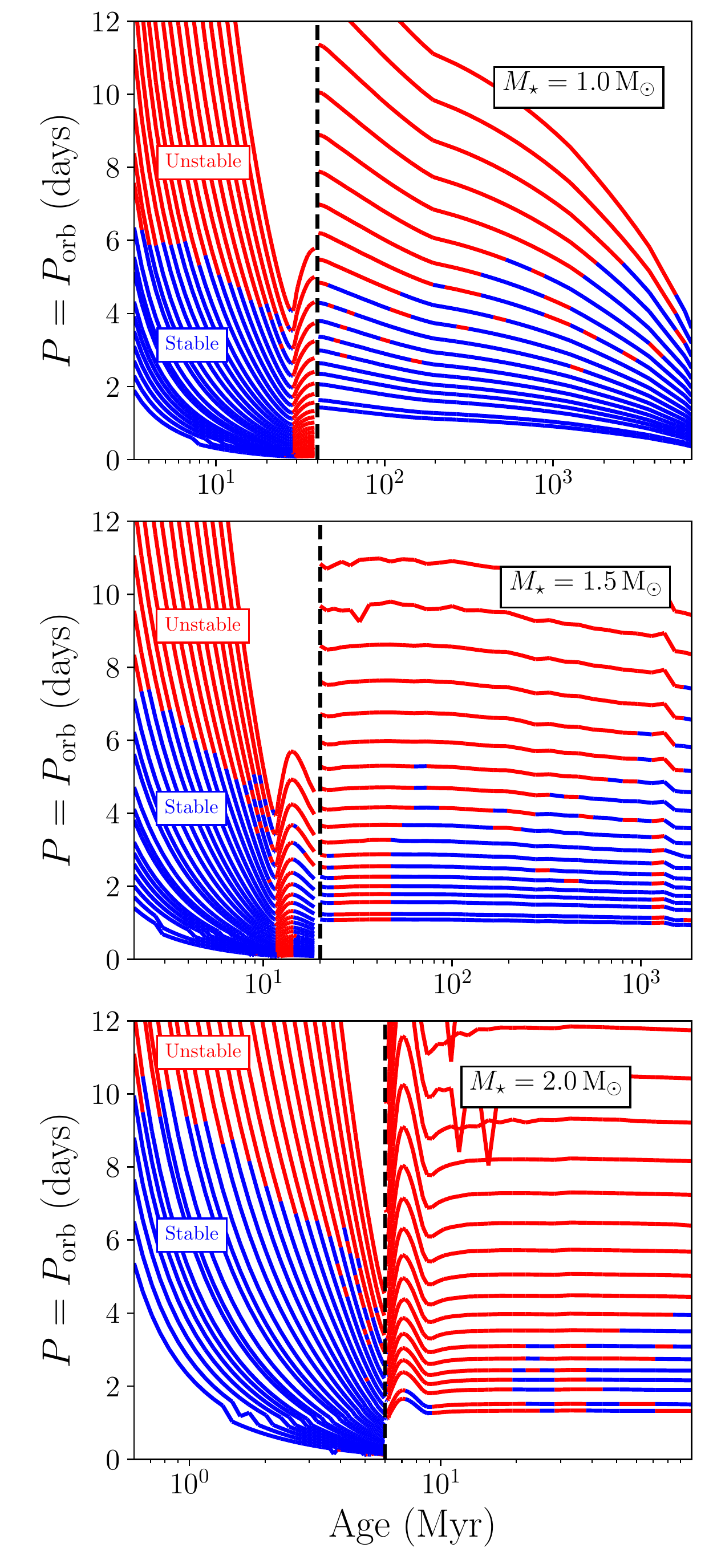}
\caption{
G-mode periods 
%$P_\ag = 2\pi/\om_\ag$ 
as functions of stellar age, 
for primary stars with masses $\Ms$ {as indicated}. Stellar evolution generally leads to decrease in g-mode periods, facilitating resonance locking with tidal forcing. 
The line colors denote stable (blue, $t_{\rm orb}^{\min} <t_{\rm evol}$) 
and unstable (red,  $t_{\rm orb}^{\rm min} > t_{\rm evol}$) resonance locking, with the tide excited by a low-mass companion ($0.3 \, M_\star$) at the same orbital period (component $(k,\ell,m) = (1,2,0)$) and at a modest eccentricity ($e=0.4$).   We consider g-modes with radial node orders $n = 3-200$ during the PMS (to the left the black dashed line), and $n = 50-500$ during the MS (to the right). Resonance locking occurs pervasively during the PMS, but is more sporadic during the MS.
%{\jj Prefer linear plot, because locks when $P_{\rm orb} \lesssim 1 \ {\rm day}$ not important, and linear plot shows behavior when $P_{\rm orb} \sim 5-10 \, {\rm days}$ much better}
\label{fig:ModeTracks}
%{\y maybe remove hte 'PMS/MS' words? too crowded to see, and obvious enough}
}
\end{center}
\end{figure}

We first numerically evaluate the regions of parameter space where stable locking with g-modes can occur. Figure~\ref{fig:ModeTracks} depicts the evolution of g-mode periods as the stars age, where we mark out modes that can stably lock with the $k=1$ tidal potential, excited by a lower mass companion ($0.3 \, \Ms$) at a moderate eccentricity ($e=0.4$). For this potential (the dominant tidal term), locking requires $ P = P_{\rm orb} = 2\pi/\om$.
%{\y I am confused  here -- for eq. tide we take $P = P_{orb}/2$, here we take $P=P_{orb}$. there must be a simple reason. let me know if you have a ready answer} {\jj For equilibrium tides, Goodman \& Oh (1997) takes $P_{\rm tide} = P_{\rm orb}/2$ in analogy with Zahn (1966): the important length-scale is the length a tidal disturbance propagates while interacting with a convective eddy.  Because over a single orbital period, the tidal disturbance propagates ``forward'' for $P_{\rm orb}/2$, then ``backwards'' for $P_{\rm orb}/2$, Zahn and Goodman \& Oh set $P_{\rm tide} = P_{\rm orb}/2$.  For resonance locking, the important quantity is the periodicity of the dominant orbital harmonic, because this is what becomes resonant with the g-mode's period.  Therefore, we set $P_{\rm tide} = P_{\rm g-mode} = P_{\rm orb}$.} {\y I got it. Equilibrium tide assumes $m=0$ but zero frequency tidal forcing. we have $m=0$ but }

%{\y this is material from original  appendix D} {\jj Agree material good for main text}

{
To ascertain these numerical results, we estimate the boundary for resonance locking, using the Solar model at $10^7$ yrs. To compare with Figure~\ref{fig:ModeTracks}, we only consider the $k=1$  orbital harmonics, and approximate $X_{120}(e) \simeq 3e/2$, valid for $e \ll 1$ \citep{Weinberg(2012)}.
Binaries that allow resonance locking satisfy equation~\eqref{eq:stable} with the torque time taken from equation~\eqref{eq:t_aklm^min}. Using estimates in equations
~\eqref{eq:om_WKB2},
~\eqref{eq:cg_diff_est}, \eqref{eq:I_est}, and~\eqref{eq:xihnorm}, with relevant parameters taken from our \texttt{MESA} model and $t_{\rm evol}$ set to $10^7$ yrs, we find that such binaries satisfy
\begin{equation}
P_{\rm orb} \lesssim  P_{\rm lock} 
\sim 6\, {\rm days} \,\, \left(\frac{e}{{0.4}}\right)^{6/31}\, .
\label{eq:e_min}
\end{equation}
%{\y I use this form to compare against Fig. 7, which takes $e=0.4$} {\jj ok} 
This is in reasonable agreement with  results shown  in the top panel of Fig. \ref{fig:ModeTracks} when  $e \ll 1$.} 

%\be
%e_{\rm min} \sim 3.4 \times 10^{-5} \left( \frac{\Porb}{1 \, {\rm day}} \right)^{31/6}.
%\label{eq:e_min}
%\ee
%Solving for the critical orbital period $P_{\rm c}$ below which binaries have nearly circular orbits ($e_{\rm min} \lesssim 0.1$) gives $P_{\rm c} \approx 4.7 \, {\rm day}$, which 
%\x{is very close to the results from our more detailed calculations of circularization of binary populations by} 
%{\w agrees well with our more detailed calculations for} resonance locking \x{along} {\w during} the pre-main-squence (Fig.~\ref{fig:PopCirc})

So our numerical calculations show resonance locking to be  pervasive during the  PMS. Solar-type binaries with periods as long as $6$ days {(at $e=0.4$)} can find stable g-mode locking, while this period increases to $\sim 10$ days for A-type binaries ($2 \, \Msun$). This is due to the better tidal coupling and weaker mode damping for PMS stars -- their g-mode at a given frequency has a lower radial order. For stars we study, one expects significant orbital evolution over the PMS phase.

In contrast, resonance locking during the MS is much less extensive.  Figure~\ref{fig:ModeTracks} shows that only solar-type binaries can sustain significant resonance locking. At the same orbital period, the g-modes of relevance are much higher in radial order in the MS compared to that in PMS, leading to weaker tidal coupling. This is even more exasperated for stars above the so-called `Kraft break' \citep[$M_\star \gtrsim 1.4\, \Msun$,][]{Kraft(1967)}.  These stars have no surface convection zones, eliminating the evanescent tails through which g-modes couple effectively to the tidal potential.  Resonance locking, at least with the dominant $\{k,\ell\}=\{1,2\}$ potential, is impossible for these stars.

%Yanqin: I removed this footnote because I did not understand what you are trying to say here.

%\footnote{The so-called heart-beat stars, also A/F-type stars, can sustain resonance locking \citep{FullerLai(2012b),Burkart(2012),Fuller(2017),Fuller+(2017)}. This is because they are in very eccentric binaries. High order orbital harmonics ($k \gg 1$) is not diminishingly small and lock onto g-modes with low radial orders.}

\subsection{Orbital Evolution of Binaries}
\label{sec:Results_OrbEv}

\begin{figure}
\begin{center}
\includegraphics[scale=0.77]{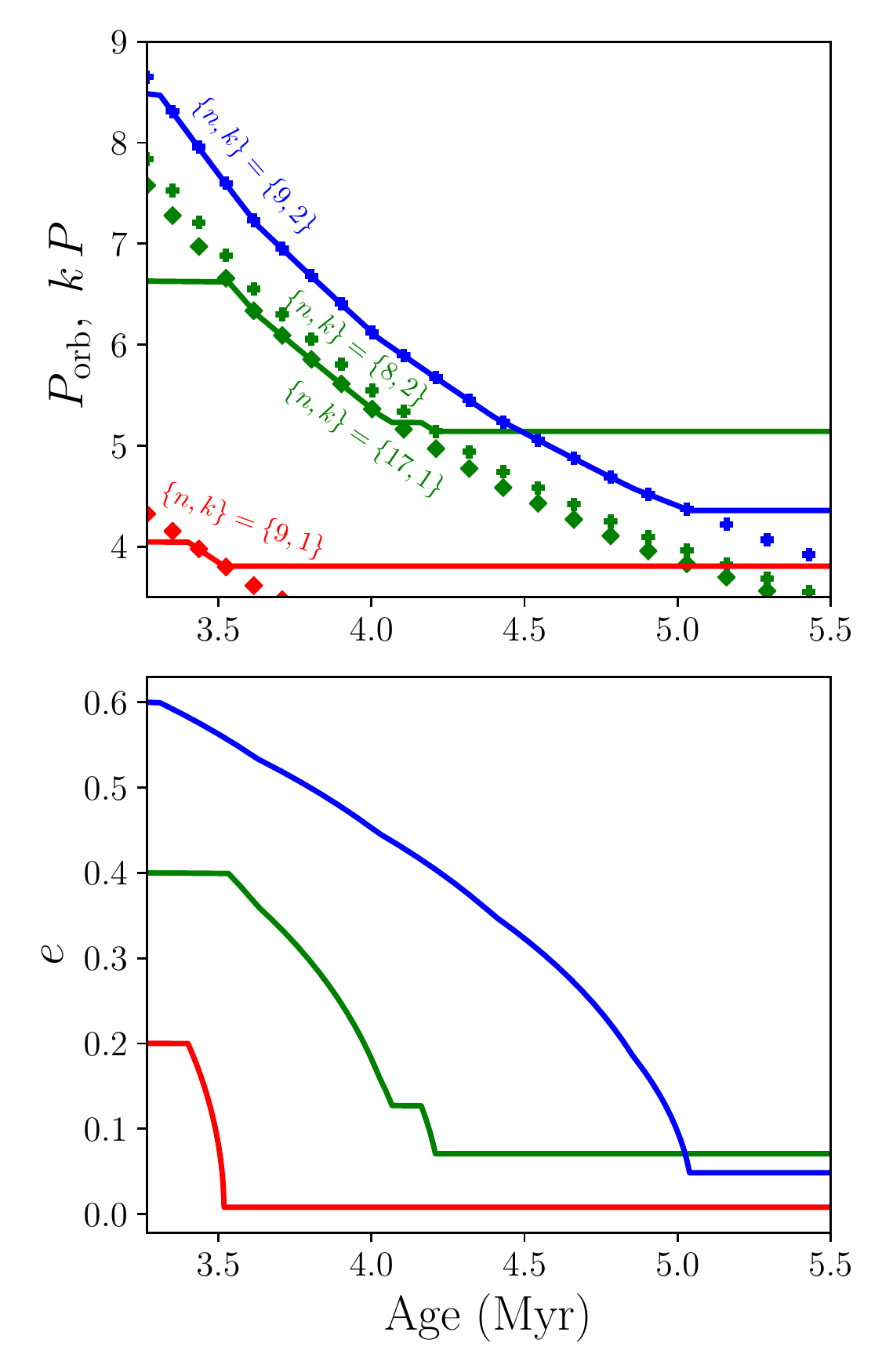}
\caption{ Orbital evolution of three solar-type binaries during the PMS, as a result of resonance locking. Starting from initial periods $P_{\rm orb}(0)=4.0,6.6,8.5$ days, and initial eccentricities $e(0)=0.2,0.4,0.6$, the orbital evolution proceeds as the solid curves. The symbols represent g-mode periods (or their $k$ harmonics) during the same period, with the corresponding ${n,k}$ values.  Resonance locking requires that $k \Omega = \omega$, or $P_{\rm orb} = k P$. Orbital evolution follows the mode evolution while resonance locking is effective. The $n=9$ g-mode is responsible for circularizing two of the systems, forced by two different orbital harmonics ($k=1$ and $2$), while the middle binary system locks onto two different g-modes in turn.
After a few million years, resonance locking no longer  operates and the binary evolution is stalled. 
}
\label{fig:CircTime}
\end{center}
\end{figure} 

 We now use the above results to calculate a binary's orbital evolution, adopting a simplified approach. Instead of directly integrating  equations~\eqref{eq:dotE_ModeLock}-\eqref{eq:dedt} to calculate the binary evolution, we simply look for g-modes with frequencies $\om_\ag \simeq k \Om$ at some point in the host star's evolution.  We use equation~\eqref{eq:stable} to determine if the $\{k,\ell,m\}$ component of the tidal potential can lock onto mode $\ag$.  If so, we assume that the orbital period evolves in sync with the mode evolution 
 ($\Om \simeq \om_\ag/k$), and the orbit circularizes at a constant angular momentum (eq.~[\ref{eq:Lam_km}]).   If resonance locking is not possible, the orbit does not evolve, while the star continues to age.

Figure~\ref{fig:CircTime} displays a few examples of this scheme.  Three solar-type  PMS binaries, starting at 4.0, 6.6 and 8.5 day orbital periods and different eccentricities, are found to stably
lock onto a few g-modes of $n\sim 10$. The tidal potential that is reponsible for locking has harmonics value $k=1$ or $2$.
As the stars contract along their  PMS tracks, the g-mode periods shorten, forcing the orbital period to shorten in step. The associated energy dissipation brings about orbital circularization, which in turn reduces the tidal forcing over time and increases the timescale for orbital evolution (eq.~[\ref{eq:t_aklm^min}]). For the cases considered here, resonance locking becomes ineffective after only a few million years, but by then, significant orbital circularization (and shrinking) has occurred.
 %\x{$t_\ag$ ($\Gamma_{\ag k \ell m}^{\rm orb} < 1$, see eq.~[\ref{eq:Gamma_def}]), which occurs when $e$ becomes sufficiently small. } 
It is interesting to notice that the binary that starts with the widest orbit  ($P_{\rm orb}=8.5$ days) actually progresses further than the next widest one ($P_{\rm orb}=6.6$ days).

Figure \ref{fig:CircTime} also shows that multiple g-modes and orbital harmonics can work together to circularize a binary. For the binary with an initial orbital period of 6.6 days, after locking between the $n=17$ g-mode and the $k=1$ harmonic stops, the $\{n,k\} = \{8,2\}$ mode ``picks-up'' the orbit, and causes a brief episode of additional circularization.  There appears little shortage of appropriate g-modes for resonance locking, at least during the PMS stage.

%\subsection{Binary Circularization}
%\label{sec:Results_Pop}

\begin{figure}
\begin{center}
\includegraphics[scale=0.7]{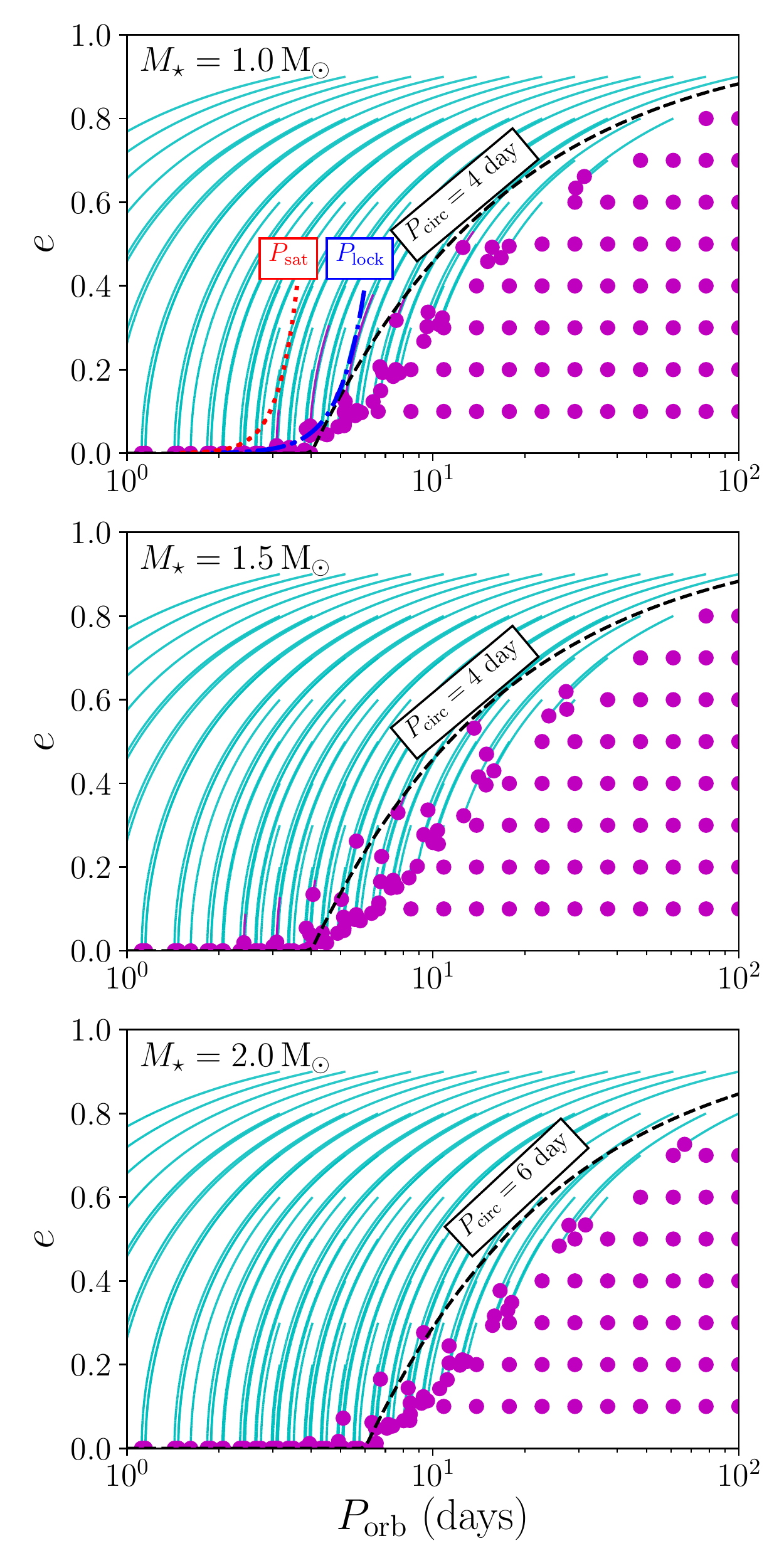}
\caption{
Evolutionary tracks (in lines) and final states (as magenta points) of binary circularization by resonance locking, over the host star's lifetime.
Starting from an initially  uniform grid in $\log P_{\rm orb} - e$, cyan lines trace binary evolution during the pre-main sequence, while magenta lines that during the main-sequence -- the latter are barely visible in most cases. 
Here, we have included the $(\ell,m)=(2,0)$ tidal potential with orbital harmonics $k=1,2,3,5,7,10,13,17,22,27,33$ (integers roughly uniformly spaced in logarithmic values). 
Different panels are for the three host star masses $\Ms$ as indicated. 
The dashed curves crudely describe the upper bounds of binary orbits, as a result of resonance locking. They are orbits at constant periaps ($a(1-e)$), with $P_{\rm circ}$ indicating the period where $e=0$. {The maximum period out to which linear resonance locking operates (eq. [\ref{eq:e_min}]) is displayed by the dot-dashed blue line, while estimated range for saturated resonance locking is shown by the  black dotted line  (eq.~[\ref{eq:e_sat}]).
%{\y can we change simple to $P_{\rm lock}$ and $P_{\rm sat}$? also, I change my mind (yet again) on plotting the saturated one. thanks.} {\jj changed, good now?}} 
%{\jj will do, curves have $P_{\rm orb} = 3.5,3.5,5.4 \ {\rm day}$ when $e=0$}
%{\y Can we remove the equilibrium tide curve? I am not sure how it is aquired, given that we only have $\tau_{\rm circ}$... too much to explain, Also the text 'resonance locking' changed into $3.5$, $3.5$, $5.4$ days}
}
\label{fig:PopCirc}
}
\end{center}
\end{figure}

\begin{figure}
\begin{center}
\includegraphics[width=\columnwidth]{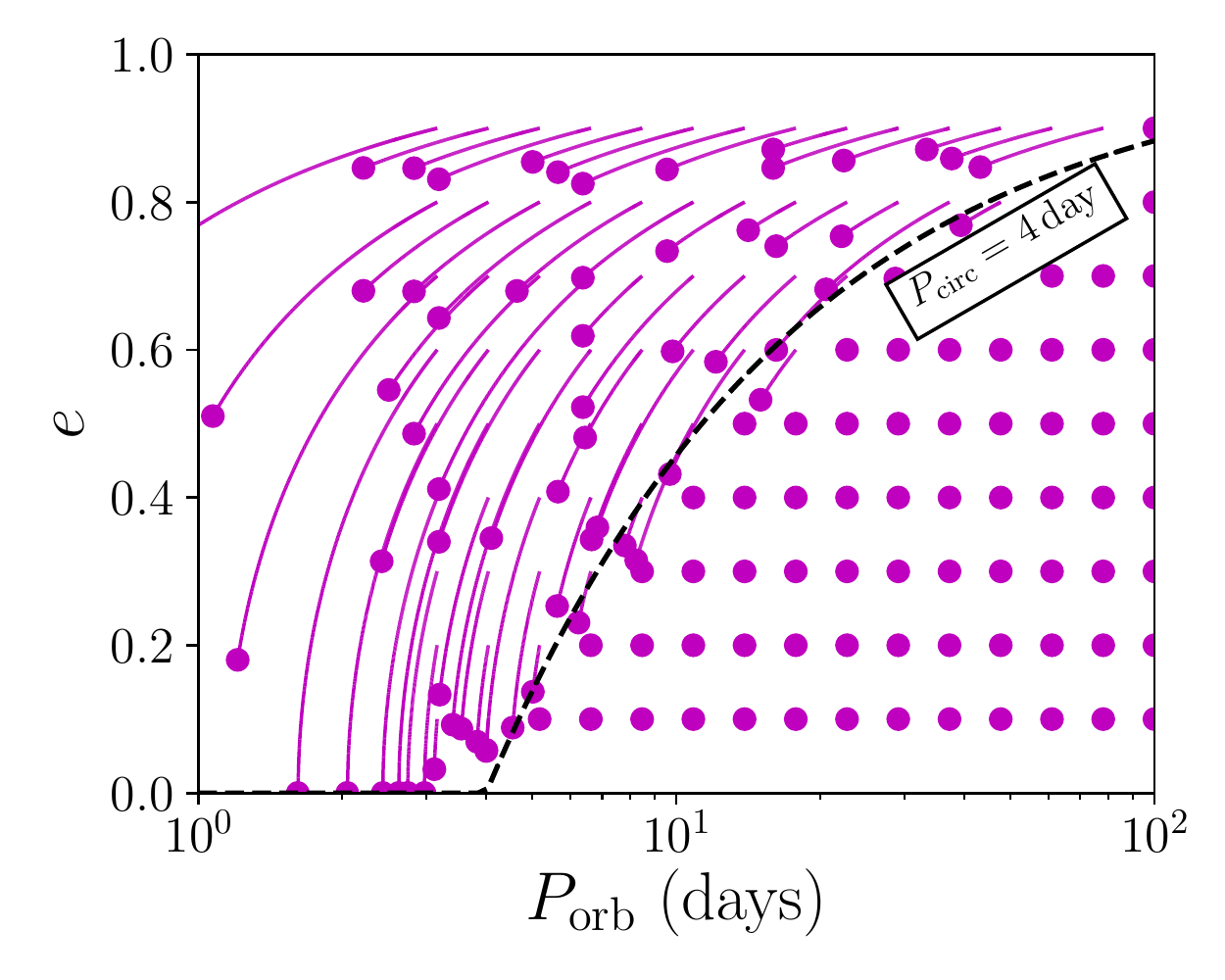}
\caption{
Same as the top panel of Figure~\ref{fig:PopCirc} (host star mass $\Ms = 1.0 \, \Msun$), except we integrate binary circularization tracks starting at $t = 4 \times 10^7 \, {\rm yrs}$ (i.e., zero-age-main-sequence).
While some circularization occurs, many binary systems remain  eccentric above the dashed curve, the rough  boundary of circularization due to PMS resonance locking.}
\label{fig:PopCirc_MS}
\end{center}
\end{figure}

 In the following, we report the results of orbital evolution for a population of binaries.
Figure~\ref{fig:PopCirc} displays the binary circularization tracks  and final orbits, 
 starting from a regular grid of (logarithmic) orbital periods and eccentricities,
  for three different host star masses $\Ms$. 

It appears that resonance locking can effectively circularize all binaries with periapses smaller than some critical values, corresponding to  circular orbits that range from $\sim 4$ days for solar-type to $\sim 6$ days for A-type binaries.\footnote{In \citet{ZanazziWu(2020)}, we provide a more rigorous measure for the circularization periods.} Binaries with periaps smaller  than these values should be circular. Even very wide orbits can circularize, as long as they are sufficiently eccentric (e.g.,  $P_{\rm orb} = 100$ days requires $e \gtrsim 0.8$). This is due to the effective resonance locking between their high-order tidal harmonics (large $k$) and medium order g-modes ($n \sim 10$).

We find that almost the entirety of orbital evolution in Figure~\ref{fig:PopCirc} is solely due to resonance locking during the PMS phase.  Little to no resonance locking is found to occur during the MS (with some minor exceptions in solar-type binaries, see top panel of Fig. \ref{fig:PopCirc}).  To illustrate this issue more clearly, we repeat our calculations for solar-type binaries, but without the benefits of the PMS phase.
 This is shown in Figure~\ref{fig:PopCirc_MS}.
 Many eccentric binaries with short orbital periods ($\Porb \lesssim 5 \, \days$) remain eccentric, some highly so. 
 
The inefficiency of tidal circularization during the MS stage can be attributed to two factors.  First, the stronger stable stratification during the MS (Fig. \ref{fig:N2_ev}) shifts g-mode frequencies upward. So the resonant g-modes have in general much higher radial orders. This leads to both weaker tidal coupling (Fig. \ref{fig:I_lm}), and stronger damping (Fig. \ref{fig:ModeProps}). Second, the MS stage is largely static, compared to the PMS stage. G-mode periods vary only mildly, if at all (Fig. \ref{fig:ModeTracks}). Many binaries could not find suitable g-modes to lock onto. Both factors are more severe for more massive stars that have fully radiative envelopes.

{
\section{The Role of non-linearity}
\label{sec:nonlinearity}

The calculations presented above  assume that the gravity-waves remain linear in its propagation and damping. 
However,  we are prompted by the referee to consider the role of non-linearity in resonance locking: if the tidal displacement becomes non-linear, the wave can overturn the stratification and  dissipate energy rapidly. This physics complicates the resonant locking picture substantially.

%When amplitude of a primary mode becomes sufficiently large, it can excite daughter-mode pairs that are higher in radial order and lower in frequency via the so-called parametric instability \citep[e.g.][]{KumarGoodman(1996),WuGoldreich(2001),Weinberg(2012)}.   This in turn damps the primary mode.  
In stars with radiative interiors,
as gravity-waves propagate towards the centre, their wavelengths shorten and their amplitudes rise. This is most extreme near  the inner turning  point (where $\omega \approx N$), where the background stratification can  be overturned by  the large amplitude wave, akin to the physics of Kelvin-Helmholtz instability  \citep[e.g.][]{GoodmanDickson(1998),BarkerOgilvie(2010),BarkerOgilvie(2011),Barker(2011),Weinberg(2012)}.  
For stars like the Sun, \citet{GoodmanDickson(1998),BarkerOgilvie(2010)} showed that gravity-waves excited  by  stellar or planetary companions can find themselves to be nonlinear on their first travel inward to the core (we call this non-resonant excitation). The non-linearity is stronger for longer period companions, because they excite lower frequency waves, which have shorter wavelengths, and can penetrate into more central regions. Given  this, it is natural to ask whether similar physics operates for pre-main-sequence binaries, and if resonantly excited modes, with their large amplitudes, can in  fact exist.

\subsection{Magnitudes of non-linearity}

Non-linearity for non-resonantly excited waves is insignificant for pre-main-sequence stars. Inserting  values relevant for the solar model at 10 Myrs into  equation~(14) of \citet{GoodmanDickson(1998)} (see also \citealt{OgilvieLin(2007),BarkerOgilvie(2010)}),
we find that at the lower turning point ($r=r_m$), a $({k,\ell})=(1,2)$ gravity-wave
reaches a dimensionless amplitude
\be
|\kr \xir|_{r=r_{\rm m}}^{\rm non-res} \sim 6 \times 10^{-4} \left(\frac{e}{{0.4}}\right)\, \left( \frac{\Porb}{1 \, {\rm day}} \right)^{1/6}\,  ,
\ee
or, non-resonant gravity-waves are linear for systems of interest (eq. [\ref{eq:e_min}]).
This non-linearity is about $10^6$ weaker than that in a Solar model (eq. [34] in  that work). This mostly results from the reduced stratification in PMS cores,  which moves the inner turning point outward (eq.~[\ref{eq:rmin}]).

In contrast, resonantly excited gravity-modes can reach much higher amplitudes. In Appendix~\ref{subsec:nonestimate}, we estimate for the same gravity-wave  but now resonantly excited to find (eq.~[\ref{eq:nonlin_max}])
\be
|\kr \xir|_{r=r_{\rm m}}^{\rm max} \sim 1.4 \times 10^4 \left( \frac{e}{0.4} \right) \left( \frac{1 \, {\rm day}}{\Porb} \right)^{7/2}.
\ee
For all systems of interest (eq.~[\ref{eq:e_min}]),  non-linearity is severe.

So in conclusion, during  the PMS phase, tidally excited gravity-waves can still form discrete modes (not mere running  waves), but they could not reach the amplitudes expected from (linear) resonant driving  due to nonlinear breaking. The efficacy of resonant locking is severely compromised by the  presence of non-linearity.}

{
\subsection{Saturated Resonant Lock?}

What is the consequence of this compromise? While resonance lock still operates, the amplitude a mode can reach is now limited to a lower value than our linear estimates, leading to a reduced rate of tidal dissipation. This leads us to consider a new scenario, saturated resonant lock.

We discuss this scenario in detail in  Appendix~\ref{app:nonlin_amp} . Here is a brief re-cap.
Consider a gravity-mode growing in amplitude due to resonant tidal forcing. It reaches 
the threshold amplitude for breaking, in a time shorter than it takes to reach the linear estimate. At this point, its energy may quickly be converted into heat in the stellar core. Dividing this energy by the growth time,  
 we obtain an  effective rate of energy  dissipation. We then compare the rate of orbital evolution with the rate of stellar evolution to find a new condition for resonance lock. Inserting values appropriate for the Sun at 10 Myrs, we find that resonance locking is successful (satisfying  eq.~[\ref{eq:stable}]) for binaries with
\be
P_{\rm orb} \lesssim P_{\rm sat} 
\sim 3.6 \, {\rm days}\, \left(\frac{e}{ {0.4}}\right)^{6/41}\, .
%{e_{\rm sat} \sim 6.2 \times 10^{-5} \left( \frac{\Porb}{1 \, {\rm day}} \right)^{41/6}\, .}
\label{eq:e_sat}
\ee
This is  $\sim 35\%$ shorter than  the estimate based on linear resonance lock (eq. \ref{eq:e_min}), but still much  beyond what equilibrium tides can  accomplish (see below).

Whether resonance lock in the presence of nonlinear saturation  operates the way we speculate here is unclear.  A  more careful look, involving perhaps hydrodynamical simulations, is needed.
}

{
\subsection{Nonlinear Wave Breaking during the main-sequence}

Although the non-linearity of non-resonant dynamical tides is small during the PMS phase, this  is not so during the main-sequence, and \cite{GoodmanDickson(1998)} argued that non-resonant dynamical tides should circularize two solar-type binaries with ages $t$ out to orbital periods \citep[see also][]{BarkerOgilvie(2010),Barker(2020)}
\begin{equation}
P_{\rm orb} \approx 6.4 \left(\frac{t}{ {10^{10} {\rm yrs}}}\right)^{1/7} \, {\rm days}    \, .
\label{eq:Goodman6.4}
\end{equation}
Combining our picture of PMS resonant locking  with  the \cite{GoodmanDickson(1998)} picture, we predict  that  solar-type binaries should be circularized out to $\sim 3$  days by the end of the PMS, and exhibit a slow rise of the circularization period out to  $\sim 6$  days, as binary ages approach 10 Gyrs.  Stars above the Kraft break ($\Ms \gtrsim 1.4 \, \Msun$), in  contrast, lack radiative interiors and should not display additional circularization with age. Both these predictions can be tested.
}

\section{Discussions}
\label{sec:discuss}

 In the following, we first  compare resonance locking with other proposed tidal mechanisms. We then discuss uncertainties in our calculations and possible improvements.

\subsection{The Equivalent $Q$-value {for Resonance Locking}}

\begin{figure}
\begin{center}
\includegraphics[width=\linewidth]{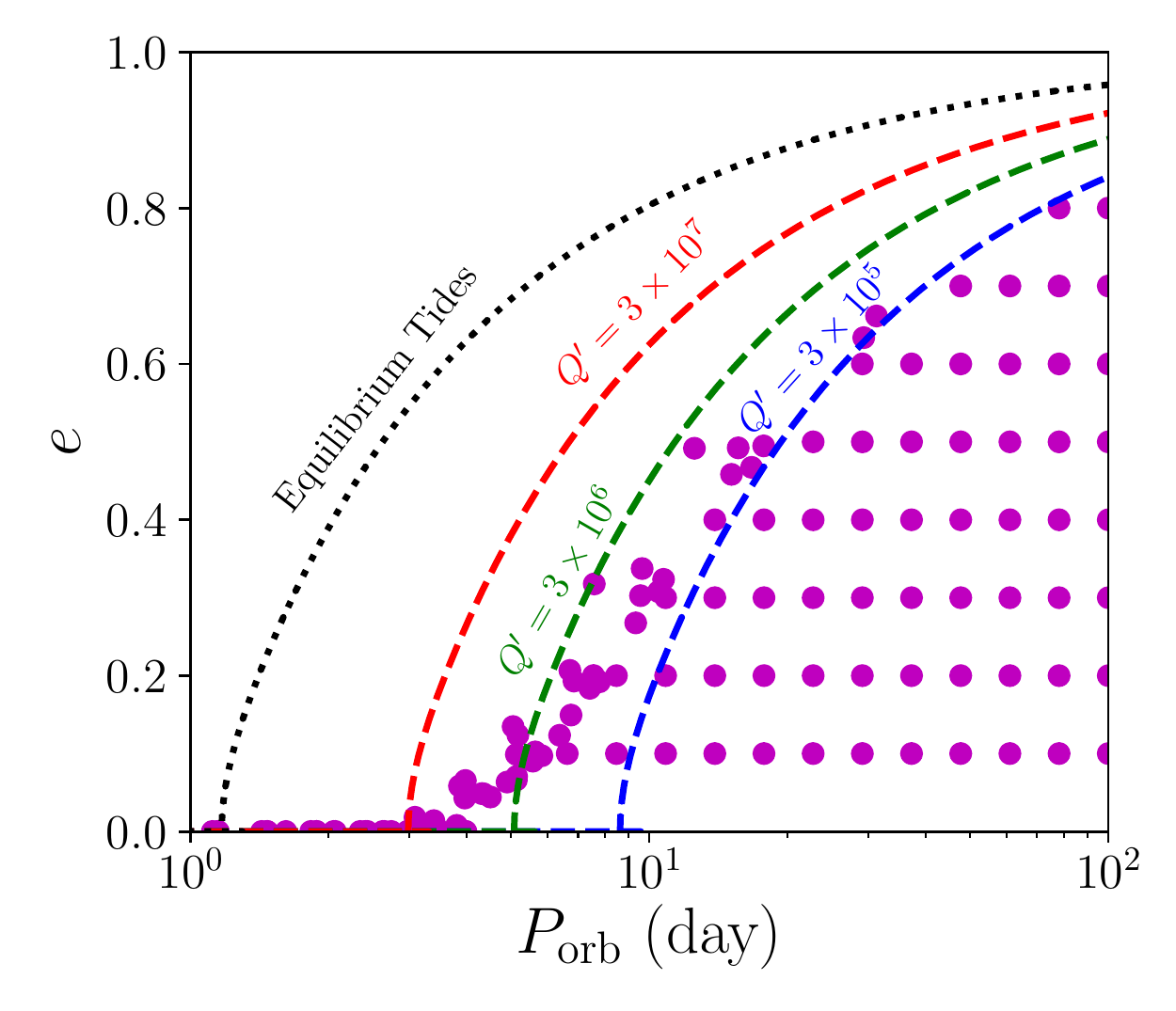}
\caption{Tidal circularization by resonance locking in solar-type binaries (magenta dots, from the top panel of Fig. \ref{fig:PopCirc}), are compared with circularization expected from the equilibrium tide (dotted line), and 
that obtained using a parametrized tidal-Q model (colored curves standing for different $Q'$ values). Tidal circularization by the equilibrium tide has comparable contribution from both the PMS and the MS, and we have also 
adopted for it an eccentricity dependence as that in equation~\eqref{eq:Qmodel}. The tidal-Q model is evaluated for the PMS phase, setting $R_\star =  2 \, \Rsun$. The curve with an effective $Q' = (3/2) k_{2,\star} Q \sim 3\times 10^6$, corresponding to a circularization time that is comparable to the PMS lifetime ($\tau_{\rm circ} \sim 4\times 10^7$ yrs), provides a good resemblance to the results of resonance locking. 
%our circularization calculations for G-type host stars with equivalent circularization time tracks from constant time-lag tidal theory.  The colored dashed lines display tracks with a constant circualrization time $\tau_{\rm circ}$, for different modified quality factor values $Q' \simeq \frac{10}{3} Q$ (for PMS stars).  The dotted black curve shows the predicted circularization time from equlibrium tidal dissipation (eq.~[\ref{eq:tau_eq}]) over both the PMS and MS evolution.  The circularization of G-type stars due to resonance locking during the PMS is fit reasonably well by $Q \sim 1 \times 10^7$ (assuming $\Rs = 2 \, \Rsun$ and $q = 0.3$), and drives much more circularization over the host's lifetime compared with equilibrium tides.
\label{fig:EqTideComp}
}
\end{center}
\end{figure}

A constant time-lag model is often invoked to parametrize tidal dissipation. This assumes the tidal bulge lags behind the perturber by an amount $\Delta t_\star$ that is independent of the perturbing tidal frequency. This model is convenient to integrate, explaining its outsized popularity, but does not correspond to any particular physical tidal theory.  This model is often also parameterized by the so-called tidal Q-factor, where $Q = \Dg t_\star \Omega$, or the modified tidal Q-factor, $Q' = \frac{3}{2} k_{2,\star} Q$, where $k_{2,\star}$ is the star's tidal Love number.  Here, we infer the equivalent value of $Q'$ for our resonance locking model. 

In this model, the circularization time (cf eq. \ref{eq:dedt_Z}) is expressed as \citep{Hut(1981)}
\begin{equation}
    \tau_{\rm circ} = \frac{Q'}{11 \Omega} \left( \frac{\Ms}{\Ms'} \right) \left( \frac{a}{\Rs} \right)^5 \left[ \frac{F_e(e) G(e)}{F(e)} - \frac{18}{11} G_e(e) \right]^{-1}
    \label{eq:Qmodel}
\end{equation}
where {the functions $F(e), G(e), F_e(e),  G_e(e)$ \citep[as defined in][]{Hut(1981)}}
%\begin{align}
%    F(e) &= \frac{ 1 + \frac{15}{2} e^2 + \frac{45}{8} e^4 + \frac{5}{16} e^6 }{(1 - e^2)^6}, \\
%    G(e) &= \frac{1 + 3 e^2 + \frac{3}{8} e^4}{(1-e^2)^{9/2}}, \\
%    F_e(e) &= \frac{1 + \frac{3}{2} e^2 + \frac{1}{8} e^4}{(1-e^2)^5}, \\
%    G_e(e) &= \frac{1 + \frac{15}{4} e^2 + \frac{15}{4} e^4 + \frac{5}{64} e^6}{(1-e^2)^{15/2}}\, .
%\end{align}
%aThese eccentricity dependencies 
account for the rise of tidal forcing with orbital 
eccentricity. We adopt a Love number $k_{2,\star} \simeq 0.2$, the value for a fully convective star \citep[e.g.,][]{Chandrasekhar(1939),BatyginAdams(2013)}, for the PMS phase.

The comparison in Fig. \ref{fig:EqTideComp} shows that, to reproduce the amount of tidal circularization achieved by resonance locking during the PMS, we require a value of  $Q \sim 3 Q' \sim 10^7$, equivalent to $\tau_{\rm circ} \sim 4\times 10^7$ yrs, the PMS lifetime for a solar-type star. We do not provide an equivalent $Q$ value for the MS phase, because there is little circularization then (Fig. \ref{fig:PopCirc_MS}). {Moreover, nonlinear effects (\S \ref{sec:nonlinearity}) increases the effective $Q$ value over that inferred here.}

We caution that the tidal circularization by resonance locking is not properly described by a constant-$Q$ (constant time lag) model. Rather it is an on-off process that depends on whether the locking condition can be satisfied or not.

\subsection{The un-importance of Equilibrium Tides}
\label{sec:EqTides}

Another physically motivated model for tidal dissipation is the equilibrium tide model
\citep[for reviews, see][]{Zahn(2008),Ogilvie(2014)}.
Instead of considering dynamical response of the star, one assumes that the tidal frequency is so low that the star adjusts instantaneously to the tidal forcing. This describes the tidal response well when the binary lies far away from any dynamical resonance, with the resulting displacements
(equilibrium tides) varying over global scales within the star,  similar to the fundamental-modes \citep[see, e.g.][]{Terquem(1998)}
This minimizes their dissipation via radiative diffusion. In stars with surface convection zones, turbulent viscosity is the dominant dissipation for these waves. Dissipation produces a phase lag between the tidal forcing and the response, which then leads to tidal circularization and synchronization.

Here, we calculate the  impacts of equilibrium tides, both during the MS and the PMS phases. This calculation shows that  the equilibrium tide is ineffective in both stages, compared to resonance locking.

In the limit of low eccentricity, the equilibrium tidal theory of \cite{Zahn(1977),Zahn(1978),Zahn(1989)} results in the orbital evolution 
\begin{align}
    \frac{1}{a} \frac{\der a}{\der t} &= - \frac{2}{\tau_{\rm circ}} e^2,
    \label{eq:dadt_Z} \\
    \frac{1}{e} \frac{\der e}{\der t} &= - \frac{1}{\tau_{\rm circ}},
    \label{eq:dedt_Z}
\end{align}
where the circularization time $\tau_{\rm circ}$ is the critical parameter 
that determines the efficiency of eccentricity damping (eq.~[21] of \citealt{Zahn(1989)}).  The derivation of equations~\eqref{eq:dadt_Z} and~\eqref{eq:dedt_Z} assumes $e \ll 1$. Circularization is expected to be faster for high eccentricity systems \citep{Hut(1981),VickLai(2020)}.  The timescale  $\tau_{\rm circ}$ is recast by \citet{GoodmanOh(1997)} as their equation~(45). We re-evaluate that equation for a binary with mass ratio $q=0.3$. Considering only tides in the primary star, this is
\begin{equation}
    \tau_{\rm circ}^{-1} =  1684\,  \left(\frac{{R_\star}}{a}\right)^8 \, \frac{1}{{M_\star R_\star^8}} \, {\int \rho \nu_T r^8 \der r}\, ,
    \label{eq:taucirc}
\end{equation}
where $\nu_T$ is the turbulent viscosity appropriate for the equilibrium tide. Over most of the convection zone, the tidal period ($P_{\rm tide}$) is short compared to the convection turn-over time. Adopting  a reduction in turbulent viscosity that scales as the ratio of the two timescales to the squared power \citep{GoldreichNicholson(1977),GoodmanOh(1997),OgilvieLesur(2012),Duguid(2020),VidalBarker(2020)}, 
%{\y added Vidal+Barker} {\jj yup, and added a few more too} 
 \citet{GoodmanOh(1997)} expressed, for a convection zone that can be approximated as an $n=1.5$ polytrope,
\begin{equation}
    \rho \nu_T = \frac{9}{{128}} \left(\frac{P_{\rm orb}}{2 \pi t_f}\right)^2 \frac{{M_\star}}{{4 \pi R_\star t_f}} x^{-3} (1-x)^{-1}\, ,
    \label{eq:rhonut}
\end{equation}
where $x = r/R_\star$, and we have taken $P_{\rm tide} = P_{\rm orb}/2$. Here, $t_f = (M_\star R_\star^2/L_\star)^{1/3}$ is an estimate for the global convection turn-over time. For the Sun, $t_f \sim 0.45$ yrs.

Assuming a thin surface convection zone, we can evaluate the integral in equation~\eqref{eq:taucirc}  and write
\begin{equation}
    \tau_{\rm circ}^{-1} \approx 
 0.05  \left(\frac{{R_\star}}{a}\right)^8 \, \frac{P_{\rm orb}^2}{t_f^3}\, .
 \label{eq:taucirc2} 
\end{equation}
In the following, we enhance the dissipation rate by a factor of $10$ to account for the possibility of a fully convective star, and for our perhaps overly severe reduction of the turbulent viscosity \citep[Fig. 2 of][]{GoodmanOh(1997)}. Expressing the stellar dynamical time as $t_{\rm dyn} = (G M_\star/R_\star^3)^{-1/2}$, the final expression is then,
\begin{eqnarray}
    \tau_{\rm circ} &\approx & 1.5\times 10^{-4} \, t_f^3\,  t_{\rm dyn}^{-16/3}\,  P_{\rm orb}^{10/3}\, \nonumber \\
    & = &  7 \times 10^{12} \, {\rm yrs} 
    \left(\frac{P_{\rm orb}}{10\, {\rm days}}\right)^{10/3} \nonumber \\
    & & \times \left(\frac{\Ms}{\Msun}\right)^{11/3}\, 
 \left(\frac{\Rs}{\Rsun}\right)^{-6}\, 
    \left({\Ls}\over{{\rm L}_\odot}\right)^{-1}\, .
\label{eq:taucirc3}   
\end{eqnarray}
%{\y JJ, I prefer the negative signs. It makes it clear that all things to be scaled are at the top and the ngative power is physically transparent.} {\jj OK}

To estimate the maximum orbital period that can become circularized over the main-sequence, we follow \citet{GoodmanOh(1997)} by setting $\tau_{\rm circ} \sim 1/3\times t_{\rm MS}$, the main-sequence lifetime. For the Sun, this yields a very tight orbit, $P_{\rm orb} \sim 1.3$ days. This is about twice shorter than the $2.5$ days estimate in \citet{GoodmanOh(1997)}, because of all the reduction factors we have included above.

As stars are puffier during their PMS phase, action during that stage can be more significant \citep{Zahn(1989)}. PMS evolution of the Sun \citep[e.g.,][]{Gough(1980)},  can be approximately fitted as
\begin{eqnarray}
    R_\star (t) & \approx & 1.26\,  R_\odot \left({t\over{10^7 {\rm yrs}}}\right)^{-0.3}\, ,\nonumber \\
    L_\star (t) & \approx & 0.4\,  L_\odot \left({t\over{10^7 {\rm yrs}}}\right)^{-0.6}\,,
    \label{eq:pms}
\end{eqnarray}
valid from $\sim 3\times 10^5$ yrs to $\sim 3\times 10^7$ yrs. To compare the relative importance of the  PMS action,  we integrate equation~\eqref{eq:dedt_Z} to obtain the relative contributions to circularization
\begin{align}
    {{{\Delta e}|_{\rm PMS}}\over{\Delta e|_{\rm MS}}} &\approx {{1.1\times 10^7 {\rm yrs}}\over{t_{\rm MS}}} \, \left({{t_0}\over{10^7 {\rm yrs}}}\right)^{-1.4}
    \nonumber \\
    &\sim 
    0.6 \left({{t_0}\over{10^5 {\rm yrs}}}\right)^{-1.4} 
\, ,
\label{eq:pms2}
\end{align}
where $t_0$ is the start of our  PMS stage, and we have evaluated the expression for  a solar-type star. So while the rate of tidal damping is enhanced during the  PMS, its short duration limits its contribution to the overall effect.\footnote{Compared to the PMS, a star on the Hayashi track is even fluffier. And it has a higher luminosity powered by accretion. The impact of this stage is unclear.}
This conclusion \citep[also see][]{GoodmanOh(1997)} differs from that obtained in \citet{Zahn(1989),ZahnBouchet(1989)}, mainly because we have adopted the sharper viscosity reduction, as is advocated by \citet{GoldreichNicholson(1977),GoodmanOh(1997)}.  

A graphical comparison of the two models, the equilibrium tide model and the resonance lock model, is presented in Fig. \ref{fig:EqTideComp}. The equilibrium tide is {un-important at all times}. 

\subsection{Examining the Assumptions}
\label{sec:TheoryUncertain}

 In this work, we  have adopted a simple approach to study resonance locking. We first calculate the properties of the gravity-modes in a given star (mass, age), then evaluate whether any component (quantum number $\{k,\ell,m=0\}$) of the tidal force can successfully lock on to a single g-mode
 (eq.~[\ref{eq:stable}]). If so, the orbital frequency  $\Omega(t)$ is assumed to evolve in step with the frequency of this mode $\omega(t)$, $\Om \simeq \om/k$, as stellar evolution modifies the mode frequency.  The orbit is assumed to circularize at a constant angular momentum  (eq.~[\ref{eq:Lam_km}]). We assume the orbital evolution stalls when there is no appropriate resonance locking. We have justified the last point by arguing that the equilibrium tide is ineffective.  Dynamical tides without resonance locking has also previously been shown to be irrelevant for solar-type stars \citep{Terquem(1998),GoodmanDickson(1998)}.
 
In these calculations, we have only considered locking with zonal ($m=0$) g-modes, and have ignored the effects of Coriolis force and nonlinear  wave interactions. We have adopted a form for the turbulent viscosity that is strongly reduced for slow tides. Here, we discuss these simplifications and assumptions in turn.

First, we are justified to consider locking with only one mode at any given time. G-modes are sparse, and the resonance width ($|\dg \om| \sim |\om - \om_{km}| \sim |\cg|$) is much smaller than the frequency difference between two consecutive g-modes ($|\Dg \om| \sim \om/n$ ; see Fig.~\ref{fig:ModeProps}). 

 We consider only the zonal modes  because of the mis-match in orbital and spin angular momenta (\S~\ref{sec:OrbModeLock}). The adoption of zonal modes means we can forgo the spin evolution.\footnote{In fact, our approach is fully  self-consistent when the star is not spinning, since zonal modes then do not carry any angular momentum.} This simplifies the evolution significantly, and it ensures that the resonance locking is stable -- 
\cite{Burkart(2014)} showed that, if both energy and angular momentum are transferred between the binary orbit and the stellar spin (e.g., by $m\neq 0$ modes or the equilibrium tide), it may be difficult to maintain resonance locking. We further justify our choice of zonal modes by investigating the inefficiency of sectoral ($m\neq 0$) modes in Appendix \ref{app:ModeLock_NotAxisSymm}. 

 However, our approach assumes that the frequencies of zonal modes are independent of the spin rate. This is only correct if  $\Omega_\star \ll \omega$. Unfortunately, gravity-modes of interest have periods up to a few days (Fig. \ref{fig:ModeTracks}), comparable in value to the spin periods of T Tauri stars \citep[e.g.][]{Bouvier(2013)}, or of stars that are spin synchronized with the orbit.
For modes in these stars, the Coriolis force provides a substantial restoring force, and their frequencies (and structure) become dependent on the spin rate \citep[for an example, see  Fig. 2 of][]{Bildsten}, in addition to their dependencies on stellar structure. 
%We write this as $\omega = \omega(\tau, \Omega_*)$. 
This could drive the system into or away from resonance locking \citep[e.g.][]{Lainey(2020)}.

%(e.g., angular momentum re-distribution inside the stars).
%However, the spin does fight back. 

We have opted not to include the Coriolis effect in this work, for two reasons that are both to do with our inadequacies. First,  including it will make g-modes much more analytically un-tractable -- while one can use the `traditional approximation' to  calculate g-modes for fully stratified stars \citep[when $\Omega_* \ll N$,][]{Chapman,Bildsten}, our stars typically have convective outer shells. Buoyancy-dominated g-modes become inertial-gravity modes and there is currently no good technique to tackle them \citep[but see attempts in e.g.][]{Lai(1997),WitteSavonije(1999a),SavonijeWitte(2002),OgilvieLin(2004),LaiWu(2006),OgilvieLin(2007),Wu(2005a),Wu(2005b),Ogilvie(2013),LinOgilvie(2017)}. Second, as the mode frequency depends on the spin rate, we have to account for all physical processes that change the spin, including magnetic braking, stellar contraction, tidal torques from both the equilibrium tides and other dynamical tides. We also have to consider the re-distribution of angular momentum inside stars. This is beyond our current capability. The neglect of Coriolis force remains {one of} our most serious caveats.

Damping of g-modes directly affects resonance locking. If the modes are more strongly damped, the  tidal torque will be smaller and resonance locks may break. We have experimented with enhancing the damping by a factor of $10$, and find that the circularization period moves inward by about a day.

%{\y maybe also something about mode adiabaticity?}
%{\jj Will put in something brief, mode adiabaticity is a good approximation}
%{\y happy to ignore too -- we have addressed it in previous section, not much new to add}

%{\y JJ, I don't know if I concur. I have taken a look to see why he got overstability, and I think it is related to spin. So it's not a concern for us. } {\jj OK, agree, have taken section out}

%{\y re-written here} {\jj nicely re-written, have taken out blue text}
{We have briefly considered the role of nonlinear wave breaking on our results.  
In stars with radiative interiors, mode interactions occur predominantly near the inner cores where the gravity-wave wavelengths are short and the amplitudes are large. We find that, even in the PMS phase, resonantly excited gravity-modes invariably reach nonlinear amplitudes, though they remain discrete modes. 
The presence of non-linearity reduces the efficacy of resonance locking. Our crude estimate reduces the critical binary period by $\sim 35\%$ (or separation by $\sim 23\%$), compared to when non-linearity is ignored. A more thorough investigation is needed.}

Concurring with \citet{GoodmanOh(1997)}, we find that the effects of equilibrium tide to be negligible in Sun-like stars, during both the PMS and MS phases. This conclusion depends on the turbulent viscosity form that we have adopted \citep{GoldreichNicholson(1977),GoodmanOh(1997)}. If the form advocated by \citet{Zahn(1966)} is adopted instead, with a less severe reduction for the cases where convective turn-over is much slower than the tide, \citet{ZahnBouchet(1989)} found that binaries out to $P_{\rm orb} \approx 8$ days can be circularized during the PMS phase. This would then make the equilibrium tide comparable in importance to resonance locking. If so, an intricate inter-play between spin and orbital evolution may result \citep[e.g.][]{WitteSavonije(2002),SavonijeWitte(2002),FullerLai(2012a),Burkart(2012)}. As the equilibrium tide drives the star towards spin synchronization \citep{Hut(1981)},  resonance locking of g-modes may be  modified or disrupted as the frequencies of these modes are altered by the spin. This is made more complicated by the fact that the equilibrium tide deposits its energy and therefore angular momentum primarily in the outer convective shell. In contrast, much of g-mode properties (frequency, structure and damping) are determined in the stellar interior. It is yet unclear how the two regions communicate.

There are a few situations where dynamical processes different what we study here may enter the play.
%{\y stars w/o convection zones, gravity-mode damping is subtle, see recent paper by Su+2020, https://arxiv.org/abs/2002.11118}
For more massive stars without surface convection zones, tidal excitation and dissipation of gravity-waves behave very differently. For instance, the absence of an evanescent tail weakens the tidal coupling of g-modes. Moreover, dissipation of gravity waves may be dominated by the so-called ``critical layers'' \citep{Su}. 
In another scenario, relevant for highly eccentric stellar binaries ($e \gtrsim 0.9$), f-modes may be stochastically excited at each periastron passage \citep[e.g.][]{PressTeukolsky(1977),Mardling(1995a),Mardling(1995b),VickLai(2018),Wu(2018)}. The diffusive growth of f-modes can efficiently extract orbital energy and circularizes the binary's orbit rapidly. \citet{Wu(2018)} estimated an effective $Q \sim 1$ during this phase.

\subsection{Observational Comparison}
\label{sec:ObsImps}

We argue in this work that most of the orbital evolution has finished by the time stars enter the  main-sequence.  Moreover, we find that the circularization periods only extend to $\sim 4$ days {or shorter} for solar-type binaries.  This runs counter to the results by \citet{MeibomMathieu(2005),Meibom(2006)}, where they showed that the longest circular orbits appear to be wider in more evolved stars and can extend to $\sim 15$ days.  In Paper II,  we will  present a comprehensive analysis of current data, and compare it against our theory.  {We suggest that  the data are more compatible with shorter circularization periods. } A brief comparison is shown in Figure~\ref{fig:multiplemodel}, showing observed binaries, alongside {the} prediction {from linear resonance lock during the PMS. Non-linearity during the PMS and the MS will modify these  predictions and are not yet included in this prediction.}

\begin{figure}
\begin{center}
\includegraphics[width=\columnwidth]{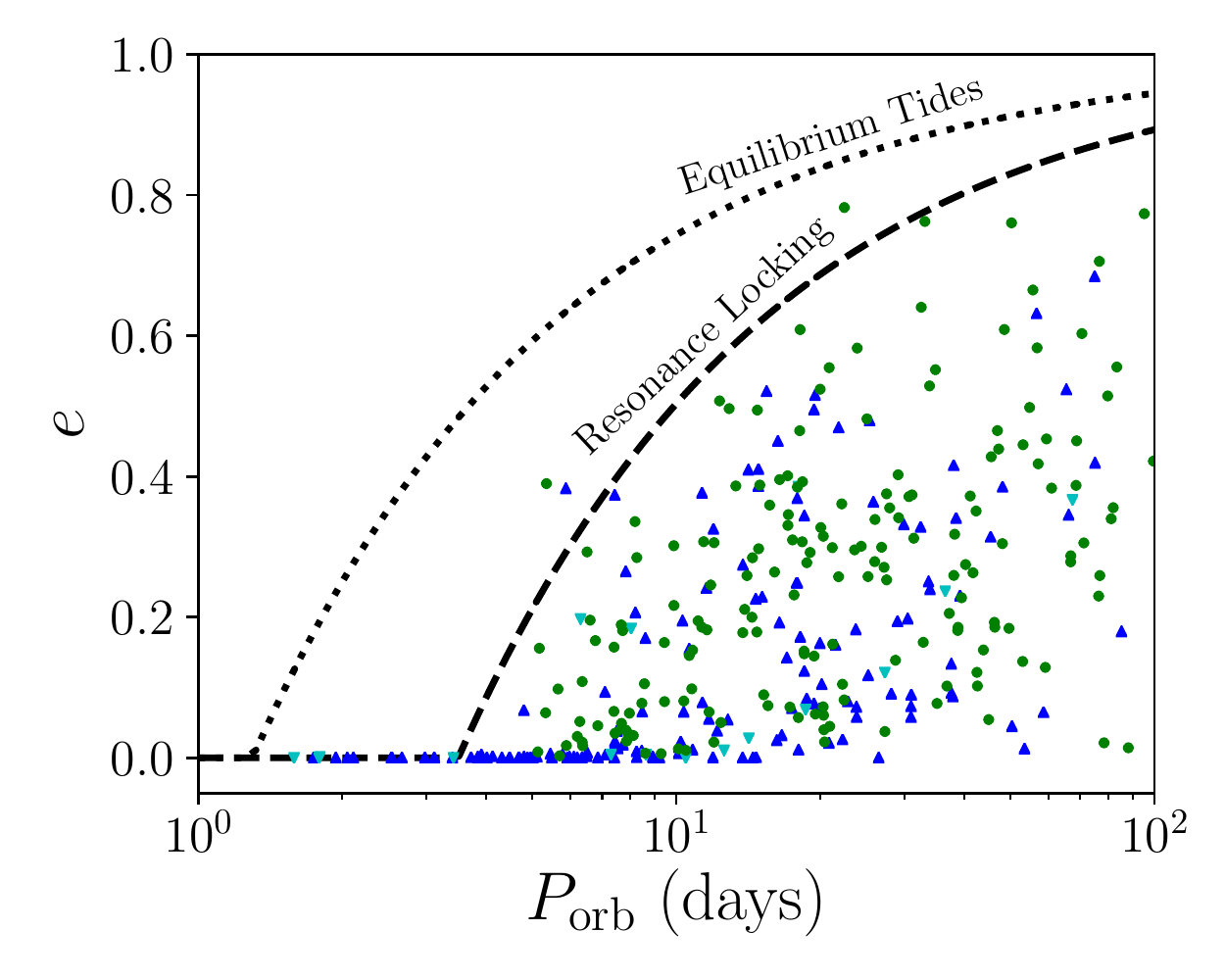}
\caption{A preview of observation vs. theory. More details see Paper II. Here, the colored dots represent  observed binaries that contain G-star primaries, with green circles being spectroscopic binaries from \cite{PriceWhelan(2020)}, and blue triangles eclipsing binaries from \cite{Windemuth(2019)}. The upside-down cyan triangles mark eclipsing binaries that display eclipse-timing variations and so may be in triple systems. We do not display data from \cite{PriceWhelan(2020)} when $P_{\rm orb} < 5 \ {\rm day}$, because the cadence of the observations likely lead to an over-estimate of $e$ at short $P_{\rm orb}$ values (see Paper II for further discussion).  The two theory lines are taken from Figure~\ref{fig:EqTideComp}, {where the line  for resonance locking  has not accounted for non-linearity.}
}
%\label{fig:ThryObsComp}
\label{fig:multiplemodel}
\end{center}
\end{figure}

Observant readers may notice that Figure~\ref{fig:multiplemodel} seems to show an excess of circular orbits out to $\sim$10 days. In paper II, we will show this excess is statisitically significant, and is not  explainable within our framework of resonance locking. Additional mechanisms, such as  formation of short-period nearly-circular binaries via disk migration \citep[e.g.][]{Kratter(2010),MoeKratter(2018),TokovininMoe(2020)}, may need to be invoked.

Another way to test resonance locking during the pre-main sequence is to observe resonance locking in action. Pulsations of resonantly forced gravity-modes may become observable in  pre-main-sequence binaries. Resonance locking theory can provide clear predictions on their amplitudes and phases \citep{Burkart(2012),FullerLai(2012b),Fuller(2017),Fuller+(2017)}.

We have skirted the issue of spin evolution in this work, by invoking tidal forcing of zonal g-modes. As such, we can not make predictions on stellar spin.
%{\y modified quite a bit here, it is still not clear what we want to say...} 
Observationally, measurements of $P_{\rm rot}$ for short-period binaries show that most binaries with $\Porb \lesssim 10 \, \days$ have nearly synchronous rotation periods, but with some notable exceptions \citep[e.g.][]{Meibom(2006),Marilli(2007),Lurie(2017)}.  It is possible 
that resonance locking is instrumental in the spin evolution, but further work is required to address this issue.

%{\y we are at tug-of-war for 'that' a lot, to me that's grammatically more comfortable} {\jj OK, I personally prefer less usage of the word `that,' but I will stop fussing over this issue} 

%{\y JJ: for those formation theories you discard (migration during MS), it's planet tide, not stellar tide that matters. So we can't say anything whatsoever. Stellar tide is important only for causing orbital decay (after the orbital eccentricity has been damped down by planet tide).
%I suggest remove this whole section.} {\jj OK, but I know for Lee \& Chiang (2017), as well as Pu \& Lai (2019), the tides on the star are super important.  But understand what you mean, especially for Lidov-Kozai migration.  Agree fine to remove}

\section{Conclusions}
\label{sec:Conc}

We have investigated the process of resonance locking to circularize binary  systems.  Tidal excitation of dynamical tides transfers orbital energy into mode energy, which is subsequently deposited in the stellar centre as heat. Where resonance between the tidal potential and an internal mode can be sustained, the orbit evolves over the same timescale as stellar evolution. We find this to be the case for pre-main-sequence binaries out to a few days. Main-sequence binaries, in contrast, rarely enjoy the benefit of resonance locking, and we find negligible tidal evolution during this phase of the stellar evolution. 

{Stars both above and below the Kraft break harbour surface convection zones during the PMS. Our calculations show that they can both be circularized to similar distances. Ignoring non-linearity, we find that as these binaries exit the PMS, their orbits should have evolved to fall under a curve of $a(1-e) \geq a_{\rm crit}$, where $a_{\rm crit}$ corresponds to a period of $4-6$ days (Fig. \ref{fig:PopCirc}).  Expressed in  units  of MS stellar radii, we have
\begin{equation}
a_{\rm crit}/\Rs \approx  10 \, .
% 11.5 if p=4 days, M_tot=1.3 M_sun, R=1 R_sun
% 9.4 if p=6 days, M_tot=2.5 M_sun, R=2 r_sun, 
\label{eq:a_crit}
\end{equation}
In contrast, equilibrium tide can only circularize these stars out to $\sim 1.3$ day periods, both during the PMS and the MS, or $a_{\rm crit}/\Rs \sim 5.5$.
} 
%We find that binaries with small peri-centre should be circularized after the pre-main-sequence. For solar-type binaries, this includes all systems with  equivalent circular orbits inward of $\sim 3.5$ days. The latter value rises to $5.4$ days for A-type binaries ($\sim 2 \, \Msun$).
%We can further quantify the  circularization limits by taking ratios between the critical peri-centre distance ($a_{\rm crit}$) and stellar radius at MS ($\Rs$). For solar-type main-sequence binaries, a $3.5$ day circular orbit corresponds to
%\begin{equation}
%a_{\rm crit}/\Rs \approx  10.5 \, .
%\label{eq:a_crit}
%\end{equation}

{Non-linearity in the stellar cores, however, is substantial for all models of interest. Our preliminary attempt to estimate its impact finds that resonance locking should still operate during the PMS, but with a lower efficiency. This process which we call `saturated resonance locking' may circularize binaries out to $a_{\rm crit}/\Rs \sim 8$ (equivalent to $\sim 35\%$ reduction in period). More work is warranted. }

These circularization limits seem to fall far short of what have been reported in the literature \citep[$\sim 10-18$ days for old stellar clusters and halo stars,][]{MeibomMathieu(2005),Meibom(2006)}. However, as we will argue in a companion paper, {re-examining these observational claims lead to very different conclusions.}

Despite our progress, there remains a number of theoretical uncertainties in the physics of resonance locking. We have shown that zonal gravity-modes (azimuthal number $m=0$) play the most important role in resonance locking. However, the frequencies and structure of these modes are affected by stellar spin, when the latter is sufficiently fast. So the spin evolution, ignored in this paper, may upset resonance locking in ways not accounted for here. Nonlinear mode interactions, which may enhance mode dissipation, could also affect our conclusions. 

\acknowledgments

We thank {our anonymous referee for insightful comments. We also thank} Eugene Chiang, Jim Fuller, Kaitlin Kratter, Dong Lai, Maxwell Moe, Gordon Ogilvie, and Michelle Vick for helpful conversations.  YW acknowledges NSERC for research funding. 

\appendix

\section{Gravity-modes: Notations and Properties}
\label{sec:gmodes}

%{\y discussion repeating others, may shrink a bit?} {\jj will shrink, but will focus on writing companion paper first}
This section reviews the main properties of g-mode oscillations \citep[see also e.g.][for reviews]{Unno(1979),Christensen-Dalsgaard(2014)}.  Because g-modes rapidly propigate across the star before energy can be diffused, they are approximatly adabatic, or $\Dg p = \cs^2 \Dg \rho$.
Here $\Dg p$ and $\Dg \rho$ are the Lagrangian perturbed pressure and density, while $\cs^2 = (\der p/\der \rho)_S$ is the adiabatic sound-speed squared.  We cut off the outer radius of the star where non-adiabatic effects are important, and set $\Rs$ to be 97\% of the outer radius \texttt{MESA} calculates ($\Rs = 0.97 \, R_{\star,{\rm MESA}}$).

To calculate g-modes in the star, we perturb the star's equilibrium state $X(r)$ by Eularian perturbations $\dg X(r) Y_{\ell m}(\theta,\vphi) \e^{-\im \om t}$,
where $Y_{\ell m}$ are spherical harmonics.  We decompose the angular dependence of the displacement vector $\bxi(\br,t)$ into vector spherical harmonics:
\begin{equation}
\bxi(\br,t) = \xi_{r}(r) \hr Y_{\ell m}(\theta,\vphi) \e^{-\im \omega t} + \xi_{\perp}(r) \, \left[r \bdel Y_{\ell m}(\theta,\vphi)\right] \e^{-\im \omega t}.
\end{equation}
Linearizing the continuity and momentum equations gives a coupled set of ordinary differential equations for a g-mode's displacement:
\begin{align}
\frac{\der}{\der r}(r^2 \xi_{r}) &= \frac{g}{\cs^2}(r^2 \xi_{r}) + \frac{r^2}{\cs^2} \left( \frac{L_\ell^2}{\omega^2} - 1 \right) \frac{\dg p}{\rho},
\label{eq:dxirdr}\\
\frac{\der}{\der r} \left( \frac{\dg p}{\rho} \right) &= \frac{(\om^2 - N^2)}{r^2}(r^2 \xi_{r}) + \frac{N^2}{g} \frac{\dg p}{\rho},
\label{eq:dhdr}
\end{align}
where $\rho$ is the star's density, while $L_\ell^2$ and $N^2$ are the squared Lamb and \brunt frequencies, respectively.
Since we will be concerned primarily with high-order g-modes, we use the Cowling approximation \citep{Cowling(1941)}.  The transverse displacement $\xi_{\perp}$ and Eularian density perturbation $\dg \rho$ are given by
\be
\xi_{\perp}(r) = \frac{1}{\om^2 r} \frac{\dg p}{\rho},
\hspace{5mm}
\dg \rho(r) = \frac{\dg p}{\cs^2} + \frac{\rho N^2}{g} \xi_{r}.
\label{eq:useful}
\ee

To solve equations~\eqref{eq:dxirdr}-\eqref{eq:dhdr}, we impose a few boundary conditions.  Regularity at the origin ($\der [r^2 \xi_{r}]/\der r|_{r=0}$ and $\der[\dg p/\rho]/\der r|_{r=0}$ finite) requires
\be
\lim_{r\to 0} (r^2 \xi_{r}) = \frac{\ell}{\om^2} Y_0 r^{\ell + 1},
\hspace{5mm}
\lim_{r\to 0} \left( \frac{\dg p}{\rho} \right) = Y_0 r^\ell,
\ee
where $Y_0$ is determined by the mode normailization.   A free boundary a the star's surface requires $\Dg p$ to vanish, or
\be
\left( \frac{\dg p}{\rho} - g \xi_{r} \right)_{r=\Rs} = 0.
\ee
We use a shooting algorithm to calculate the displacement amplitudes throughout the star \citep{Press(2002)}.

The dispersion relation of g-modes gives insight into their behavior.  Assuming $\dg p, \xi_{r} \propto \e^{\im \kr r}$ with $\kr \gg r^{-1}$, one can show using equations~\eqref{eq:dxirdr}-\eqref{eq:dhdr} that the radial wave-number $\kr$ is 
%\citep[e.g.][]{FullerLai(2012b)}
\be
\kr^2 = \frac{(L_\ell^2 - \om^2)(N^2 - \om^2)}{\cs^2 \om^2}.
\label{eq:kr}
\ee
Low-frequency g-modes propagate where $\om^2 < L_\ell^2, N^2$, and are always evanescent in the star's convective regions ($N^2 \le 0$).  Since over most of the star's radiative region ($N^2 > 0$), g-mode frequencies satisfy $\om^2 \ll L_\ell^2, N^2$, we will frequently use the approximate expression
\be
\kr \simeq {N\over{\omega}}\kp \simeq \frac{\sqrt{\ell(\ell+1)}N}{r \om} , 
\label{eq:kr_rad}
\ee
where $\kp = \sqrt{\ell(\ell+1)}/r$ is the azimuthal wavevector. This dispersion relation can in turn be used to solve for the g-mode's eigenfrequency $\om$ \citep[e.g.][]{Lai(1994),FullerLai(2012b)}:
\be
\om \simeq \frac{\sqrt{\ell(\ell+1)}}{\pi n} \int_0^{\Rs} \frac{N(r)}{r} \der r,
\label{eq:om_WKB}
\ee
where $n$ is the number of radial displacement nodes (number of locations $\xi_r[r]=0$), while the integral is taken over the star's radiative zone.  We have implicitly assumed the radial node-number is large ($n\gg 1$).  For the Solar model, a 10-day period $\ell=2$ g-mode has $n \sim 1250$. % {\y is this $n$ number ok with your model?} {\jj At late times ($> 1 \, {\rm Gyr}$), yes.}

We assume a conventional normalization \citep[e.g.][]{Unno(1979),Lai(1994)}, where the mode integral satisfies
%{\y the issue with the factor of 2 unresolved yet}
\be
2\la \bxi| \bxi \ra = 2 \int_0^{\Rs} \left[ \xi_{r}^2 + \ell(\ell+1) \xi_{\perp}^2 \right] \rho r^2 \der r = 1.
\label{eq:bxi_norm}
\ee
Here, $\la {\bm u} | {\bm v} \ra \equiv \int {\bm u}^* \bcdot {\bm v} \rho \der V$, where ${\bm u}^*$ is the complex conjugate of ${\bm u}$. 
Although we will use normailization~\eqref{eq:bxi_norm} throughout this work, because different works frequently use different mode normailizations \citep[e.g.][]{Weinberg(2012),Burkart(2012)}, it is of use to write quantities in terms of $\eps = 2 \om^2 \la \bxi | \bxi \ra$.
The quantity $\eps$ is related to the mode's energy, and with our normalization $\eps = \om^2$.

For our order-of-magnitude calculations to follow, we estimate amplitude for the normalized eigen-function at the bottom of the star's convection zone.  For a \brunt that rises with depth $z$ roughly as $N^2 \sim g/z$, the gravity-modes propagate below a depth $z_1$, where
the upper turning point $z_1 \sim\omega^2/(g k_h^2)$. For high order modes ($n \geq{  \sqrt{{\Rs}/z_{\rm cvz}} \sim {\rm few}}$  for the sun), this turning point lies above the bottom of the convection zone $z_{\rm cvz}$,  and $z_1 \sim z_{\rm cvz}$. We consider this case here.
 
In the energy integral (eq.~[\ref{eq:bxi_norm}]), every node contributes equally to the mode energy. So integration over the first node should be $1/n$. Combine this with $z_1 \sim z_{\rm cvz}$, the horizontal displacement (the dominant component) has a normalized magnitude of
\begin{equation} 
    \left|{\tilde \xi_\perp}\right|^2_{ z=z_1} \sim {1\over{n \, \ell^2 \,  M_{\rm cvz}}}\,
    \label{eq:xihnorm}
\end{equation}
where $M_{\rm cvz}$ is the mass contained within the convection zone. The odd unit for the displacement is a result of our normalization convention.  For modes with $z_1 \gtrsim z_{\rm cvz}$ (including cases where the star has no surface convection zone), one should substitute $M_{\rm cvz}$ by $M_{z_1}$, the mass above the first turning point.  We do not consider modes in stars without  surface convection zones. 

Inside the propagating cavity, according to the WKB analysis, the total energy flux carried by the mode is constant across all radii,
\begin{equation}
    r^2 \, \rho \xi_\perp^2\,  v_{\rm group,r} \simeq {\rm constant}\, ,
    \label{eq:WKBflux}
\end{equation}
where the radial group speed $v_{\rm group,r} = {{\partial \omega}\over{\partial k_r}} \simeq \omega/k_r$. Using expression~\eqref{eq:kr_rad}, we obtain the following radial dependence for the envelope of the pulsation amplitude (the WKB envelope),
\begin{equation}
    \left|{\bar \xi}_{\perp}(r)\right|^2  \sim {{R_\star^2 \rho(z_1) \left|{\bar \xi_\perp}(z=z_1)\right|^2 \,  z_1}\over{r^2 \rho(r)/k_r (r)}} \sim {{k_r(r)}\over{n\ell^2\, r^2 \rho(r)}} 
    \simeq {1\over{\omega n \ell}}\,  {N(r)\over{\rho r^3}}\, .
    \label{eq:WKBenv}
\end{equation}
This is also consistent with the statement that every radial node contributes the same amount of energy. 
%The envelope for $\xi_r(r)$ can be derived by assuming that the g-mode is fairly incompressible, and $k_r \xi_r \approx k_h \xi_\perp$, or $\xi_r (r) \propto 1/\sqrt{\rho r^3 N}$. 
Moreover, the above expression gives $(k_h \xi_\perp)^2 \propto N/(\rho r^5)$. This exhibits a steep rise toward the stellar centre.

\section{Derivation of mode amplitude evolution and damping rate}
\label{app:Schenk_rev}

%{\y didn't check -- its importance is not very clear to me, but perhaps as an appendix it doesn't matter.}
In this appendix, we review the derivation of equation~\eqref{eq:c_a_general} derived by \cite{Schenk(2002)}, and perturbative calculation for the damping rate~\eqref{eq:cg_ag}. We do so to ensure the perturbative expression~\eqref{eq:cg_ag} holds when the Corioulus force is included.  We begin with the tidal oscillations in the star's rotating frame:
\be
\ddot \bxi + \cC \bcdot \dot \bxi + \cR \bcdot \bxi = {\bm f}_{\rm ext},
\label{eq:bxi_ev_gen}
\ee
where ${\bm f}_{\rm ext}(\br,t)$ is an arbitrary external force, while
\be
\cC \bcdot \dot \bxi = - 2 {\bm \Omega}_\star \btimes \dot \bxi
\ee
is an operator for the Corioulus force, with $\cR$ defined as before.  When $\Omega_\star \ne 0$, eigenfunctions $\bxi_\ag(\br) \e^{-\im \lam_\ag t}$ are not orthogonal, or $\la \bxi_\ag | \bxi_\bg \ra \ne 0$ when $\ag \ne \bg$.  They do however satisfy
\be
\la \bxi_\ag | \im \cC \bcdot \bxi \ra + (\lam_\ag + \lam_\bg) \la \bxi_\ag | \bxi_\bg \ra = 0
\label{eq:bxi_orth}
\ee
when $\ag \ne \bg$.  Because of this, it is non-trivial to determine $\bxi(\br,t)$ in terms of the star's natural oscillation modes $\bxi_\ag \e^{-\im \lam_\ag t}$.  To find a decomposition, we re-write equation~\eqref{eq:bxi_ev_gen} in a higher-dimensional vector space.  Defining a new vector
\be
\bzg(\br,t) = 
\left[
\begin{array}{c}
\bxi \\
\dot \bxi + \frac{1}{2} \cC \bcdot \bxi
\end{array}
\right],
\ee
equation~\eqref{eq:bxi_ev_gen} may be re-written as
\be
\dot \bzg + \cM \bcdot \bzg = {\bm F}(\br,t),
\label{eq:bzg_ev}
\ee
where
\be
\cM = 
\left[ \begin{array}{cc}
- \frac{1}{2} \cC & 1 \\
-\cR + \frac{1}{4} \cC^2 & - \frac{1}{2} \cC
\end{array} \right],
\hspace{5mm}
{\bm F}(\br,t) = 
\left[ \begin{array}{c}
0 \\
\bF_{\rm ext}(\br,t)
\end{array} \right].
\ee
We then decompose $\bzg(\br,t)$ as
\be
\zg(\br,t) = \sum_A c_A(t) \bzg_A(\br) \e^{-\im \lam_A t},
\label{eq:bzg_exp}
\ee
where $\bzg_A(\br)$ satisfies
\be
(\cM + \im \lam_A) \bcdot \bzg_A = 0.
\ee
As in \cite{Schenk(2002)}, we use the subscript $A$ instead of $\ag$, because the basis for $\bzg$ has a higher dimension than $\bxi$.  Each left eigenvector $\bzg_A$ has a right eigenvector $\bchi_A$ which satisfies
\be
(\cM^\dagger - \im \lam_A^*) \bcdot \bchi_A = 0,
\ee
where $\cM^\dagger$ is the Hermitian conjugate of $\cM$.  It may be shown \citep{Schenk(2002)} $\bchi_A$ can be re-written as
\be
\bchi_A = 
\left[ \begin{array}{c}
\im \lam_A^* \btau_A - \frac{1}{2} \cC \bcdot \btau_A \\
\btau_A
\end{array} \right],
\ee
where $\btau_A$ may be written in terms of $\bxi_A$ using
\be
\btau_A = - \frac{1}{b_A^*} \bxi_A,
\ee
where the constant $b_A$ has yet to be determined.  Notice because
\be
\la \bchi_A | \bzg_B \ra = \frac{1}{b_A} \big[ \la \bxi_A | \im \cC \bcdot \bxi_B \ra - \im (\lam_A + \lam_B) \la \bxi_A | \bxi_B \ra \big],
\ee
$\bchi_A$ is orthogonal to $\bzg_B$.  We choose $b_A$ so $\bchi_A$ is orthonormal to $\bzg_B$ ($\la \bchi_A | \bzg_B \ra = \dg_{AB}$):
\be
b_A = \la \bxi_A | \im \cC \bcdot \bxi_A \ra + 2 \lam_A \la \bxi_A | \bxi_A \ra.
\label{eq:b_A}
\ee
Inserting equation~\eqref{eq:bzg_exp} into~\eqref{eq:bzg_ev}, and using the orthonormality of $\bzg_A$ with respect to $\bchi_A$, we obtain equation~\eqref{eq:c_ag_ev}.

Our calculations include weak dissipative forces $\cR^\one$ in $\cR$, which we assume are much smaller in magnitude than $\cR^\zero = \cR - \cR^\one$.  To calculate the effects of the dissipative forces, we decompose $\cM = \cM^\zero + \cM^\one$, where
\be
\cM^\zero =
\left[ \begin{array}{cc} 
- \frac{1}{2} \cC & 1 \\
- \cR^\zero + \frac{1}{4} \cC^2 & -\frac{1}{2} \cC 
\end{array} \right],
\hspace{5mm}
\cM^\one = 
\left[ \begin{array}{cc}
0 & 0 \\
-\cR^\one & 0
\end{array} \right].
\ee
Expanding $\bzg_A = \bzg_A^\zero + \bzg_A^\one + \dots$, $\lam_A = \lam_A^\zero + \lam_A^\one + \dots$, and matching terms of like order, we find at order (1)
\be
( \cM^\zero + \im \lam_A^\zero ) \bcdot \bzg_A^\one = - (\cM^\one + \im \lam_A^\one ) \bcdot \bzg_A^\zero.
\label{eq:bzg_A_one}
\ee
Because both $\bzg_A^\one$ and $\lam_A^\one$ are undetermined in equation~\eqref{eq:bzg_A_one}, we may choose a solution $\bzg_A^\one$ which satisfies
\be
\la \bchi_A^\zero | \cM^\zero \bcdot \bzg^\one \ra + \im \lam^\zero \la \bchi_A^\zero | \bzg^\one \ra = 0.
\ee
Hence $\lam_A^\one$ is given by
\be
\lam_A^\one = -\im \la \bchi_A^\zero | \cM^\one \bcdot \bzg_A^\zero \ra = \frac{1}{b_A} \la \bxi_A^\zero | \cR^\one \bcdot \bxi_A^\zero \ra. 
\label{eq:lam_A}
\ee
Dropping the zero superscript for $\bxi_A$, letting $\om_\ag = \lam_A^\zero$, $\im \cg_\ag = \lam_A^\one$, and $\bF_{\rm diss} = \cR^\one \bcdot \bxi_A^\zero$, we obtain equation~\eqref{eq:cg_ag} when $\cC = 0$.

\section{Damping rates for gravity Modes}
\label{sec:gamma}

In this appendix, we estimate the damping rates $\cg$ (eq.~ [\ref{eq:cg_ag}]) for g-mode oscillations due to three different drag forces ${\bF}$: radiative diffusion in the star's radiative regions, turbulent diffusivity in the star's convective region, and wave leakage from the upper atmosphere.
The damping rates we are after are the time-averaged rates of energy dissipation
\begin{equation}
\gamma \equiv  - {{\la{\dot E}\ra}\over E} = -  {{\omega}\over{2\pi}} {1\over E} \oint \der t {{\der E}\over{\der t}}\, ,
    \label{eq:definegamma}
\end{equation}
where our convention is that positive rate for damping. 
In the following, we present both formal integrals for these rates, as well as their order-of-magnitude estimates. 
This latter exercise is necessary as it acts both as a sanity check for our numerical results (presented in Fig. \ref{fig:ModeProps}), and it guides our understanding on how these rates depends on global properties (e.g., stellar luminosity, mass, mode frequency, etc).  Moreover, we take the solar model as a reference when estimating the rates: this assumes a surface convection zone of depth $z_{\rm cvz}$ and mass $M_{\rm cvz}$. We also discuss cases where a surface convection zone is absent.

%{\y this is seriously repeat with Fig. 2. Can we just leave it out? or add the info to Fig. 2 if really necessary?}
%{\jj have removed this figure, it was mainly put in to show estimates agree well with observations, but absolutely not necissary}

\subsection{Radiative Diffusion Damping}
\label{subsec:radiative}

The radiative diffusion arises when heat is lost from the pulsational motion via photo diffusion. This could occur anywhere inside the star, even in the convective region, as long as radiative diffusion is carrying some flux. In the following, we focus on the radiative zone and follow \cite{GoldreichWu(1999)} for much of the discussions.  We will frequently use the fact that radiative diffusion damps g-modes primarily in the radiative zone where $\kr r \gg 1$.

Radiative diffusion causes departure from adiabaticity and removes mode energy at a rate of $T \der s/\der t$, where $s$ is entropy and is related to the flux variation by the equation of energy conservation
\begin{equation}
T {{\der \Delta s}\over{\der t}} = - {1\over\rho} {\der\over{\der r}} \Delta F_{\rm rad} \simeq - {{F_{\rm rad}}\over{\rho}} {\der \over{\der r}} \left({{\Delta F_{\rm rad}}\over{F_{\rm rad}}}\right)\, ,  \label{eq:dsdt}
\end{equation}
where the flux variation can be related to other pulsational quantities by the equation of radiative diffusion and is \citep[see][]{GoldreichWu(1999)}
\begin{equation}
{{\Delta F_{\rm rad}}\over{F_{\rm rad}}} 
= - (1+\kappa_\rho) {{\Delta \rho}\over\rho} 
+ (4 - \kappa_T) {{\Delta T}\over T} 
- {{\der \xi_r}\over{\der r}} 
+ \left({{\der\ln T}\over{\der r}}\right)^{-1} {\der \over{\der r}}\left({{\Delta T}\over{T}}\right) 
\simeq \left( \frac{\der \ln T}{\der r} \right)^{-1} \frac{\der}{\der r} \left( \frac{\Dg T}{T} \right)
- \frac{\der \xir}{\der r}
\, .
    \label{eq:deltaF}
\end{equation}
Here $\kappa_\rho = \der \ln \kg/\der \ln \rho$ and $\kappa_T = \der \ln \kg/\der \ln T$ are the opacity dependence on density and temperature, neglected from here on out.  Further using $\Dg \rho/\rho = -\bdel \bcdot \bxi \simeq - \xir/H$ everywhere for g-modes (where $H = \cs^2/g$ is the scaleheight, valid when $\om^2 \ll L_\ell^2$), 
\be
\frac{\der \ln T}{\der r} = - \frac{\der \ln T}{\der \ln \rho} \frac{g}{c_{\rm e}^2},
\ee
where $c_{\rm e}^2 = \der p/\der \rho$ is the equilibrium sound speed, and the quasi-adiabatic approximation ($\Dg \rho/\rho = [\der \ln \rho/ \der \ln T]_S [\Dg T/T]$), the flux perturbation becomes
\be
\frac{\Dg F_{\rm rad}}{F_{\rm rad}} \simeq \im \kr \left[  \left( \frac{\der \ln \rho}{\der \ln T} \right)_S \frac{c_{\rm s}^2}{g} - \left( \frac{ \der \ln \rho}{\der \ln T} \right) \frac{c_{\rm e}^2}{g} \right] \frac{\Dg T}{T}.
\ee
In addition, we make the crude approximation
\be
\left( \frac{\der \ln \rho}{\der \ln T} \right)_S \frac{c_{\rm s}^2}{g} - \left( \frac{ \der \ln \rho}{\der \ln T} \right) \frac{c_{\rm e}^2}{g}
\approx 
\left( \frac{\der \ln \rho}{\der \ln T} \right)_S \frac{c_{\rm s}^2}{g}.
\ee
Although this approximation will overestimate the magnitude of $\cg_{\rm diff}$, it simplifies its numerical calculation, and still gives damping rates within a factor of $\sim 2-3$ of more detailed calculations \citep[e.g.][Fig. 2]{Fuller(2017)}.  Inserting equation~\eqref{eq:dsdt} into~\eqref{eq:definegamma}, we obtain
\be
\cg_{\rm diff} = \frac{\om}{2\pi \eps} \oint \der t \int_{\rm rad} \left( \frac{\Dg T}{T} \right) \frac{\der \Dg F_{\rm rad}}{\der r} r^2 \der r \simeq \frac{1}{\eps} \int_{\rm rad} F_{\rm rad} (\Gamma_3 - 1) H \kr^2 \left| \frac{\Dg \rho}{\rho} \right|^2 r^2 \der r,
\label{eq:cg_diff_num}
\ee
where the adiabatic exponent $\Gamma_3 - 1 = (\der \ln T/\der \ln \rho)_S$.  Equation~\eqref{eq:cg_diff_num} is the expression we use to numerically calculate $\cg_{\rm diff}$ for our stellar models.

% Here is new estimate, very similar to old estimate with minor modifications in the arguments

To estimate the value of $\cg_{\rm diff}$ analytically, we use $|\Dg \rho/\rho| \simeq |\xir/H| \simeq |(\kp/\kr)(\xip/H)|$ and $r^2 F_{\rm rad} \sim \Ls$ in the outer regions of the star's radiative core ($r \gtrsim 0.2 \, \Rsun$ for PMS and MS models).  Then $\cg_{\rm diff}$ becomes
\be
\cg_{\rm diff} \sim \frac{\Ls}{\eps} \int_{\rm rad} \frac{\kp^2 \xip^2}{H} \der r.
\ee
Because $r^2 F_{\rm rad} \approx {\rm constant}$ only applies to the outer regions of the radiative core, and $r^2 F_{\rm rad} \to 0$ as the core's center is approached, the integrand remains finite even as $\kp^2 \xip^2/H$ diverges as $r \to 0$.  To evaluate this integral, we use our estimate for the WKB envelope of $|\xip|^2$ in the star's radiative zone (eq.~[\ref{eq:WKBenv}]), $\kr \simeq (N/\om) \kp \sim (N/\om) (\ell/r)$, and the fact every radial node of $\xip$ ha s a width $\Delta r \sim \kr^{-1} \sim \der r$ to obtain
\be
\cg_{\rm diff} \sim \frac{\Ls}{\eps} \left[ n \left( \frac{r \om}{\ell N} \right) \frac{1}{H} \left( \frac{\ell^2}{r^2} \right) \left( \frac{1}{\om n \ell} \frac{N}{\rho r^3} \right) \right]_{\rm rad} \sim \frac{\Ls}{\Ms \Rs^2 \om^2} \Lam_{\rm core},
\label{eq:cg_diff_est_app}
\ee
where the quantity
\be
\Lam_{\rm core} = \left( \frac{\Ms \Rs^2}{\rho H r^4} \right)_{r=r_{\rm c}}
\ee
measures the mass concentration in the star's radiative core at the characteristic radius $r_{\rm c}$, and has values of order $\Lam_{\rm core} \sim 10^2 - 10^3$ ($10^3 - 10^4$) during the PMS (MS), and increases as the star evolves and becomes more centrally concentrated.  The value of $\Lam_{\rm core}$ diverges near the center, but does not overwhelm the contribution to the integral for $\cg_{\rm diff}$ is because $r^2 F_{\rm rad}$ vanishes there.

Evaluating this for a 10 day period g-mode in the sun, we obtain $\gamma_{\rm rad} \sim 2 \times 10^{-3} \ {\rm day}^{-1} \sim 2 \times 10^{-8} \ {\rm s}^{-1}$. This agrees with the numerical calculation in Table 1 of \citet{Terquem(1998)} and the illustrative estimate in equation~(21) of \citet{GoodmanDickson(1998)}. 

%High-order g-modes suffer strongly from diffusion in the deep radiative interior.
%For modes with period $\sim 10$ days, $\gamma \sim \omega/n$ and these should be considered as a continuum, as opposed to discreet modes \citep{GoodmanDickson(1998)}. Wave trains propagating into the interior are not reflected backward, but lose most of their energy.}
%{\y this is another condition for resonance locking to work}

We ignore radiative diffusion from inside the convection zone. This is justified because radiation carries only a tiny fraction of the total flux there. Our estimated damping rate (eq.~[\ref{eq:cg_diff_est_app}]) does not apply to stars without a surface convection zone, since most of the diffusion occurs in the deep interior.

\subsection{Turbulent Damping}

Our treatment of the force acting on g-modes from turbulent eddies in the star's convective zone is more heuristic.  We approximate the viscous force (which acts on the gradient of perturbed velocity) as 
\be
\bF_{\rm turb} = - \frac{\im \omega}{r^2 \rho} \frac{\pd}{\pd r} \left( r^2 \rho \nu_{\rm turb} \frac{\pd \bxi}{\pd r} \right)\, ,
\label{eq:nuturb}
\ee
where $\nu_{\rm turb}$ is the effective turbulent viscosity. To account for the two extremes,  when the turn-over time is much shorter or much longer than the mode period, we adopt the following prescription for $\nu_{\rm turb}$ \citep{GoodmanOh(1997)},
\be
\nu_{\rm turb} = \frac{H_{\rm p} v_{\rm cv}}{1 + (\om t_{\rm cv}/2\pi)^s}.
\label{eq:nu_ag}
\ee
Here $H_{\rm p}= (\der \ln p/\der r)^{-1} \sim z$ is the local pressure scale-height, and $z$ the depth from the surface. $v_{\rm cv}$ is the convective velocity, and $t_{\rm cv} = H_{\rm p}/v_{\rm cv}$ is the eddy turnover time.  
%\x{ \citet{GoldreichNicholson(1977)} has argued that the exponent for the viscosity suppression factor, $s=2$, based on physical ground 
%\citep[also see][]{GoodmanOh(1997)}.  But different opinion exists \citep{Zahn(1966)} and numerical experiments have yet to definitively settle on the actual value \citep{Penev(2007),OgilvieLesur(2012)}.}
In the limit of slow convection, \citet{GoldreichNicholson(1977)} have
argued that a Kolmogorov turbulent spectrum gives rise to $s = 2$. In particular, for eddies of a range of sizes, the most relevant ones for turbulent viscosity are those
that have turn-over time comparable to the tidal period, $t_\lam = \lam/v_\lam$ ($t_\lam \om \sim 2\pi$), where $\lam$ and $v_\lam$ are their length-scales and velocities, respectively. This, also advocated later by \citet{GoodmanOh(1997)}, differs from the prescription by \citet{Zahn(1966)}, where he set $s=1$.
%{\y GoodmanOh did not give a different exponent from Goldreich} {\jj OK, I thought actually he derived $s = 5/3$, but then argued $s = 2$ was better.  I think I misinterpreted him though.}
%\x{Different suppression exponents $s$ have been derived by other authors \citep{Zahn(1966),GoodmanOh(1997)},}
Direct numerical simulations have yet to conclusively settle this issue \citep[e.g.][]{Penev(2007),OgilvieLesur(2012),Duguid(2020),VidalBarker(2020)}.

We insert this into equation \eqref{eq:cg_ag}, integrate by parts, and obtain an expression for the
turbulent damping rate,
\be
\cg_{ {\rm turb}} = - \frac{\om^2}{\eps} \int_{\rm cv} \nu_{\rm turb} \left| \frac{\pd \bxi}{\pd r} \right|^2 \rho r^2 \der r\, ,
\label{eq:cg_turb_num}
\ee
where $\epsilon$ is the mode energy, $\epsilon = 2 \omega^2 \la \bxi|\bxi \ra = \omega^2$,
for our adopted normalization.

%{\y I don't agree w/ original argument. the weighting by density is important. much changed from before. check?}
To evaluate this integral at the order-of-magnitude level, we first estimate the magnitude of $\xir$ and $\xip$ in the convective zone.  Since $N^2 \approx 0$, $H\ll r$, and $\om^2 \ll L_\ell^2$, the ODEs for $\xir$ and $\xip$ are
\be
\frac{\der \xir}{\der r} \simeq \frac{\xir}{H} + k_\perp^2 r \xip,
\hspace{5mm}
\frac{\der \xip}{\der r} \simeq \frac{\xir}{r} - \frac{\xip}{r}.
\label{eq:dxidr_cvz}
\ee
Because $H \sim z = \Rs - r \ll r$ in the convective zone, equation~\eqref{eq:dxidr_cvz} has the approximate solution
\be
\xir \approx - \frac{1}{2} \kp^2 r z \xip + \frac{A}{z^2},
\ee
where the constant $A$ is determined by regularity conditions.  Regularity as $z \to 0$ requires $A=0$, so an order-of-magnitude estimate for $\xir$ in the convective zone is
\be
\xir \sim - \kp^2 r H \xip.
\label{eq:xir_est_cvz}
\ee
Notice equations~\eqref{eq:dxidr_cvz} and~\eqref{eq:xir_est_cvz} imply $|\xir| \lesssim |\xip|$ but $|\der \xir/\der r| \gtrsim |\der \xip/\der r|$.  Hence in the convective zone, $|\pd \bxi/\pd r| \sim \kp^2 r \xip$.

In the deeper part of the convection zone, $\om t_{\rm cv} \gtrsim 2\pi$, and \begin{equation}
    \rho \nu_{\rm turb} 
    \sim {{\rho v_{\rm cv}^3}\over{\omega^2 z}} \sim {{F_{\rm cv}}\over{\omega^2 z}} \, , 
\end{equation}
where $F_{\rm cv}$ is the convective flux and is of order the total flux, $F_{\rm cv} = L_\star/4\pi R^2$. 
In contrast, near the surface, $\omega t_{\rm cv} \ll 1$, and 
\begin{equation}
    \rho \nu_{\rm turb} \sim \rho v_{\rm cv} z \sim \rho^{2/3} z F_{\rm cv}^{1/3}\, .
\end{equation}
As a result, the dominant contribution to viscous damping arises where $\omega t_{\rm cv} \sim 1$,  \citep[see also][]{WuLithwick19}, at the depth
\be
z_\nu \sim \frac{v_{\rm cv}}{\om} \sim \frac{1}{\om} \left( \frac{F_{\rm cv}}{\rho_{\rm cv}} \right)^{1/3},
\label{eq:z_nu}
\ee
where $\rho_{\rm cv} \sim M_{\rm cvz}/(\Rs^2 z_{\rm cvz})$.  Hence, the turbulent damping rate is of order
\begin{align}
\cg_{\rm turb} &\sim - \frac{\om^2}{\eps} \int_{\rm cvz} \left( \frac{F_{\rm cv}}{\om^2 z_\nu} \right) (k_\perp^2 r \xip)^2 r^2 \der r \sim \bar \cg_{\rm turb} \frac{\ell^2 \om}{n} \left( \frac{z_{\rm cvz}^{2/3} \Ls^{2/3}}{\om^2 \Rs^2 M_{\rm cvz}^{2/3}} \right)
\nonumber \\
&\sim \bar \cg_{\rm turb} \ \frac{\om}{n} \times \frac{1}{(\tau_{\rm th} \om_{\rm dyn})^{2/3}} \times \left( \frac{z_{\rm cvz}}{z_1} \right) \times \left( \frac{z_{\rm cvz}}{\Rs} \right)^{1/3},
\label{eq:cg_turb_est_app}
\end{align}
where $\tau_{\rm th} \sim M_{\rm cvz} z_{\rm cvz} g/\Ls$ is the thermal time at the base of the convection zone, and $\bar \cg_{\rm turb}$ is an order-unity coefficient used to reduce the magnitude of $\cg_{\rm turb}$ (since we evaluated $\rho \nu_{\rm turb}$ at it's maximum value).  We find the reduction in the turbulent damping estimate~\eqref{eq:cg_turb_est_app} of $\bar \cg_{\rm turb} \sim 1/3$ better matches the more detailed calculations using equation~\eqref{eq:cg_turb_num}.

For the Solar-model, $\gamma_{\rm turb}  \sim 6\times 10^{-4}(\omega/n)\sim 4 \times 10^{-7}\, {\rm day}^{-1} \sim 5 \times 10^{-12}\,  {\rm s}^{-1} $ for a $\ell=2$ mode at $10$ day period, assuming $n\sim 1000$.  Pre-main-sequence stars, with their deeper convection zones (and slower convection), experience stronger damping. In contrast, stars with shallow convection zones can't remove mode energy as efficiently.  Lastly, given the dispersion relation for g-modes ($\omega \propto l/n$), the damping rate is independent of mode radial order, $\gamma_{\rm turb} \propto \ell$.

\cite{Luan(2018)} got a different scaling with mode number ($\cg_{\rm diff} \propto n$, see their eq.~[A13]), because the turbulent dissipation rate $\rho \nu_{\rm turb}$ was evaluated at the base of the convection zone ($z_{\rm cvz}$), rather than where turbulence peaked within the convection zone ($z_\nu$).  The numerical calculations in this work support our estimate for $\cg_{\rm turb}$.

 %{\y not sure what to do with this. didn't chk Luan: Our scaling of $\cg_{\ag,{\rm cv}}$ differs from \cite{Luan(2018)}  -- equation~(A4) of \cite{Luan(2018)}.}

\subsection{Wave Leakage}
\label{subsec:leakage}

Even for stars with a surface convection zone, the atmosphere above the photosphere is stratified. This creates a g-mode cavity through which the internal g-modes that are trapped below the surface convection zone can tunnel through and escape. We proceed to estimate the associated wave damping rate. Some of these calculations were also detailed in \citet{WuLithwick19} in the context of f-modes.

In this region, 
 the stratification $N \sim \sqrt{g/H}$, where 
 $H$ is the photospheric scale height, and the sound speed  $c_s \sim \sqrt{g H} \sim N H$. The dispersion relation is 
 \begin{equation}
     k_z = {1\over 2}{{\Gamma_1}\over H} \pm {1\over 2} \sqrt{\left({{\Gamma_1}\over{H}}\right)^2 - 4 \left[ \kp^2 \left({{N^2}\over{\omega^2}}-1\right) + {{\omega^2}\over{c_s^2}}\right]} \, .
     \label{eq:kz}
 \end{equation}
 So propagative waves have either frequencies higher than the acoustic cut-off frequency $\omega_{\rm ac} = {\Gamma_1 c_s}/(2H)$, or frequencies lower than 
 \begin{equation}
 \omega_{\rm ac,2} = \omega_{\rm ac} {H\over \Rs} = \Gamma_1 {{c_s}\over {2 \Rs}}  \, ,
     \label{eq:omegaac2}
 \end{equation}

Low frequency gravity-waves with $\omega < \omega_{\rm ac,2} \sim 2\pi/(5 \, {\rm days})$ (evaluated with solar values, $H = 20$km and $T = 5800$K) can leak out, with a simplified dispersion relation of the form
\begin{equation}
k_z = \kp {N\over \omega} \approx {{\ell}\over{H}} \left({{\omega_{\rm ac,2}\over{\omega}}}\right)\, .
    \label{eq:dispersionH}
\end{equation}

The rate of energy leakage is determined by the flux of kinetic energy (dominated by the horizontal velocity) at the photosphere
\begin{equation}
    {\dot E}_{\rm leak} = \Rs^2 \times \left[\rho \omega^2 \ell(\ell+1) \left|{\xi_\perp}\right|^2 \times v_{\rm group,z}\right]_{\rm ph}\, ,
    \label{eq:Edotleak}
\end{equation}
where $V_{\rm group,z}\approx \omega/k_z$ is the wave's vertical group velocity. This leads us to an estimate for the damping rate
\begin{equation}
    \gamma_{\rm leak}\equiv  {{{\dot E}_{\rm leak}}\over{E}} = 
    \Rs^2 \rho_{\rm ph} \left|{\tilde \xi}_\perp\right|^2 {\omega\over{k_z}} 
\approx  {\omega\over {n \ell}} \times \left( {{M_{\rm ph}}\over{M_{\rm cvz}}}\right) \times \left({\omega\over{\omega_{\rm ac,2}}}\right) \, ,
\end{equation}
where the photospheric mass $M_{\rm ph} \approx \Rs^2 \rho_{\rm ph} H$, and we have used equation~\eqref{eq:dispersionH} above, and equation~\eqref{eq:xihnorm} for the normalized horizontal displacement, assuming that there is little gradient in the horizontal displacement across the convection zone. 

For the Solar model, $M_{\rm ph}/M_{\rm cvz} \sim 10^{-9}$, and $\gamma_{\rm leak} \sim 10^{-9}(\omega/n\ell) $ for a mode near $\omega_{\rm ac,2}$, and is much below the convective damping estimated above. However, leaking becomes increasingly more important as the convection zone thins. 
In the limit of a vanishing convection zone, g-modes with $\omega < \omega_{ac,2}$ are no longer trapped inside the star and are instead running waves with $\gamma_{\rm leak} \sim \omega/n$.

\section{Adiabadic Evolution of The Amplitude Equation}
\label{app:adiabatic}

%{\y YOU ARE ON YOUR OWN HERE, DIDN'T EDIT/CHECK, also I CHANGED notation for $c_\ag$ in main text, and get rid of $\ag$} {\jj OK, still want to keep this section, got rid of all $\ag$ subscripts and changed $t_\ag$ to $t_{\rm evol}$.}
This appendix derives the regime of validity for the mode amplitude $c(t)$ to be given by equation~\eqref{eq:c_aklm}.  We start with the evolution equation for $c$ excited by the tidal component $\{k,\ell,m\}$:
\be
\dot c + (\im \om + \cg) c = \im \om U_{k \ell m} \e^{-\im \om_{km} t}.
\label{eq:c_aklm_evolution}
\ee
Equation~\eqref{eq:c_aklm_evolution} has the general solution \citep[e.g.][]{FullerLai(2012a),Vick(2019)}
\be
c(t) = c(0) \e^{-\int_0^t (\im \om + \cg) \der t'} + \e^{-\int_0^t (\im \om + \cg) \der t'} \int_0^t \im \om U_{k \ell m}\e^{\int_0^{t'} \left[ \im (\om - \om_{km}) + \cg \right]\der t''} \der t'
\label{eq:c_a_general}
\ee
The first and second terms in equation~\eqref{eq:c_a_general} are homogeneous and inhomogeneous solutions to equation~\eqref{eq:c_aklm_evolution}, respectively.  To show how equation~\eqref{eq:c_a_general} reduces to~\eqref{eq:c_aklm}, we first notice for the systems of interest, their damping timescales are much shorter than their ages ($\cg^{-1} \ll t$), which implies the homogeneous solution quickly damps by age $t$ to give
\be
c(t) \simeq \int_0^t \im \om U_{k \ell m} \e^{-\int_{t'}^t (\im \om + \cg) \der t''} \e^{-\im \int_0^{t'} \om_{km} \der t''} \der t'.
\label{eq:c_a_inhomo}
\ee
We further note that binaries typically have mode damping times much shorter than mode evolution times ($\cg^{-1} \ll t_{\rm evol}$), which implies all expressions outside the damped exponential
\be
\exp \left( - \int_{t'}^t \cg \der t'' \right) \simeq \exp\left[ - \cg(t)(t-t') \right]
\ee
contribute negligibly to the integral from $0$ to $t$ in~\eqref{eq:c_a_inhomo} when $t' \ll t$.  Also assuming $\om, \om_{km} \gg t_{\rm evol}^{-1}$, we obtain equation~\eqref{eq:c_aklm}.  The key condition for solution~\eqref{eq:c_aklm} to hold is the mode damping rate be sufficiently high ($\cg \gg t^{-1},t_{\rm evol}^{-1}$). 

\cite{Burkart(2014)} derived a different condition for adiabadicity:
\be
\cg^2 \gtrsim \om/t_{\rm evol}. 
\label{eq:Burkart_ad}
\ee
We briefly repeat the derivation of equation~\eqref{eq:Burkart_ad}.  When $|c| \ll 1$, $\om = \om_{km}$, $\om \gg t_{\rm evol}^{-1}$, and crucially $\cg^{-1} \gg t$ (where $t$ is the age of the star), equation~\eqref{eq:c_a_general} reduces to
\be
c(t) \simeq \im \om t U_{k \ell m} \e^{-\im \om t}.
\label{eq:c_a_Burkart}
\ee
The time derivative of equation~\eqref{eq:c_a_Burkart} is assumed to be the maximum evolution rate a mode near resonance could evolve as:
\be
\left. \frac{\der |c|}{\der t} \right|_{\rm max} = \om U_{k \ell m}.
\label{eq:dotc_a_max}
\ee
When equation~\eqref{eq:dotc_a_max} is less than
\be
\left. \frac{\der |c|}{\der t} \right|_{\rm ev} \sim \frac{\om^2}{(\om - \om_{km})^2 - \cg^2} \frac{U_{k \ell m}}{t_{\rm evol}},
\label{eq:dotc_a_ev}
\ee
which is roughly the time derivative of the in-homogeneous mode amplitude~\eqref{eq:c_aklm}, the locked mode can no longer keep up with the systems evolution, and equation~\eqref{eq:c_aklm} no longer holds.  However, a crucial step in deriving solution~\eqref{eq:c_a_Burkart} for $c(t)$ at resonance is the assumption $\cg^{-1} \gg t$.  Since most astrophysical systems satisfy $\cg^{-1} \ll t$, equation~\eqref{eq:Burkart_ad} is usually an invalid condition for adiabadicity.  
%{\y what is $t$?} {\jj the age of the star, I have now restated this at the beginning of this paragraph}

Physically, the value of $\der |c|/ \der t|_{\rm max}$ given by equation~\eqref{eq:dotc_a_max} is roughly the rate a mode out of equilibrium ``catches up'' to its equilibrium amplitude~\eqref{eq:c_aklm} when mode damping $\cg$ is negligible.  However, if the mode damping rate is sufficiently high, any memory of a previous mode amplitude is erased, and remains approximately equal to its adiabatic amplitude~\eqref{eq:c_aklm}.

\section{resonance locking from sectoral Modes}
\label{app:ModeLock_NotAxisSymm}

This section looks at resonance locking from stellar evolution due to sectoral modes ($m \ne 0$), which exert a torque on the host star driving the evolution of the stellar spin frequency $\Oms$.  As was stated in Section~\ref{sec:OrbModeLock}, these $m \ne 0$ modes are either very difficult for binaries to lock onto, or will drive negligible inward migration/circularization compared to zonal modes.  This Appendix discusses why these modes are unstable to resonance locking.

%{\w sectoral, zonal} {\jj have changed, hope you agree}

% Moved this section to the main text, like we discussed
%\subsection{Inefficiency of $m \ne 0$ mode circularization}

%{\y not in-efficiency, but in-capability for locking?}

%Because the host star's moment of inertia $\Is$ is much smaller than the companions orbital inertia $\Ms a^2$, a simple argument show locks onto $m \ne 0$ modes drive negligible inward migration (and hence circularization).  Specifically, since torque balance implies $\dot \Om \sim [\Is/(\Ms' a^2)]\dot \Omega_\star$, a locked mode evolves as
%\be
%\dot \omega_\ag \simeq k \dot \Om - m \dot \Om_\star \simeq - m \dot \Om_\star,
%\ee
%since $\Is \ll \Ms a^2$ typically.  Hence $m \ne 0$ modes circularize over a timescale
%\be
%\left| \frac{1}{a} \frac{\der a}{\der t} \right| \sim \frac{\Is}{\Ms a^2} \frac{1}{|t_\ag|},
%\ee
%while $m = 0$ modes circularize over a timescale
%\be
%\left| \frac{1}{a} \frac{\der a}{\der t} \right| \sim \frac{1}{|t_\ag|},
%\ee
%showing modes with $m \ne 0$ circularize over timescales much longer than modes with $m=0$.

\subsection{Stable resonance locks with evolving stellar spin}
\label{app:nonsec_evolspin}

When a non-zero torque is exerted on the host star, both $\Om$ and $\Oms$ evolve as the binary tidally migrates inward.  The tidal torque may be written as \citep{Burkart(2014),Fuller(2017)}
\be
\left. \frac{\der \Oms}{\der t} \right|_{\rm tide} = - \frac{m}{k \Om \Is} \dEorb = \frac{m \mus'}{2 k \kgs} \left( \frac{a}{\Rs} \right)^2 \Om \frac{\dEorb}{\Eorb},
\label{eq:dOmsdt_tide}
\ee
where $\kgs = \Is/(\Ms \Rs^2)$ and $\mus' = \Ms'/(\Ms + \Ms') = (1+q)^{-1}$.  Assuming tidal torques dominate the spin evolution ($\Oms/t_\star \simeq \der \Oms/\der t|_{\rm tide}$, see eqs.~[\ref{eq:dotOms_norm}]-[\ref{eq:Gamma_km^s}] below), equation~\eqref{eq:dotE_ModeLock} may be solved for the binary orbital evolution during resonance locks:
\be
\frac{\dot E_{\rm orb}}{\Eorb} = \frac{2}{3 t_{\rm mode}} \left( 1 - \frac{m \Oms}{k \Om} \right) \Bigg/ \left[ 1 - \frac{m^2 \mus' B}{3 k^2 \kgs} \left( \frac{a}{\Rs} \right)^2 \right].
\label{eq:dotE_ModeLock_Spin}
\ee
Because the orbital evolution at resonance is (using eq.~[\ref{eq:dotEorb_full}] with $\sum c \simeq C^{\rm max}$)
\be
\frac{\dot E_{{\rm orb};k \ell m}^{\rm max}}{\Eorb} = \frac{4 k^2 \eps {\rm Re}(U_{k \ell m})}{\mus' \cg a^2 \Ms} \left( 1 - \frac{m \Oms}{k \Om} \right),
\label{eq:dotEorb_max}
\ee
resonance locking is unstable when the denominator of equation~\eqref{eq:dotE_ModeLock_Spin} is negative, since the resonant mode locations evolve $\Eorb$ in the opposite direction as tidal angular momentum and energy dissipation (eq.~[\ref{eq:dotEorb_max}]; see Fig.~\ref{fig:ModeLock}, case $t_{\rm evol}$ [c]).  Here ${\rm Re}(X)$ denotes the real part of $X$.

Stable resonance locks require the denominator of equation~\eqref{eq:dotE_ModeLock_Spin} to be positive.  Solving for $k = k_{\rm c}$ when the denominator of equation~\eqref{eq:dotE_ModeLock_Spin} is zero gives
\be
k_{\rm c} = \pm m \sqrt{\frac{B \mus'}{3\kgs}} \left( \frac{a}{\Rs} \right).
\label{eq:kc}
\ee
The critical orbital harmonic $k_{\rm c}$ was also derived by \cite{FullerLai(2012a),Burkart(2014)}, where orbital harmonics $|k| < |k_{\rm c}|$ cannot stably resonance lock.  Since $k_{\rm c} \gg 1$ for typical binary star parameters, $m \ne 0$ tidal components lock less easily than $m = 0$ tidal components.

\subsection{Stable resonance locks with near-synchronous stellar spin}
\label{app:nonsec_syncspin}

Instead of assuming $\Oms$ evolves with the locked mode as in the previous section, this section considers a situation where $\Oms$ evolves to an equilibrium state over a timescale much faster than the binary orbital evolution (since $\Is \ll \Ms a^2$ typically).  The stellar spin $\Oms(t)$ for our model (eq.~[\ref{eq:dOmsdt}]) is then set by the balance of tidal torques, stellar winds and contraction/expansion:
\be
\frac{\der \Oms}{\der t} = \left. \frac{\der \Oms}{\der t} \right|_{\rm tide} + \left. \frac{\der \Oms}{\der t} \right|_{\rm wind} + \left. \frac{\der \Oms}{\der t} \right|_{\rm inert} = 0.
\label{eq:dOmsdt}
\ee
The first, second, and third terms in equation~\eqref{eq:dOmsdt} represent stellar spin up/down from the tidal torque from mode $\ag$ and potential component $\{k,\ell,m\}$, stellar winds, and stellar contraction/expansion, respectively.  Equations~\eqref{eq:dotE_ModeLock} and~\eqref{eq:dOmsdt_tide} may be shown to reduce to (when $t_\star \to \infty$)
\be
\left. \frac{\der \Oms}{\der t} \right|_{\rm tide} = \frac{m \mus'}{3 k \kg_\star} \left( \frac{a}{\Rs} \right)^2 \left( 1 - \frac{m \Om}{k \Oms} \right) \frac{\Om}{t_{\rm mode}}
\label{eq:dOmsdt_tide_eq}
\ee
We parameterize stellar wind spin-down with \citep{Skumanich(1972)}
\be
\left. \frac{\der \Oms}{\der t} \right|_{\rm wind} = -\ag_\star \Oms^3,
\label{eq:dOmsdt_wind}
\ee
where $\ag = 1.5 \times 10^{-14}$ for a solar-type star \citep[e.g.][]{VerbuntZwaan(1981),BarkerOgilvie(2009)}.  Spin up (down) from stellar contraction (expansion) is
\be
\left. \frac{\der \Oms}{\der t} \right|_{\rm inert} = - \frac{\Oms}{\tau_\star},
\label{eq:dOmsdt_inert}
\ee
where $\tau_\star = \Is (\der \Is/\der t)^{-1}$ is the spin up/down timescale from changes in $\Is$.  The $\dot \Omega_\star$ contributions from stellar winds and contraction/expansion may be normalized to the tidal torque acting on the host star:
\be
\left. \frac{\der \Oms}{\der t} \right|_{\rm wind} = \frac{m \mus'}{3 k \kgs} \left( \frac{a}{\Rs} \right)^2 \frac{\Omega}{t_{\rm mode}} \Gamma_{km}^{\rm w} \bar \Omega_\star^3,
\hspace{5mm}
\left. \frac{\der \Oms}{\der t} \right|_{\rm inert} = \frac{m \mus'}{3 k \kgs} \left( \frac{a}{\Rs} \right)^2 \frac{\Omega}{t_{\rm mode}} \Gamma_{km}^\star \bar \Omega_\star,
\label{eq:dotOms_norm}
\ee
where $\bar \Om_\star = \Om_\star/\Om$,
\begin{align}
    \Gamma_{km}^{\rm w} &= \frac{3 k \kg_\star}{m \mus'} \left( \frac{\Rs}{a} \right)^2 t_{\rm mode} \ag_\star \Om^2 = 0.0395 \frac{k}{m} \left( \frac{\kg_\star}{0.05} \right) \left( \frac{0.3}{\mus'} \right) \left( \frac{\Rs}{a} \right)^2 \left( \frac{t_{\rm mode}}{10^6 \, {\rm yrs}} \right) \left( \frac{\ag_\star}{1.5 \times 10^{-14} \, {\rm yrs}} \right) \left( \frac{1 \, {\rm day}}{\Porb} \right)^2,
    \label{eq:Gamma_km^w} \\
    \Gamma_{km}^\star &= \frac{3 k \kg_\star}{m \mus'} \left( \frac{\Rs}{a} \right)^2 \frac{t_{\rm mode}}{\tau_\star} = 0.5 \frac{k}{m} \left( \frac{\kg_\star}{0.05} \right) \left( \frac{0.3}{\mus'} \right) \left( \frac{\Rs}{a} \right)^2 \frac{t_{\rm mode}}{\tau_\star}.
    \label{eq:Gamma_km^s}
\end{align}
Since $a > \Rs$ and $\tau_\star \gtrsim t_{\rm evol}$, we typically have $\Gamma_{km}^{\rm w}, \Gamma_{km}^\star \ll 1$, so tidal torques dominate the $\Oms$ evolution.

Equations~\eqref{eq:dotOms_norm}-\eqref{eq:Gamma_km^s} may be re-arranged to give an equation for the equilibrium $\Oms$:
\be
1 - \frac{m}{k} \bar \Om_\star - \Gamma_{km}^{\rm w} \bar \Om_\star^3 - \Gamma_{km}^\star \bar \Om_\star = 0.
\label{eq:dbOmsdt=0}
\ee
The general analytic solution to the cubic equation~\eqref{eq:dbOmsdt=0} is complicated, but since $\Gamma_{km}^{\rm w}, \Gamma_{km}^\star \ll 1$ for typical binary star parameters, defining
\be
\Dg \bar \Om_{km} \equiv 1 - \frac{m}{k} \bar \Om_\star,
\ee
and assuming $\Dg \bar \Om_{km} \ll 1$, we get the analytic approximate solution
\be
\Dg \bar \Om_{km} \simeq \frac{k}{m} \left( \Gamma_{km}^\star + \frac{k^2}{m^2} \Gamma_{km}^{\rm w} \right).
\ee
Thus the semi-major axis evolution may be calculated using
\be
\Om(a) = \frac{\om}{k \Dg \bar \Om_{km}} \simeq \frac{m \om}{k} \left( \Gamma_{km}^\star + \frac{k^2}{m^2} \Gamma_{km}^{\rm w} \right)^{-1}.
\ee
When $m \ne 0$, resonance locking is only possible for high-order modes ($\om \ll k\Om/m$), which will have higher damping rates and lower overlap integrals than $m = 0$ modes (since $\om \simeq k \Om$; see Fig.~\ref{fig:ModeProps}).  Therefore, $m \ne 0$ resonance locks with nearly synchronous spins are significantly less stable than $m = 0$ resonance locks, contributing far less to the circularization of binary star populations.

\section{Non-linearity and dissipation of resonantly excited gravity-modes}
\label{app:non-lin}

{

\subsection{Estimate of Core Non-linearity}
\label{subsec:nonestimate}

This section estimates the degree of non-linearity for resonantly driven g-modes, as they  propagate into the stably stratified core of a young star. For definiteness, we focus on a 10 Myr old model of the Sun, with $\Rs \approx 1.2 R_\odot$, and a core density $\rho_{\rm c} \approx
10 \, {\rm g}/{\rm cm}^3$. Estimates using a 3 Myr model lead to similar conclusions.

Near the centre, the stratification runs as (Fig. \ref{fig:N2_ev})
\be
N(r) \approx 2 \left( \frac{r}{\Rs} \right) \left( \frac{G \Ms}{\Rs^3} \right)^{1/2}\, ,
\ee
so the lower resonant cavity of a g-mode, where $\omega = N(r)$, is
\be
r_{\rm m} \approx 0.5 \left( \frac{\om^2 \Rs^3}{G \Ms} \right)^{1/2} \approx 0.075 \, k \left( \frac{1 \, {\rm day}}{\Porb} \right) \Rsun\, ,
\label{eq:rmin}
\ee
{or,  almost $100$ times further out than that in a 5 Gyrs old Sun (eq.~[32] in \citealt{GoodmanDickson(1998)}).}

Using the WKB envelope  for the horizontal displacement (eq.~[\ref{eq:WKBenv}]), the maximal non-linearity of a resonantly locked g-mode ($\om \simeq k \Om$, and $\Delta = |\omega - \omega_{km}|/\omega \leq \gamma/\omega$) is reached near $r_{\rm m}$,
\be
|\kr \xir|_{r=r_{\rm m}}^{\rm max} \simeq |\kp \xip|_{r=r_{\rm m}}^{\rm max} = C^{\rm max}| \kp \tilde \xi_\perp |_{r=r_{\rm m}} \sim \left( \frac{\om}{\cg} \right) U_{k\ell m} \frac{\ell}{r_{\rm m}} \left( \frac{1}{n\ell} \frac{1}{\rho_{\rm c} r_{\rm m}^3} \right)^{1/2}.
\label{eq:G83}
\ee

The dispersion relation for a $\{\ell,m\} = \{2,0\}$ mode
in such a model is (eq.~[\ref{eq:om_WKB}]),
%\be
$n \approx 10 \, k^{-1} ({\Porb}/{1 \, {\rm day}})$.
%\ee
We also obtain the following numerical values from our calculations
(see Figs. \ref{fig:ModeProps} - \ref{fig:I_lm}, eqs.~[\ref{eq:cg_diff_est}] \&~[\ref{eq:I_est}])
\begin{eqnarray}
    \frac{\cg}{\om} &\approx & \frac{\cg_{\rm diff}}{\om} \approx 1.2 \times 10^{-8} \ k^{-3} \left( \frac{\Porb}{1 \, {\rm day}} \right)^3, \nonumber \\
    I_{20} &\approx & 5 \times 10^{-5} \ k^{5/2} \left( \frac{1 \, {\rm day}}{\Porb} \right)^{5/2} \Msun^{-1/2} \Rsun^{-1}\nonumber \\
    \epsilon &\approx & 0.013 \ k^2 \left( \frac{1 \, {\rm day}}{\Porb} \right)^2 \frac{G \Msun}{\Rsun^3},\nonumber \\
    \frac{a}{\Rs} &\approx & 3.9 \left( \frac{\Porb}{1 \, {\rm day}} \right)^{2/3} \nonumber \\
    U_{k \ell m} &\approx &  1.3 \times 10^{-5} \ X_{k 20}(e) k^{1/2} \left( \frac{1 \, {\rm day}}{\Porb} \right)^{5/2} \Msun^{1/2} \Rsun. 
\end{eqnarray}
 Putting everything together, we find the dimensionless non-linearity parameter to be
\be
|\kr \xir|_{r=r_{\rm m}}^{\rm max} \sim 2.4 \times 10^5 \ X_{k20}(e) k^{3/2} \left( \frac{1 \, {\rm day}}{\Porb} \right)^{7/2}.
\label{eq:nonlin_max}
\ee
For our discussion, we set the threshold for strong non-linearity to be $|k_r \xi_r|_{r = r_m} = 10$, on account of the fact that WKB estimate fails near the turning point for which the above estimate is made \citep{GoodmanDickson(1998)}. For $k=1$ and $e \ll 1$,  $X_{120}  \simeq 3e/2$,  we find that non-linearity is important if the binary period  
%and recast the above expression as a condition on \x{the binary period. Non-linearity is important {\w for binaries closer than}
%\begin{equation}
%\frac{\Porb}{1 \, {\rm day}}\,
%\lesssim 18 \ k^{3/7} X_{k 20}^{2/7}(e).
%\label{eq:porb-nonlinear}
%\end{equation}
\be
%e_{\rm nl} \sim 2.7 \times 10^{-5} \left( \frac{\Porb}{1 \, {\rm day}} \right)^{7/2}.
P_{\rm orb} \lesssim P_{\rm nl} \sim 15.5 \, {\rm days}\, \left({e\over{0.4}}\right)^{2/7}\, .
\label{eq:e_nl}
\ee
%{\y changed the form, not value, easier to communicate this way} {\jj agree, have taken out blue coloring}
%When $e \gtrsim e_{\rm nl}$, one must take into account how non-linearity affects the growth of standing modes in the star. 
This is longer than the limit for resonance locking (eq. [\ref{eq:e_min}]), indicating that  
 for the Sun at 10 Myrs old, the efficacy of resonance locking with gravity-modes is severely compromised by non-linearity.

\subsection{Mode amplitude at nonlinear Breaking And Effective Dissipation}
\label{app:nonlin_amp}

Here, we estimate the  amplitude of a g-mode  when wave-breaking becomes important ($C=C_{\rm sat}$). Scaling from equation~\eqref{eq:nonlin_max}, an estimate for $C = C^{\rm max}$, 
we find a threshold amplitude 
($|k_r \xi_r (r = r_m)| \approx 10$)  to be
\be
\frac{|C^{\rm sat}|}{|C^{\rm max}|}
%= \frac{1}{\Dg_{\rm sat}} \left( \frac{\cg_{\rm diff}}{\om} \right) 
\sim 4.2 \times 10^{-5} \, X_{k20}^{-1}(e) k^{-3/2} \left( \frac{\Porb}{1 \, {\rm day}} \right)^{7/2}\, .
\label{eq:Csat}
\ee
%{\y this is a factor of $3.8$ higher than your original. all  changes below follow from here} {\jj ok have taken out blue text}
%To reach such an amplitude, the g-mode can be forced very much off resonance,  $|\om-\om_{km}| \gg \cg_{\rm diff}$.

%\subsection{Effective dissipation including nonlinear damping}
%\label{app:nonlin_damp}

 We speculate that the following dynamical picture determines the effective dissipation.  Without nonlinear breaking, the g-mode is pumped by tidal forcing to reach its maximum amplitude ($C^{\rm max}$) after a time $\Delta t  \sim \gamma_{\rm diff}^{-1}$. With its damping dominated by  linear dissipation, the rate of energy dissipation is ${\dot E}^{\rm max}_{\rm orb} =  \epsilon |C^{\rm max}|^2/\Delta t$.

However, if non-linearity sets in earlier at a smaller amplitude $C^{\rm sat}$, the effective damping is reduced. The wave reaches this amplitude within a shorter time, of order $\Delta t' \sim \gamma_{\rm diff}^{-1} |C^{\rm sat}|/|C^{\rm max}|$.
And within a few radial propagation times ($t_{\rm group} \sim \Rs [\pd \om/\pd \kr]^{-1} \sim n/\omega \ll \Delta t'$), most wave flux carried to the centre is converted to heat  by overturning. This  is consistent with results from \citet{BarkerOgilvie(2010),Su} where they found 
that wave breaking reduces the amplitude of the outgoing  wave to be a fraction of the ingoing wave (of order $20\%-50\%$).
%ingoing wave $|A^{\rm in}|$ is expected to be much larger than the amplitude of the outgoing wave $|A^{\rm out}|$ (simulations find $|A^{\rm out}| \approx 0.2 - 0.5 |A^{\rm in}|$, e.g. 
%Because the radial communication time is typically much shorter than the time the mode takes to reach its maximum amplitude ($\Dg t_{\rm max} \gg t_{\rm group}$), the mode amplitude saturates at $|C^{\rm sat}|$.  causing the amplitude of the resonantly-excited standing mode to quickly decrease from $|C^{\rm sat}|$ to a minimum value $|C^{\rm min}| \ll |C^{\rm sat}|$.  Using equation~\eqref{eq:dotc_a_max}, we can calculate the time it takes for the mode amplitude to ``cycle'' from the minimum amplitude $|C^{\rm min}|$ to its saturated value $|C^{\rm sat}|$:
%\be
%\Dg t_{\rm cyc} \simeq \frac{|C^{\rm sat}| - |C^{\rm min}|}{\omega U_{k \ell m}} \approx \Dg_{\rm sat}^{-1} \om^{-1} = \cg^{-1} \frac{|C^{\rm sat}|}{|C^{\rm max}|}.
%\label{eq:Dt_cyc}
%\ee
%We used equation~\eqref{eq:dotc_a_max} in~\eqref{eq:Dt_cyc}, since the timescale which the mode amplitude decreases, then grows, is much smaller than the mode damping timescale ($t_{\rm group},\Dg t_{\rm cyc} \ll \cg^{-1}$).  This yields an effective dissipation rate $\dot E_{\rm orb}^{\rm sat} \approx 2 \epsilon |C^{\rm sat}|^2/\Dg t_{\rm cyc}$.  Comparing to $\dot E_{\rm orb}^{\rm max}$ using equation~\eqref{eq:Csat}, we find
The effective rate of energy  loss is then $\dot E_{\rm orb}^{\rm sat} \sim \epsilon |C^{\rm sat}|^2/\Delta t'$ and is reduced from the linear form as
\be
\frac{\dot E_{\rm orb}^{\rm sat}}{\dot E_{\rm orb}^{\rm max}} \approx \frac{|C^{\rm sat}|}{|C^{\rm max}|} 
%= \frac{1}{\Dg_{\rm sat}} \frac{\cg}{\om} 
\sim  4.2 \times 10^{-5} \, X_{k20}^{-1}(e) k^{-3/2} \left( \frac{\Porb}{1 \, {\rm day}} \right)^{7/2}.
\label{eq:dEsatdt}
\ee

Following the procedure that leads to equation~\eqref{eq:e_min}, 
we estimate a new binary period out to which (saturated) resonance locking can be effective in circularizing the binary. Setting again $t_{\rm  evol} \sim 10^7 \ {\rm yrs}$, but now increasing $t_{\rm orb}$  in equation~\eqref{eq:t_aklm^min} by  the inverse of the above ratio, we find that $t_{\rm orb} \leq t_{\rm evol}$ when 
\be
P_{\rm orb} \leq P_{\rm sat} 
\sim 3.6 \, {\rm days}\, \left({e\over {0.4}}\right)^{6/41}\, .
%{e_{\rm sat} \sim 6.2 \times 10^{-5} \left( \frac{\Porb}{1 \, {\rm day}} \right)^{41/6}\, .}
\label{eq:e_sat2}
\ee
Compared to the criterion for linear resonance locking to be effective (eq.~[\ref{eq:e_min}]), saturated resonance  locking can only circularize  binaries out to a shorter orbital period.}


\begin{thebibliography}{}

\bibitem[Barker(2020)]{Barker(2020)} Barker, A.~J.\ 2020, \mnras, 498, 2270. doi:10.1093/mnras/staa2405


\bibitem[Barker(2011)]{Barker(2011)} Barker, A.~J.\ 2011, \mnras, 414, 1365
% 3D simulations of IGW breaking in solar-type stars

\bibitem[Barker \& Ogilvie(2011)]{BarkerOgilvie(2011)} Barker, A.~J., \& Ogilvie, G.~I.\ 2011, \mnras, 417, 745
% Linear Stability of non-linearly breaking IGWs

\bibitem[Barker \& Ogilvie(2010)]{BarkerOgilvie(2010)} Barker, A.~J., \& Ogilvie, G.~I.\ 2010, \mnras, 404, 1849
% Paper looking at IGW non-linearly breaking it's wave near the center of a solar-type star (with a radiative core)

\bibitem[Barker \& Ogilvie(2009)]{BarkerOgilvie(2009)} Barker, A.~J., \& Ogilvie, G.~I.\ 2009, \mnras, 395, 2268
% Nice paper, where get magnetic spin-down equation from (in modern notation)

\bibitem[Batygin \& Adams(2013)]{BatyginAdams(2013)} Batygin, K. \& Adams, F.~C.\ 2013, \apj, 778, 169
% Nice paper get indices for constants from polytropes

\bibitem[Bildsten et al.(1996)]{Bildsten} Bildsten, L., Ushomirsky, G., \& Cutler, C.\ 1996, \apj, 460, 827
% traditional approximation

\bibitem[Bluhm et al.(2016)]{Bluhm(2016)} Bluhm, P., Jones, M.~I., Vanzi, L., et al.\ 2016, \aap, 593, A133
% Paper on circularization in binaries with giants

\bibitem[Bouvier(2013)]{Bouvier(2013)} Bouvier, J.\ 2013, EAS Publications Series, 143
% Review paper on rotational evolution of (single) stars

\bibitem[Burkart et al.(2014)]{Burkart(2014)} Burkart, J., Quataert, E., \& Arras, P.\ 2014, \mnras, 443, 2957
% Technical paper on resonance locking get a lot of results from, investigate in detail stability of resonance locks

\bibitem[Burkart et al.(2012)]{Burkart(2012)} Burkart, J., Quataert, E., Arras, P., et al.\ 2012, \mnras, 421, 983
% Paper looking at resonance locking in Kepler System, type of resonance locking involving stellar spin

\bibitem[Chandrasekhar(1939)]{Chandrasekhar(1939)} Chandrasekhar, S.\ 1939, Chicago, Ill., The University of Chicago press [1939]
% Review paper on stellar structure, includes a lot of stuff on polytropes

\bibitem[Chapman \& Lindzen(1970)]{Chapman} Chapman, S., \& Lindzen, R.\ 1970, Dordrecht: Reidel
%reference from geophysics for traditional approximation

\bibitem[Christensen-Dalsgaard(2014)]{Christensen-Dalsgaard(2014)} Christensen-Dalsgaard, J.\ 2014, Lecture Notes on
Stellar Oscillations, Fifth Edition, http://astro.phys.au.dk/jcd/oscilnotes/\\Lecture\_Notes\_on\_Stellar\_Oscillations.pdf
% Review of g-modes, use a lot

\bibitem[Cowling(1941)]{Cowling(1941)} Cowling, T.~G.\ 1941, \mnras, 101, 367


\bibitem[Darwin(1879)]{Darwin(1879)} Darwin, G. H.\ 1879, Phil. Trans. Roy. Soc., 170, 1
% Classic paper deriving equilibrium tides

\bibitem[Duquennoy \& Mayor(1991)]{DuquennoyMayor(1991)} Duquennoy, A., \& Mayor, M.\ 1991, \aap, 500, 337
% One of first papers to mesure circularization of binary star populations

\bibitem[Duguid et al.(2020)]{Duguid(2020)} Duguid, C.~D., Barker, A.~J., \& Jones, C.~A.\ 2020, \mnras, 491, 923


\bibitem[Friedman \& Schutz(1978a)]{FriedmanSchutz(1978a)} Friedman, J.~L., \& Schutz, B.~F.\ 1978a, \apj, 221, 937
% First of two papers, deriving among other things the energy and angular momentum of a stellar oscillation mode (including Corioulus forces)

\bibitem[Friedman \& Schutz(1978b)]{FriedmanSchutz(1978b)} Friedman, J.~L., \& Schutz, B.~F.\ 1978b, \apj, 222, 281
% Second of two papers, showing what happens when you include dissipation in the system

\bibitem[\protect\citeauthoryear{Fuller et al.}{2017}]{Fuller+(2017)} Fuller J., Hambleton K., Shporer A., Isaacson H., Thompson S., 2017, MNRAS, 472, L25


\bibitem[Fuller(2017)]{Fuller(2017)} Fuller, J.\ 2017, \mnras, 472, 1538
% Paper deriving resonance locking formalism used

\bibitem[Fuller et al.(2016)]{Fuller(2016)} Fuller, J., Luan, J., \& Quataert, E.\ 2016, \mnras, 458, 3867
% Paper first discussing "new" type of resonance locking

\bibitem[Fuller \& Lai(2012a)]{FullerLai(2012a)} Fuller, J., \& Lai, D.\ 2012a, \mnras, 420, 3126
% Paper looking at stellar-spin resonance locking

\bibitem[Fuller \& Lai(2012b)]{FullerLai(2012b)} Fuller, J., \& Lai, D.\ 2012b, \mnras, 421, 426
% Complicated tidal response in WDs in binaries with G-modes, use a lot of technical review results in this work

\bibitem[Goldreich \& Nicholson(1977)]{GoldreichNicholson(1977)} Goldreich, P., \& Nicholson, P.~D.\ 1977, \icarus, 30, 301
% Nice paper estimating turbulent viscosity factor

\bibitem[Goldreich \& Wu(1999)]{GoldreichWu(1999)} Goldreich, P., \& Wu, Y.\ 1999, \apj, 511, 904
% Classic g-mode paper, where get scaling for Eularian density perturbation in convective zone from

\bibitem[Goodman \& Dickson(1998)]{GoodmanDickson(1998)} Goodman, J., \& Dickson, E.~S.\ 1998, \apj, 507, 938
% First paper to estimate non-linear wave breaking in stellar binaries

\bibitem[Goodman \& Oh(1997)]{GoodmanOh(1997)} Goodman, J., \& Oh, S.~P.\ 1997, \apj, 486, 403

\bibitem[Gough(1980)]{Gough(1980)} Gough, D.\ 1980, The Ancient Sun: Fossil Record in the Earth, Moon and Meteorites, 533

\bibitem[Hut(1981)]{Hut(1981)} Hut, P.\ 1981, \aap, 99, 126
% Paper deriving constant time-lag formalism

\bibitem[Ivanov \& Papaloizou(2007)]{IvanovPapaloizou(2007)} Ivanov, P.~B. \& Papaloizou, J.~C.~B.\ 2007, \mnras, 376, 682

\bibitem[Kratter et al.(2010)]{Kratter(2010)} Kratter, K.~M., Matzner, C.~D., Krumholz, M.~R., et al.\ 2010, \apj, 708, 1585


\bibitem[Kraft(1967)]{Kraft(1967)} Kraft, R.~P.\ 1967, \apj, 150, 551
% Paper discovering the Kraft break

\bibitem[Kumar \& Goodman(1996)]{KumarGoodman(1996)} Kumar, P., \& Goodman, J.\ 1996, \apj, 466, 946
% First paper to look at amplitude saturation by parametric instability

\bibitem[Lai(1997)]{Lai(1997)} Lai, D.\ 1997, \apj, 490, 847
% Nice paper using traditional approximation to calculate how rotation modified g-mode frequencies

\bibitem[Lai(1994)]{Lai(1994)} Lai, D.\ 1994, \mnras, 270, 611
% Nice paper on resonant dynamical tides, use a lot of technical results here

\bibitem[Lai \& Wu(2006)]{LaiWu(2006)} Lai, D., \& Wu, Y.\ 2006, \prd, 74, 024007
% Calculation of resonant excitation of tidal pertubations in coalescing NS binaries

\bibitem[Lainey et al.(2020)]{Lainey(2020)} Lainey, V., Casajus, L.~G., Fuller, J., et al.\ 2020, Nature Astronomy, doi:10.1038/s41550-020-1120-5


\bibitem[Latham et al.(1992)]{Latham92} Latham, D.~W., Mathieu, R.~D., Milone, A.~A.~E., et al.\ 1992, Evolutionary Processes in Interacting Binary Stars, 151, 471

\bibitem[Latham et al.(2002)]{Latham(2002)} Latham, D.~W., Stefanik, R.~P., Torres, G., et al.\ 2002, \aj, 124, 1144
% One of first studies to measure binary circularization of field stars

\bibitem[Luan et al.(2018)]{Luan(2018)} Luan, J., Fuller, J., \& Quataert, E.\ 2018, \mnras, 473, 5002
% Argue Cassini can differentiate between inertial wave dissipation vs. resonant resonance locking tidal migratin of Saturn's satellites.  Compare my damping rate estimates to those constained within this work.

\bibitem[Lurie et al.(2017)]{Lurie(2017)} Lurie, J.~C., Vyhmeister, K., Hawley, S.~L., et al.\ 2017, \aj, 154, 250
% Recent paper calculating rotatinal periods for a number of binary stars, show a large number of stars rotate non-synchronously

\bibitem[Lin \& Ogilvie(2017)]{LinOgilvie(2017)} Lin, Y., \& Ogilvie, G.~I.\ 2017, \mnras, 468, 1387
% Obliquity tides in solar-type stars causing tidal evolution of HJs

\bibitem[Mardling(1995a)]{Mardling(1995a)} Mardling, R.~A.\ 1995a, \apj, 450, 722
% First of two papers on tidal excitation of f-modes in highly eccentric binaries, show when f-modes are chaotically excited, don't assume all orbital energy is damped away immediately

\bibitem[Mardling(1995b)]{Mardling(1995b)} Mardling, R.~A.\ 1995b, \apj, 450, 732
% Second of two papers on tidal excitation of f-modes in highly eccentric binaries, look at how orbits are circularized

\bibitem[Marilli et al.(2007)]{Marilli(2007)} Marilli, E., Frasca, A., Covino, E., et al.\ 2007, \aap, 463, 1081
% Paper looking at rotation periods of PMS binaries, find a lot of binaries don't rotate pseudo-synchronously

\bibitem[Mathieu et al.(2004)]{Mathieu2004} Mathieu, R.~D., Meibom, S., \& Dolan, C.~J.\ 2004, \apjl, 602, L121

\bibitem[Mayor \& Mermilliod(1984)]{MayorMermilliod(1984)} Mayor, M., \& Mermilliod, J.~C.\ 1984, Observational Tests of the Stellar Evolution Theory, 411
% One of first studies to measure circularizatin in binary population

\bibitem[Mazeh(2008)]{Mazeh(2008)} Mazeh, T.\ 2008, EAS Publications Series, 1
% Very nice observational review paper, point out a lot of difficulties reconciling synchronization predictins of tidal theories with observations

\bibitem[Meibom et al.(2006)]{Meibom(2006)} Meibom, S., Mathieu, R.~D., \& Stassun, K.~G.\ 2006, \apj, 653, 621
% Key paper showing many short-period binaries in young stellar systems don't rotate synchronously

\bibitem[Meibom \& Mathieu(2005)]{MeibomMathieu(2005)} Meibom, S., \& Mathieu, R.~D.\ 2005, \apj, 620, 970
% Classic observational paper showing circularization period increases with age


\bibitem[Milliman et al.(2014)]{Milliman(2014)} Milliman, K.~E., Mathieu, R.~D., Geller, A.~M., et al.\ 2014, \aj, 148, 38
% Newer observational survey looking at older cluster NGC 6819 (2.5 Gyr), find P_circ ~ 6 days.

\bibitem[Moe \& Kratter(2018)]{MoeKratter(2018)} Moe, M., \& Kratter, K.~M.\ 2018, \apj, 854, 44
% Try to form binaries during PMS phase with companions.  Among other things, use dynamical tide formalism of Press & Teukolsky (1977).

\bibitem[Ogilvie(2014)]{Ogilvie(2014)} Ogilvie, G.~I.\ 2014, \araa, 52, 171
%Tides review

\bibitem[Ogilvie(2013)]{Ogilvie(2013)} Ogilvie, G.~I.\ 2013, \mnras, 429, 613
% Nice analytic work showing among other things how inertial waves depend on the star's radiative core size.

\bibitem[Ogilvie \& Lesur(2012)]{OgilvieLesur(2012)} Ogilvie, G.~I., \& Lesur, G.\ 2012, \mnras, 422, 1975
% Simulations of turbulent interactions between convective eddies and tidal disturbance, find quadratic viscosity supression

\bibitem[Ogilvie \& Lin(2007)]{OgilvieLin(2007)} Ogilvie, G.~I., \& Lin, D.~N.~C.\ 2007, \apj, 661, 1180
% Paper looking at inertial waves in solar-type stars

\bibitem[Ogilvie \& Lin(2004)]{OgilvieLin(2004)} Ogilvie, G.~I., \& Lin, D.~N.~C.\ 2004, \apj, 610, 477
% Paper showing large amount of tidal dissipation can be generated inside Jupiter with inertial waves

\bibitem[Paxton et al.(2019)]{Paxton(2019)} Paxton, B., Smolec, R., Schwab, J., et al.\ 2019, \apjs, 243, 10
% MESA paper V

\bibitem[Paxton et al.(2018)]{Paxton(2018)} Paxton, B., Schwab, J., Bauer, E.~B., et al.\ 2018, \apjs, 234, 34
% MESA paper IV

\bibitem[Paxton et al.(2015)]{Paxton(2015)} Paxton, B., Marchant, P., Schwab, J., et al.\ 2015, \apjs, 220, 15
% MESA paper III

\bibitem[Paxton et al.(2013)]{Paxton(2013)} Paxton, B., Cantiello, M., Arras, P., et al.\ 2013, \apjs, 208, 4
% MESA paper II

\bibitem[Paxton et al.(2011)]{Paxton(2011)} Paxton, B., Bildsten, L., Dotter, A., et al.\ 2011, \apjs, 192, 3
% Paper going over how MESA works

\bibitem[Penev et al.(2007)]{Penev(2007)} Penev, K., Sasselov, D., Robinson, F., et al.\ 2007, \apj, 655, 1166
% Paper conducting 3D simulations of turbulent interactions of tidal disturbence with convective eddies in a star, find viscosity supression closer to Zahn's original estimate


\bibitem[Press et al.(2002)]{Press(2002)} Press, W.~H., Teukolsky, S.~A., Vetterling, W.~T., et al.\ 2002, Numerical recipes in C++ : the art of scientific computing.
% Numerical Recipies

\bibitem[Press \& Teukolsky(1977)]{PressTeukolsky(1977)} Press, W.~H., \& Teukolsky, S.~A.\ 1977, \apj, 213, 183
% Classic paper calculating the tidal excitation of f-modes by periastron passages from a perturber

%\bibitem[Price-Whelan \& Goodman(2018)]{Price-WhelanGoodman(2018)} Price-Whelan, A.~M., \& Goodman, J.\ 2018, \apj, 867, 5
% Observatinal/Theoretical paper on binary star circuarization for red giants
\bibitem[Price-Whelan \& Goodman(2018)]{Price2} Price-Whelan, A.~M., \& Goodman, J.\ 2018, \apj, 867, 5

\bibitem[Price-Whelan et al.(2020)]{PriceWhelan(2020)} Price-Whelan, A.~M., Hogg, D.~W., Rix, H.-W., et al.\ 2020, \apj, 895, 2

% First papers on resonance locking

\bibitem[Savonije \& Papaloizou(1983)]{SavonijePapaloizou(1983)} Savonije, G.~J. \& Papaloizou, J.~C.~B.\ 1983, \mnras, 203, 581

\bibitem[Savonije \& Papaloizou(1984)]{SavonijePapaloizou(1984)} Savonije, G.~J. \& Papaloizou, J.~C.~B.\ 1984, \mnras, 207, 685


\bibitem[Savonije \& Papaloizou(1997)]{SavonijePapaloizou(1997)} Savonije, G.~J., \& Papaloizou, J.~C.~B.\ 1997, \mnras, 291, 633
% Detailed look at g and i-modes in inertial regime, take Corioulus force into account fully

\bibitem[Savonije \& Witte(2002)]{SavonijeWitte(2002)} Savonije, G.~J., \& Witte, M.~G.\ 2002, \aap, 386, 211
% Resonance locking consisting of 1 Msun host star with companion, have detailed look at resonance locking

\bibitem[Schenk et al.(2002)]{Schenk(2002)} Schenk, A.~K., Arras, P., Flanagan, {\'E}. {\'E}., et al.\ 2002, \prd, 65, 024001
% Derived mode excitation equaiton used so often

\bibitem[Skumanich(1972)]{Skumanich(1972)} Skumanich, A.\ 1972, \apj, 171, 565
% Skumanich Law paper

\bibitem[Su et al.(2020)]{Su} Su, Y., Lecoanet, D., \& Lai, D.\ 2020, \mnras, 495, 1239

\bibitem[Terquem et al.(1998)]{Terquem(1998)} Terquem, C., Papaloizou, J.~C.~B., Nelson, R.~P., et al.\ 1998, \apj, 502, 788
% Nice technical paper use a lot of formalism from, on resonant g-modes in solar-type stars

\bibitem[Tokovinin \& Moe(2020)]{TokovininMoe(2020)} Tokovinin, A. \& Moe, M.\ 2020, \mnras, 491, 5158


\bibitem[Triaud et al.(2017)]{Triaud(2017)} Triaud, A.~H.~M.~J., Martin, D.~V., S{\'e}gransan, D., et al.\ 2017, \aap, 608, A129
% Nice observational paper on the circularization of binary star populations, point out a lot of puzzles missed by other works, use a lot

\bibitem[Unno et al.(1979)]{Unno(1979)} Unno, W., Osaki, Y., Ando, H., et al.\ 1979, Tokyo: University of Tokyo Press
% Classic textbook on tides everyone cites

\bibitem[Verbunt \& Phinney(1995)]{Verbunt} Verbunt, F., \& Phinney, E.~S.\ 1995, \aap, 296, 709
% Tides in post-MS stars

\bibitem[Verbunt \& Zwaan(1981)]{VerbuntZwaan(1981)} Verbunt, F., \& Zwaan, C.\ 1981, \aap, 100, L7
% Magnetic breaking paper, one of first to give relation between alpha_star and stellar properties

\bibitem[Vick \& Lai(2020)]{VickLai(2020)} Vick, M. \& Lai, D.\ 2020, \mnras, 496, 3767
% Equilibrium tidal dissipation at high eccentricities

\bibitem[Vick et al.(2019)]{Vick(2019)} Vick, M., Lai, D., \& Anderson, K.~R.\ 2019, \mnras, 484, 5645
% Nice paper showing how eccentric HJs are circularized

\bibitem[Vick \& Lai(2018)]{VickLai(2018)} Vick, M., \& Lai, D.\ 2018, \mnras, 476, 482
% Paper self-consistently circularizing HJs with f-modes

\bibitem[Vidal \& Barker(2020)]{VidalBarker(2020)} Vidal, J., \& Barker, A.~J.\ 2020, \apjl, 888, L31

\bibitem[Van Eylen et al.(2016)]{VanEylen(2016)} Van Eylen, V., Winn, J.~N., \& Albrecht, S.\ 2016, \apj, 824, 15
% Interesting paper measuring eccentricities of eclipsing binaries, find tentative dependence on host star's effective temperature/mass

\bibitem[Weinberg et al.(2012)]{Weinberg(2012)} Weinberg, N.~N., Arras, P., Quataert, E., et al.\ 2012, \apj, 751, 136
% Non-linear dynamical tides, get a lot of technical results from here

\bibitem[Windemuth et al.(2019)]{Windemuth(2019)} Windemuth, D., Agol, E., Ali, A., et al.\ 2019, \mnras, 2075
% Nice recent observational paper on binary star populations obtained from just eclipsing binaries using photometry.

\bibitem[Witte \& Savonije(2002)]{WitteSavonije(2002)} Witte, M.~G., \& Savonije, G.~J.\ 2002, \aap, 386, 222
% Resonance locking in solar-type stellar binaries and planetary systems

\bibitem[Witte \& Savonije(2001)]{WitteSavonije(2001)} Witte, M.~G., \& Savonije, G.~J.\ 2001, \aap, 366, 840
% Look at resonance locking in system of two 10 Msun binary stars, with rotational evolution taken into account fully

\bibitem[Witte \& Savonije(1999a)]{WitteSavonije(1999a)} Witte, M.~G., \& Savonije, G.~J.\ 1999, \aap, 341, 842
% Detailed look at g and r-mode resonances in 10 Msun host star in binary, take Corioulus force into account fully

\bibitem[Witte \& Savonije(1999b)]{WitteSavonije(1999b)} Witte, M.~G., \& Savonije, G.~J.\ 1999, \aap, 350, 129
% Resonance locking in system with 10 Msun host star with 1.4 Msun companion, take Corioulus force into account fully

\bibitem[Wu(2018)]{Wu(2018)} Wu, Y.\ 2018, \aj, 155, 118
% Yanqin's paper on how to circularize highly eccentric HJs with excitation of f-modes during periastron passages

\bibitem[Wu(2005a)]{Wu(2005a)} Wu, Y.\ 2005a, \apj, 635, 674
% First of Yanqin's papers to look at inertial waves, focused on mathematical properties

\bibitem[Wu(2005b)]{Wu(2005b)} Wu, Y.\ 2005b, \apj, 635, 688
% Second of Yanqin's papers to look at inertial waves, applied previous paper to see if Jupiter's low Q could be explained

\bibitem[Wu \& Lithwick(2019)]{WuLithwick19} Wu, Y., \& Lithwick, Y.\ 2019, \apj, 881, 142
% damping of fundamental modes in fully convective bodies, including wave leakage


\bibitem[Zahn(2008)]{Zahn(2008)} Zahn, J.-P.\ 2008, EAS Publications Series, 67
% Nice review by Zahn on all of his tidal theories

\bibitem[Zahn(1989)]{Zahn(1989)} Zahn, J.-P.\ 1989a, \aap, 220, 112
% Paper looking again at the turbulent viscosity coefficient with equilirbium tides, also reviews equilibrium tide concepts

\bibitem[Zahn(1978)]{Zahn(1978)} Zahn, J.-R.\ 1978, \aap, 67, 162
% Publication of errors in previous paper

\bibitem[Zahn(1977)]{Zahn(1977)} Zahn, J.-P.\ 1977, \aap, 500, 121
% Classic equilibrium tides paper, derive equations used

\bibitem[Zahn(1975)]{Zahn(1975)} Zahn, J.-P.\ 1975, \aap, 41, 329
% Classic dyanmical tides paper

\bibitem[Zahn(1966)]{Zahn(1966)} Zahn, J.~P.\ 1966, Annales d'Astrophysique, 29, 489
% Paper which derives linear viscosity supression factor

\bibitem[Zahn \& Bouchet(1989)]{ZahnBouchet(1989)} Zahn, J.-P., \& Bouchet, L.\ 1989, \aap, 223, 112
% Paper showing equilibrium tides during PMS can circularize binaries out to ~8 days.

\bibitem[Zanazzi \& Wu(2020)]{ZanazziWu(2020)} Zanazzi, J.-J., \& Wu, Y.\ 2020,  in prep

\end{thebibliography}
\end{document}